\documentclass[12pt,preprint,number,sort&compress,lontitle,times]{elsarticle}
\usepackage{natbib}
\usepackage{graphicx}
\usepackage{epsfig}
\usepackage{lineno}
\usepackage{multirow}
\usepackage{fourier}
\usepackage{color}
\usepackage{amsmath}
\usepackage{pgf}
\usepackage [english]{babel}

\RequirePackage{wrapfig}
\RequirePackage{rotating}
\RequirePackage{shadow}

\usepackage{amssymb}

\newcommand{\dif}{\mathrm{d}}
\newcommand{\nn}{\mathbf{n}}
\newcommand{\nnl}{\tilde{\mathbf{n}}}
\newcommand{\uu}{\mathbf{u}}

\journal{Astroparticle Physics}

\begin{document}

\begin{frontmatter}



\title{Measurements and implications of cosmic ray anisotropies from TeV to trans-EeV energies}

\author{O. Deligny}
\address{Institut de Physique Nucl\'eaire, CNRS-IN2P3, Univ. Paris-Sud, Universit\'e Paris-Saclay, 91406 Orsay Cedex, France}

\begin{abstract}
Important observational results have been recently reported on the angular distributions of cosmic rays at all energies, calling into question the perception of cosmic rays a decade ago. These results together with their in-progress interpretations are summarised in this review paper, covering both large-scale and small-scale anisotropies from TeV energies to the highest ones. While the magnetic field in the Galaxy has long been considered as an external data imprinting a quasi-random walk to particles and thus shaping the angular distributions of Galactic cosmic rays through the induced average density gradient, the information encompassed in the angular distributions in the TeV--PeV energy range appear today as a promising tool to infer some properties of the local magnetic field environments. At the highest energies, the extragalactic origin of the particles has been recently determined observationally. While no discrete source of ultrahigh-energy cosmic rays has been identified so far, the noose is tightening around nearby extragalactic objects, and some prospects are discussed. 
\end{abstract}


\end{frontmatter}


\section{Introduction}

The origin of cosmic rays (CRs) remains an enduring question in astrophysics. A time-honoured paradigm is that sources of the bulk of these particles could be supernova remnants in the Galaxy. This is mainly because the intensity of cosmic rays observed on Earth can be produced by making use of $\simeq 10$\% of the energetics of these astrophysical objects~\cite{Zwicky1934}, and because the diffusive shock acceleration has been shown to be a mechanism able to convert kinetic energy of the expanding supernova blast wave into accelerated particles~\cite{Bell1978a,Bell1978b,Blanford1987}. However, alternative scenarios with sources related to transient events connected to the death of short-lived massive stars have also been put forward, such as in~\cite{LoebWaxman2006} for instance. 

The arrival directions of these particles are highly isotropic. This is expected from the propagation of charged particles in the interstellar medium where the directions of the particle momenta are randomized over time by the effective scattering in the encountered magnetic fields. Despite the scrambling action of these fields, searches for small anisotropy contrasts at large scales have been scrutinised over many decades as reviewed for instance in~\cite{Greisen1962,LinsleyWatson1977,DiSciascio2013}. During the past decade, multiple observatories located in both hemispheres have reported significant observations of large-scale and small-scale anisotropies in the TeV--PeV energy band. These results have challenged the long-standing description of CR propagation in terms of a typical spatial diffusion process from stationary sources located preferentially in the disk of the Galaxy, leading to a dipole moment only in the direction of the CR gradient and with an amplitude steadily increasing with the energy. The current picture is much more complex and elaborate~\cite{AhlersMertsch2016}.

Galactic CRs are thought to be retained by the Galactic magnetic field as long as the size of their Larmor orbit diameter is much less than the thickness of the Galactic disk. Since the strength of the magnetic field is of the order of microgauss, Galactic CRs might be confined in the Galactic disk up to energies of $100~Z~$PeV, with $Z$ the charge of the particles. Once particles are not confined anymore, the time they spent in the disk tends to the constant free escape time due to the direct escape from the Galaxy. The observed intensity should then be naturally much stronger towards the disk compared to other directions. Due to their high level of isotropy, CRs with energies in excess of $\simeq 1~$EeV have thus long been thought to be of extragalactic origin. In addition, even in the presence of efficient magnetic field amplification at the supernova remnant shock, accelerating intermediate or even heavy nuclei at EeV energies is very challenging~\cite{Blasi2013}. On the other hand, Hillas pointed out the plausible classes of astrophysical objects in which Fermi acceleration could perform up to 100~EeV or so through the essential requirement that the particle Larmor radius must be smaller than the size scale of the acceleration region~\cite{Hillas1984}. Thanks to the jump in statistics as well as to the improved instrumentation experienced in the past decade with the Pierre Auger Observatory, CRs with energies in excess of $\simeq 8~$EeV have indeed been recently observed to originate from extragalactic galaxies~\cite{AugerScience2017}. The exact sources remain, however, unknown since the first detection of a particle with energy in excess of 100~EeV by Linsley at the Volcano Ranch in 1963~\cite{Linsley1963a}. 

The intervening magnetic fields in extragalactic space and in the Galaxy are uncertain, although the understanding of the magnetic fields in the Milky Way has developed over many decades and has allowed for quantitatively-constrained models to emerge~\cite{JanssonFarrar2012}. The uncertainties remain however too large to firmly predict the deflections that ultra-high energy cosmic rays (UHECRs) should undergo from each line of sight outside from the Galaxy. The effect of the turbulent component of the field is particularly uncertain~\cite{FarrarSutherland2017}. The expected order of magnitude for the deflections is thought to behave as $\simeq 3^\circ Z(E/100~\mathrm{EeV})^{-1}$. With such an order of magnitude, magnetic deflections could be small enough to allow for mirroring to some extent the distribution of sources in the sky. Moreover, the horizon of the highest energy particles ($\gtrsim 60~$EeV) is limited as compared to that of particles of lower energies, because the thresholds are then reached of interactions with background radiations filling the Universe and leading to large energy losses. This is the ``GZK effect''~\cite{GZK}, which allows that only the foreground sources are expected to populate the observed sky maps at these energies. But the small intensity combined to the potential absence of particles with low electric charge at these energies still prevents such a ``charged-particle astronomy'' with current data. 

All these topics are addressed in detail in this review under the prism essentially of the results obtained during the last decade. This review is meant to be an introduction to the main analysis techniques as well as to the formalisms needed to interpret the results. In this sense, and to allow an introduction to the latest theoretical advances, many classical results are developed from the first principles by reviewing the main steps to derive them. After introducing the basic quantities of interest to decipher the underlying angular distributions of CRs from ground-based experiment data in~\S~\ref{sec:observations}, the guiding thread of this review is to characterize anisotropies from large to small scales, by presenting the experimental results and their interpretations as a function of energy. Thus, harmonic analysis methods in right ascension, traditionally focused on the first harmonic, are first approached in~\S~\ref{sec:harmonic} and their astrophysical consequences discussed in~\S~\ref{sec:astro}. The 3D reconstruction of the intensity on the sphere and the characterization of the anisotropies in terms of power spectrum are the subject of the next two sections, reviewing the analysis techniques in~\S~\ref{sec:3dreco} and the interpretations in~\S~\ref{sec:highorder}. Finally, \S~\ref{sec:xgal} is devoted to the highest energies, because of the specific techniques, which can in particular involve external information such as catalogs of extragalactic astrophysical objects.

\section{Sky surveys from ground-based observatories}
\label{sec:observations}

The aim of CR anisotropy studies is to reconstruct the intensity from each direction of the sky. From the all-directional flux of particles $I_0$, the intensity in each celestial direction, $I(\nn)$, is defined as the overall flux per steradian weighted by a directional-dependent factor characterising the anisotropy:
\begin{eqnarray}
\label{eqn:intensity}
I(\nn)=\frac{I_0}{4\pi}\left(1+\delta I\right(\nn)).
\end{eqnarray}

\begin{wrapfigure}{L}{8. cm}
{\includegraphics[width=0.5\textwidth]{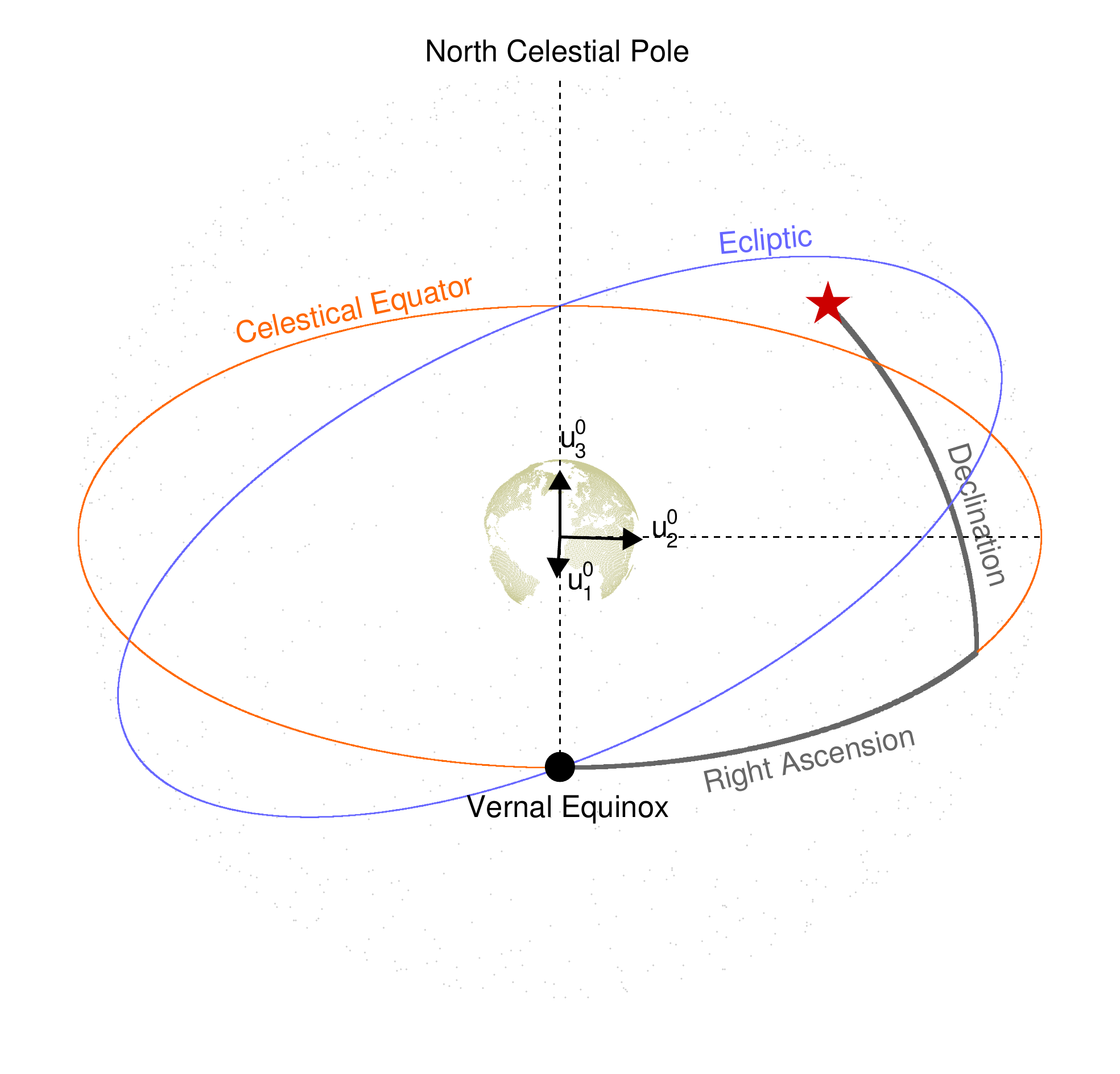}}
\caption{Equatorial coordinate system.}
\label{fig:eq_coord}
\end{wrapfigure}
Ground-based observatories have access, however, to a limited part of the sky in a highly non-uniform way. This section is dedicated to introduce the basic ideas and techniques that allow for estimating the directional exposure of any experiment. 

\subsection{Coordinate systems}
\label{subsec:coord}

Let us first define the notations used throughout this review for the relevant coordinate systems and review the rules of transformation between local and equatorial coordinates. Equatorial coordinates are the most natural ones to characterise the directional data of ground-based observatories. The projection on the celestial sphere of the equator of the Earth is used as a reference plane. This projection is the celestial equator, which divides the sky into two hemispheres, each of which has as its reference axis the projection of a terrestrial pole perpendicular to the celestial equator. From this division, the system makes it possible to establish two angular coordinates: the right ascension $\alpha$ and the declination $\delta$, which are longitude-like and latitude-like coordinates. Conventionally, right ascension is measured eastward along the celestial equator from the vernal equinox to the hour circle\footnote{The hour circle is the great circle through the object and the celestial poles of the Earth.} of the point in question. Declination is measured perpendicularly from the celestial equator to the observed celestial object, positive for objects in the northern hemisphere and negative for those in the southern hemisphere. The corresponding right-handed rectangular basis of unit vectors are denoted throughout this review as $(\uu_1^0,\uu_2^0,\uu_3^0)$, so that any unitary vector $\nn$ can be expressed as $\nn=\cos{\delta}\cos{\alpha}~\uu_1^0+\cos{\delta}\sin{\alpha}~\uu_2^0+\sin{\delta}~\uu_3^0$. A sketch of the considered geometry is shown in figure~\ref{fig:eq_coord}.

\begin{wrapfigure}{L}{8. cm}
{\includegraphics[width=0.5\textwidth]{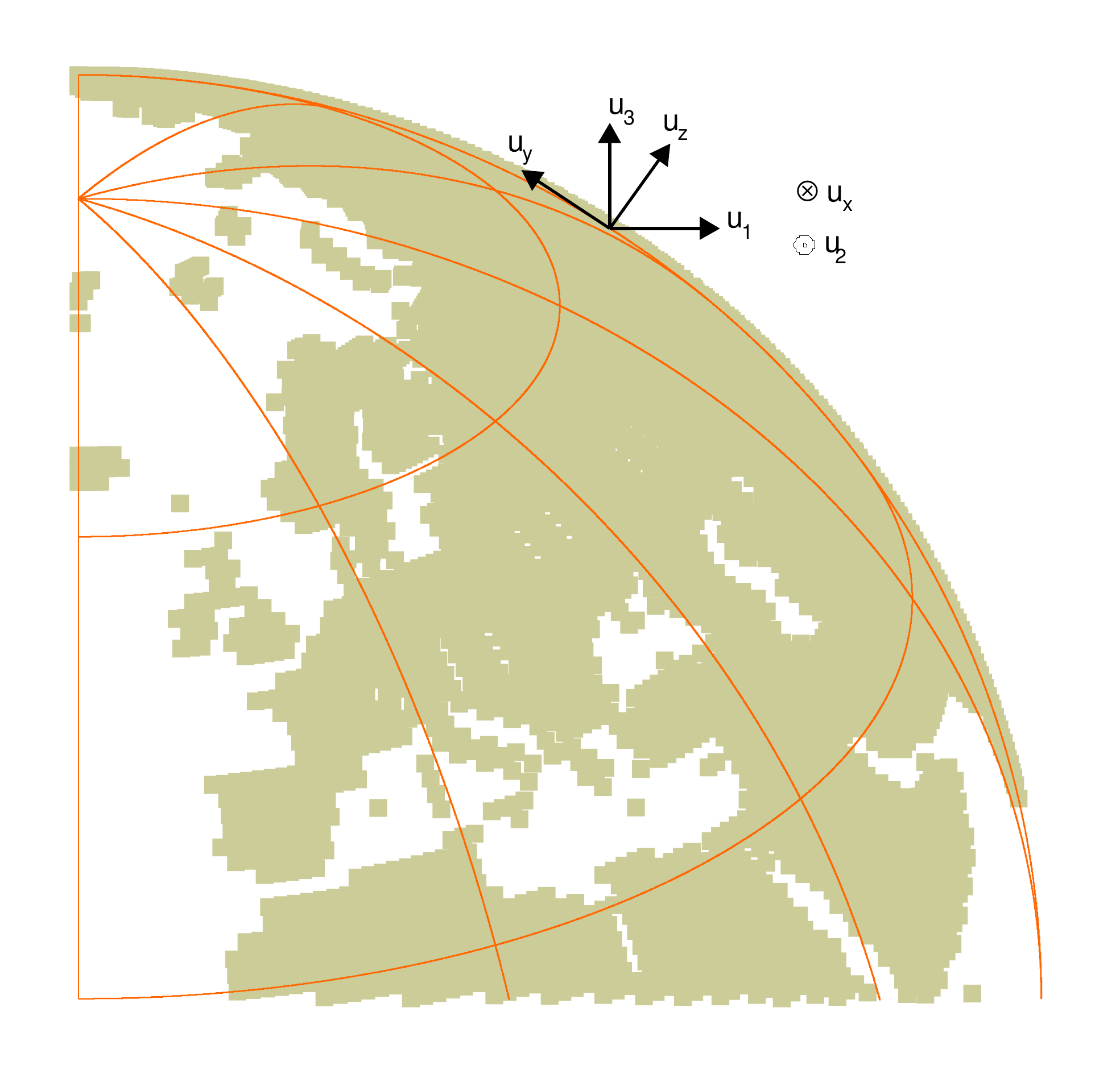}}
\caption{Local coordinate systems used throughout this review.}
\label{fig:local_coord}
\end{wrapfigure}

Let us now proceed to the local tracking of a point for an observer located at a geographic latitude $\lambda$ and longitude $l$ on Earth. It is then convenient to introduce the two coordinate systems depicted in figure~\ref{fig:local_coord}, where the observer is located in the city of Oulan-Bator ($\lambda\simeq47^\circ, l\simeq106^\circ$) for exemplify purpose: let $(\uu_x,\uu_y,\uu_z)$ be a right-handed basis of orthonormal vectors tied to the observer such that any unitary vector $\nn$ can be characterised by a zenith angle $\theta$ and an azimuthal angle\footnote{The azimuth $\varphi$ is here defined relative to the geographic East direction, measured counterclockwise.} $\varphi$ as $\nn=\sin{\theta}\cos{\varphi}~\uu_x+\sin{\theta}\sin{\varphi}~\uu_y+\cos{\theta}~\uu_z$; and let $(\uu_1,\uu_2,\uu_3)$ be a left-handed basis of orthonormal vectors in a coordinate system also tied to the observer such that $\uu_3=\uu_3^0$, $\uu_2=-\uu_x$, and $\nn=\cos{\delta}\cos{h}~\uu_1+\cos{\delta}\sin{h}~\uu_2+\sin{\delta}~\uu_3$. 

The angle $h$ can be expressed in terms of $\alpha$ by first mirroring the vector $\nn(\delta,h)$ at the $(\uu_2,\uu_3)$ plane to account for the transition from a left-handed to a right-handed coordinate system, and then by rotating the mirrored coordinate system around $\uu_3$ by an angle $\alpha_0(t)=\omega_{\text{sid}}t$, corresponding to the local mean sidereal time. The local sidereal time corresponds to the right ascension on the meridian of the observer. It is thus conventionally zero when the vernal point passes in the meridian plane of the considered location. The angular frequency $\omega_{\text{sid}}$ is $2\pi/T_{\text{sid}}$, with $T_{\text{sid}}$ the  time it takes for the Earth to complete one rotation relative to the vernal equinox. The result of the mirroring and of the rotation is $\nn=\cos{\delta}\cos{(\alpha_0-h)}~\uu_1+\cos{\delta}\sin{(\alpha_0-h)}~\uu_2+\sin{\delta}~\uu_3$. The identification of $(\alpha^0-h)$ with $\alpha$ allows then the definition of $h$, the hour angle, as the angular distance on the celestial sphere measured westward along the celestial equator from the meridian to the hour circle passing through a point. 

The $(\uu_x,\uu_y,\uu_z)$ basis is most often considered to track a point in local coordinates. The conversion from local angles $(\theta,\varphi)$ to equatorial ones $(\delta,\alpha)$ and local mean sidereal time $\alpha_0$ follows from a similar procedure as above. The expression of the vectors $(\uu_1,\uu_2,\uu_3)$ in the basis $(\uu_x,\uu_y,\uu_z)$ is $\uu_1=\cos{\lambda}~\uu_z-\sin{\lambda}~\uu_y$, $\uu_2=-\uu_x$, and $\uu_3=\cos{\lambda}~\uu_y+\sin{\lambda}~\uu_z$. On inserting these expressions into any vector $\nn=\cos{\delta}\cos{h}~\uu_1+\cos{\delta}\sin{h}~\uu_2+\sin{\delta}~\uu_3$, it is then possible to express $(\theta,\varphi)$ as a function of $(\delta,h)$ by identifying the components of $\nn$:
\begin{eqnarray}
\label{eqn:local-eq-trans1}
\sin{\theta}\cos{\varphi}&=&-\cos{\delta}\sin{h},  \\
\label{eqn:local-eq-trans2}
\sin{\theta}\sin{\varphi}&=&\cos{\lambda}\sin{\delta}-\sin{\lambda}\cos{\delta}\cos{h}, \\
\label{eqn:local-eq-trans3}
\cos{\theta}&=& \cos{\lambda}\cos{\delta}\cos{h}+\sin{\lambda}\sin{\delta}.
\end{eqnarray}

Finally, Galactic coordinates will also be used when presenting some results. Galactic coordinates locate a point in the sky with the aid of a latitude $b$ and a longitude $\ell$ defined in such a way that the Galactic plane corresponds to the equator and the origin of the longitudes corresponds to the Galactic center. The coordinates $(\ell,b)$ and $(\alpha,\delta)$ are connected by a simple rotation.

\subsection{Directional exposure}
\label{subsec:direxpo}

The directional exposure $\mu(\nn)$ of any observatory is the essential feature to characterise  directional data. It provides the effective time-integrated collecting area for a flux of CRs from each direction of the sky, in units of surface and time. At any time, the effective collecting area  is controlled by the directional aperture $A(\nnl)$ of the array and by the detection efficiency function $\epsilon(E,\nnl,t)$ at energy $E$, where, for the sake of clarity, $\nnl$ stands for a unit vector expressed in local coordinates while $\nn$ is expressed in equatorial ones. Denoting $A_0$ the surface of the ground array and $\nn_\perp$ the normal vector to the array, the directional aperture is $A(\nnl)=A_0\nnl\cdot\nn_\perp$, which is generally well approximated by $A(\nnl)=A_0\cos\theta$. In this way, $\mu(\alpha,\delta)$ generally reads as
\begin{eqnarray}
\label{eqn:direxpo}
\mu(\alpha,\delta;E)=\int_{\Delta t}\dif t~A_0\cos{(\theta(\alpha,\delta,t))}\times\epsilon(E,\nnl(\alpha,\delta,t),t),
\end{eqnarray}
with $\theta$ and $\varphi$ given by the set of equations~(\ref{eqn:local-eq-trans1},\ref{eqn:local-eq-trans2},\ref{eqn:local-eq-trans3}). To probe large-scale anisotropy contrasts at the $10^{-4}-10^{-2}$ level, the challenge consists in estimating the detection efficiency down to the required accuracy. As discussed below, this is most of the time, unfortunately, out of reach and approximation methods have to be used. 

An example of interest is that of full efficiency for triggering for zenith angles ranging from 0 to $\theta_{\text{max}}$. In this case, the integration in equation~(\ref{eqn:direxpo}) turns out to lead to a simple analytical expression~\cite{Sommers2001}. For $\epsilon=1$ indeed, the integrand depends only on the difference $\alpha_0(t)-\alpha$ so that the integration over time $t$ can be substituted for an integration over the hour angle $h$:
\begin{eqnarray}
\label{eqn:direxpo_sat}
\mu(\alpha,\delta)=\frac{A_0\Delta t}{\omega_{\text{sid}}T_{\text{sid}}}\int_{-h_{\text{m}}}^{h_{\text{m}}}\dif h~(\cos{\lambda}\cos{\delta}\cos{h}+\sin{\lambda}\sin{\delta}),
\end{eqnarray}
with $\omega_{\text{sid}}T_{\text{sid}}=2\pi$, and $h_{\text{m}}\equiv h_{\text{m}}(\lambda,\delta;\theta_{\text{max}})=\arccos{((\sin{\lambda}\sin{\delta}-\cos{\theta_{\text{max}}})/\cos{\lambda}\cos{\delta})}$ the maximum hour angle in the field of view at declination $\delta$ determined by the latitude of the site and the maximum zenith angle under consideration. This leads to
\begin{eqnarray}
\label{eqn:direxpo_sat_bis}
\mu(\alpha,\delta)=\frac{A_0\Delta t}{\pi}(\cos{\lambda}\cos{\delta}\sin{h_{\text{m}}}+h_{\text{m}}\sin{\lambda}\sin{\delta}),
\end{eqnarray}
an expression widely used in the literature.

In general, however, it is extremely difficult, if not impossible, to control the detection efficiency with sufficient precision, and the same is true for the small experimental instabilities that also lead to temporal variations in the directional aperture function. An approximation method, known as the direct-integration or time-scrambling procedure, is then used~\cite{Alexandreas1993,Atkins2003}. To understand this procedure, it is useful to start by inserting an additional integration in  equation~(\ref{eqn:direxpo}) over the solid angle in local coordinates\footnote{For convenience and simplicity, the energy dependence is dropped when not useful.}:
\begin{eqnarray}
\label{eqn:direxpo_bis}
\mu(\nn)=\int_{\Delta t}\dif t\int_{4\pi}\dif\nnl~A(\nnl,t)\epsilon(\nnl,t)\delta(\nnl,\nnl(\nn_,t)).
\end{eqnarray}
The argument in the Dirac function guarantees that the direction $\nnl$ in local coordinates considered throughout the integration corresponds to the equatorial direction $\nn$ seen at time $t$. Assuming that the detection efficiency function can be factorised into two different functions, $\epsilon(\nnl,t)\simeq\epsilon_1(\nnl)\times\epsilon_2(t)$, and considering that the same holds for the directional aperture, $A(\nnl,t)\simeq A_1(\nnl)\times A_2(t)$, the principle of the procedure consists in substituting the underlying efficiency and aperture functions by the observed event  arrival distribution and rate:
\begin{eqnarray}
\label{eqn:replacement1}
A_1(\nnl)\times \epsilon_1(\nnl)&\rightarrow& \dif N/\dif\nnl,\\
\label{eqn:replacement2}
A_2(t)\times \epsilon_2(t)&\rightarrow& \dif N/\dif t.
\end{eqnarray}
One immediate limitation of these approximations is that the resulting right-ascension-integrated exposure follows, for each declination, the right-ascension-integrated arrival directions of the events, $\int\dif\alpha~\mu\propto\int\dif\alpha~\dif N/\dif\nn$. This is because the time integration of the observed arrival directions in local coordinates allows for scanning the range of right ascensions only, keeping fixed the declination. Since the most general anisotropy decomposition follows from $\delta I(\nn)=\delta I_1(\alpha,\delta)+\delta I_2(\delta)$, it means that the approximations allow for a reconstruction of $\delta I_1(\nn)=\delta I(\nn)-\delta I_2(\delta)$ only. This is a limitation to keep in mind.

Based on equation~(\ref{eqn:direxpo_bis}), different strategies can then be adopted to compute $\mu(\nn)$. The principle is exemplified hereafter using the time-scrambling (or ``shuffling'') technique, but other strategies would yield equivalent results. From the actual event rate $\dif N/\dif t=\sum_i\delta(t,t_i)$ and event arrival directions $\dif N/\dif\nnl=\sum_i\delta(\nnl,\nnl_i)$, the procedure consists in sampling by Monte-Carlo equation~(\ref{eqn:direxpo_bis}) with $N_{\mathrm{sh}}$ different realisations of event sets obtained through random permutations of event times and arrival directions\footnote{This uses the fact that $\delta(\nnl,\nnl(\nn,t))=\delta(\nn,\nn(\nnl,t))$, given that $\nnl$ and $\nn$ are just related through a (time-dependent) rotation.}:
\begin{eqnarray}
\label{eqn:direxpo_sh}
\mu^{(0)}(\nn)\simeq\frac{1}{N_{\mathrm{sh}}}\sum_{j=1}^{N_{\mathrm{sh}}}\sum_{i=1}^N\delta(\nn(\nnl_i,t_{\sigma_j(i)}),\nn),
\end{eqnarray}
where the subscript $\sigma_j(i)$ stands for the random permutation of each element $i$. Through the substitutions in equations~(\ref{eqn:replacement1}) and~(\ref{eqn:replacement2}), a normalisation factor $1/N$ has been introduced, and the dimension of $\mu$ has been changed from time$\times$surface to inverse steradian, but the notation is kept identical for convenience since both quantities are just related by a factor $A_0\Delta t\Delta\Omega$, with $\Delta\Omega$ the solid angle in which the directional exposure is non zero. The reason for the superscript $(0)$ will become clear below. 

\begin{figure}[!h]
\centering\includegraphics[width=0.8\textwidth]{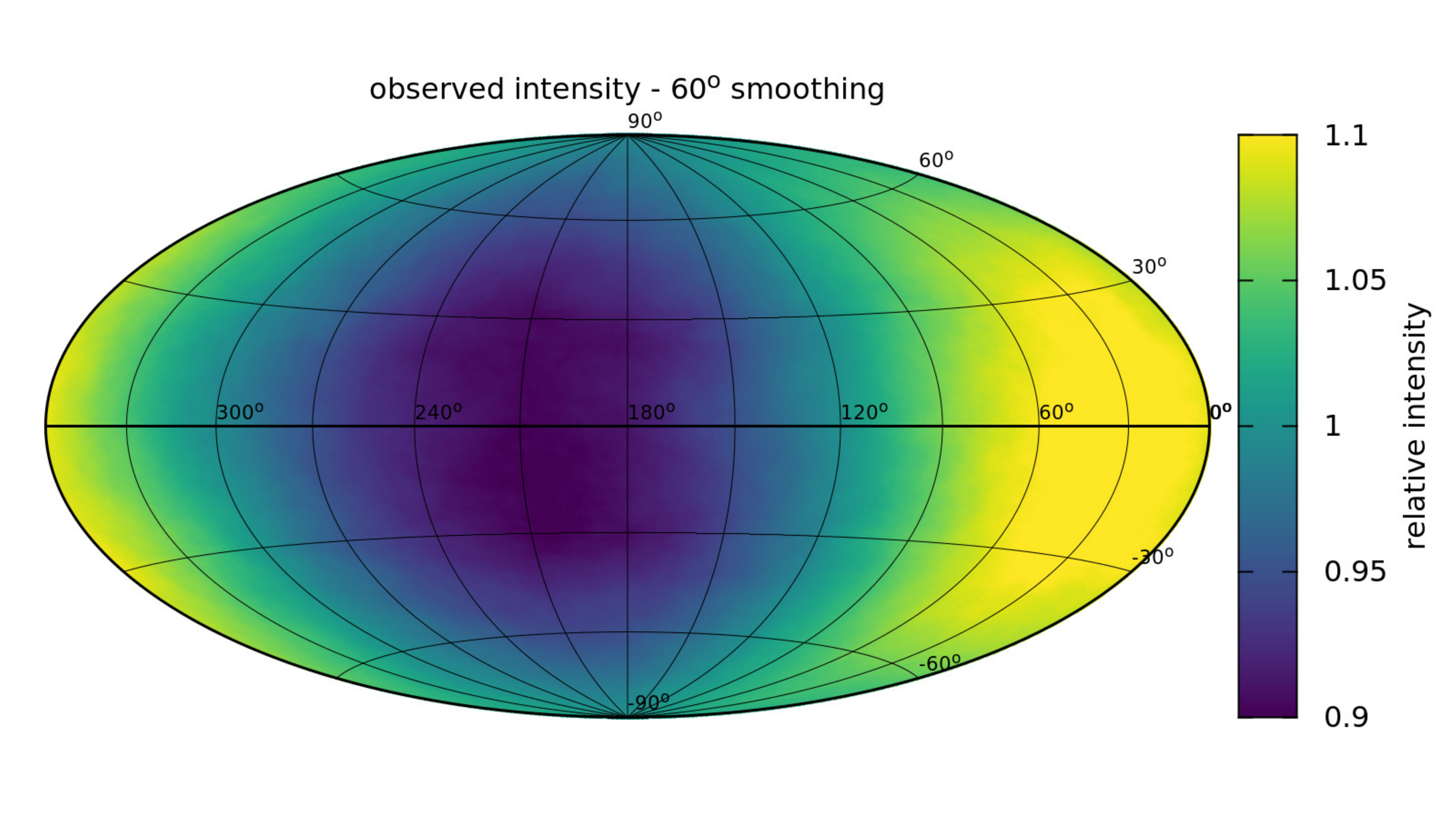}
\caption{``Raw'' intensity observed by ignoring the changes of directional exposure with right ascension due to variations of detection efficiency with time. The injected dipole of 10\% amplitude pointing to $(\alpha,\delta)=(130^\circ,0^\circ)$ is overwhelmed by the spurious one.}
\label{fig:shuff_mapintensity_iter0}
\end{figure}
To see the procedure at work, let us consider a toy model with exaggerated time variations of the detection efficiency that lead to a spurious pattern in equatorial coordinates whose amplitude overwhelms a genuine anisotropy, and let us see how the genuine pattern is recovered through the procedure. To produce an anisotropy contrast of experimental origin with amplitude $a$, the detection efficiency is chosen to follow $\epsilon(t)\propto 1+2a\cos{(2\pi t/T_{\mathrm{day}}+\varphi_1)}\cos{(2\pi t/T_{\mathrm{yr}}+\varphi_2)}$, with $T_{\mathrm{day}}$ ($T_{\mathrm{yr}}$) the durations of one solar day (year), and $\varphi_1$ and $\varphi_2$ some random phases. The genuine anisotropy is, on the other hand, shaped by a dipole vector $\mathbf{d}$ with amplitude $d$ and directions $(\alpha_{\mathrm{d}},\delta_{\mathrm{d}})=(130^\circ,0^\circ)$, that is\footnote{Note that in this case, $\delta I_1(\nn)=\delta I(\nn)$.} $\delta I_1(\nn)=\mathbf{d}\cdot\nn$. The toy experiment is chosen to be located somewhere on the Earth equator and to operate with $\theta_{\mathrm{max}}=90^\circ$. In this way, the whole sky is covered, which is convenient to avoid the complications related to the reconstruction of a dipole with partial-sky coverage, complications that will be discussed in detail in~\S~\ref{sec:3dreco}. The ``raw'' sky map, that is, the intensity observed by ignoring the changes of directional exposure with right ascension due to the variations of detection efficiency with time, is shown in figure~\ref{fig:shuff_mapintensity_iter0} with $a=0.15$ and $d=0.1$. To exhibit the dipolar structure, the relative intensity which is shown (that is, $4\pi I/I_0$) is smoothed at a $60^\circ$ angular scale using a top-hat filter\footnote{For visualisation purpose, measured intensities are transformed into smoothed density functions. Any smoothed density $\mathcal{I}(\mathbf{n})$ results from the angular distribution $I(\mathbf{n})$ convolved with a filter function $\mathcal{F}_\Theta(\mathbf{n},\mathbf{n}')$, $ \mathcal{I}(\mathbf{n}) \propto \int\dif\mathbf{n}' \mathcal{F}_\Theta(\mathbf{n},\mathbf{n}')I(\mathbf{n}')$. The widely-used top-hat filter is such that $\mathcal{F}_\Theta(\mathbf{n},\mathbf{n}')=1$ if $\arccos{(\mathbf{n}\cdot\mathbf{n}')}\leq\Theta$ and 0 otherwise.}. The color scale is fixed to the contrast expected from the genuine dipole amplitude $d$. Clearly, the genuine dipole is overwhelmed by the spurious one. 

\begin{figure}[!h]
\centering\includegraphics[width=0.8\textwidth]{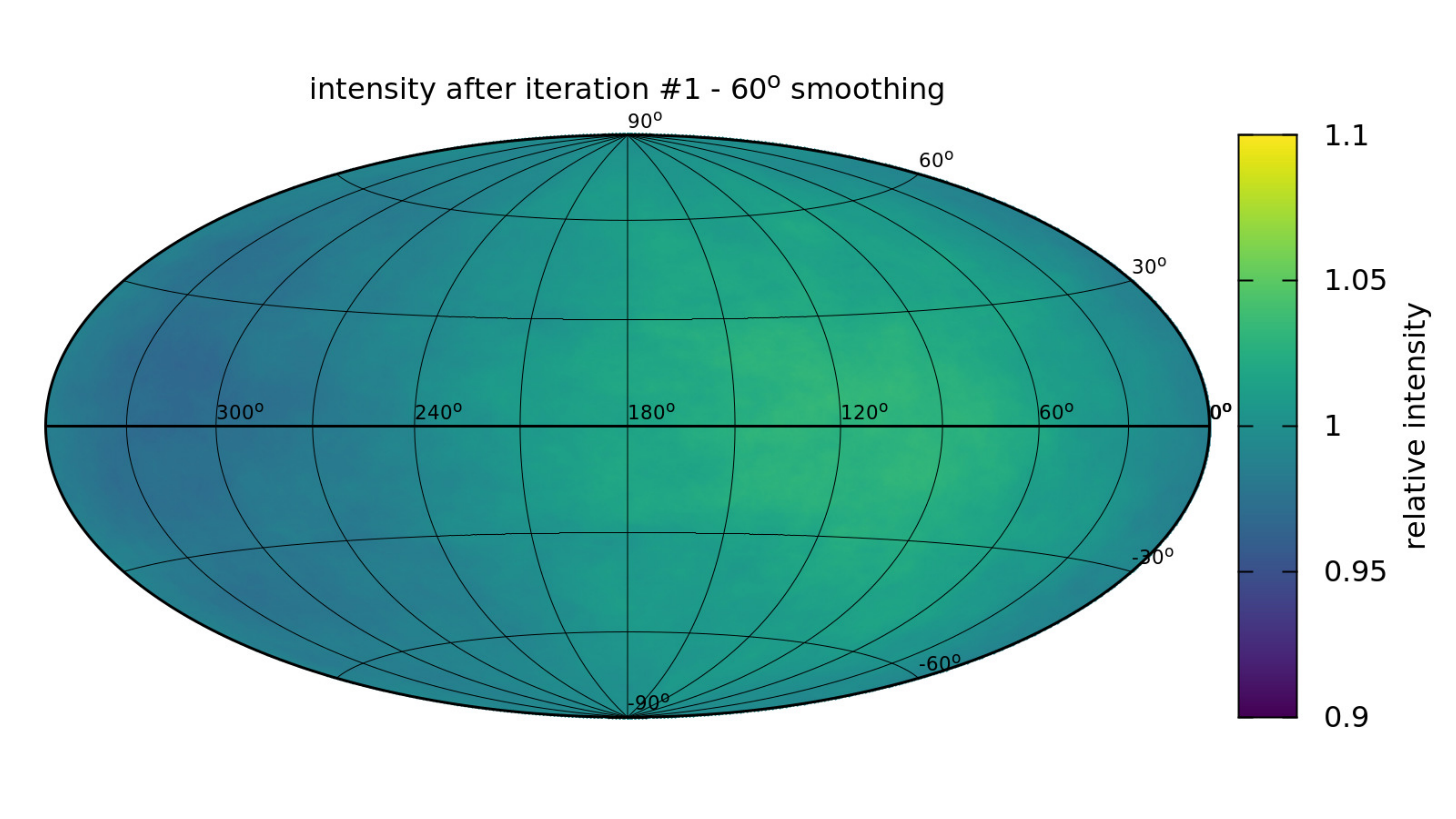}
\caption{Reconstructed intensity once the directional exposure is estimated through equation~(\ref{eqn:direxpo_sh}). The spurious dipole has been removed; but the amplitude of the reconstructed dipole still differs from the injected one, being largely attenuated.}
\label{fig:shuff_mapintensity_iter1}
\end{figure}
Applying equation~(\ref{eqn:direxpo_sh}) to estimate the directional exposure, the reconstructed intensity is shown in figure~\ref{fig:shuff_mapintensity_iter1}. The integration over solid angle of the angular distribution $\dif N/\dif\nnl$ results in giving to any equatorial direction in the field of view available at any given time $t$ an instantaneous exposure in proportion to the event rate in the corresponding direction in local angles. As a result of the integration over time of the variation of the actual event rate, the distortions of the intensity in equatorial coordinates induced by experimental variations are automatically accounted for. Hence, the spurious pattern is not observed any longer in figure~\ref{fig:shuff_mapintensity_iter1}. In contrast, the modulations in equatorial coordinates induced by the genuine anisotropy are only partially washed out, because for any given time $t$, the event time distribution is sensitive to the global intensity of CRs but not to the underlying structure in equatorial angles. An image of the genuine anisotropy is thus recovered as seen in the sky map, but with reduced amplitude. 

It is possible in this example to get a more intuitive impression of the expected outcome of the procedure depicted in figure~\ref{fig:shuff_mapintensity_iter1} and so to follow more closely the fraction of anisotropy which gets absorbed through the procedure. For a dipolar intensity, and given the specific conditions of directional aperture and detection efficiency considered here, the expected number of events per steradian and per time unit above any energy threshold $E_0$ reads, in local coordinates, as
\begin{equation}
\label{eqn:d2N_dn_dt}
\frac{\dif^2N(>E_0)}{\dif\nnl~\dif t}=A_0I_0\epsilon(t)\cos{\theta}~(1+d\nnl\cdot\nnl_{\mathrm{d}}(\nn_{\mathrm{d}},t)).
\end{equation}
For $\lambda=0$ and $\delta_{\mathrm{d}}=0$, the dipole unit vector is rotating with time in local coordinates and is expressed as $\nnl_{\mathrm{d}}(\nn_{\mathrm{d}},t)=\cos{(\alpha_0(t)-\alpha_{\mathrm{d}})}~\uu_z-\sin{(\alpha_0(t)-\alpha_{\mathrm{d}})}~\uu_x$. The rate of events in expressions~(\ref{eqn:replacement1}) and~(\ref{eqn:replacement2}) entering into equation~(\ref{eqn:direxpo_bis}) can thus be expressed by integrating equation~(\ref{eqn:d2N_dn_dt}) over time and local angles, respectively. The distribution $\dif N/\dif\nnl$ is more conveniently obtained by integrating $\dif^2N/\dif\nnl\dif t$ over local sidereal time $\alpha_0$ once the following transformation is performed:
\begin{equation}
\label{eqn:d2N_dn_dalpha0}
\frac{\dif^2N}{\dif\nnl~\dif \alpha_0}=\int\dif t\frac{\dif^2N}{\dif\nnl~\dif t}\delta(\alpha_0,\alpha_0(t)),
\end{equation}
where the Dirac function guarantees that the local sidereal time $\alpha_0(t)$ considered throughout the time integration corresponds to the direction $\alpha_0$ seen at time $t$. On inserting equation~(\ref{eqn:d2N_dn_dt}) into this expression, the angular distribution reads as
\begin{equation}
\label{eqn:dN_dnnl}
\frac{\dif N}{\dif\nnl}=\frac{A_0I_0\Delta t}{\omega_{\text{sid}}T_{\text{sid}}}\cos{\theta}\int\dif\alpha_0~\tilde{\epsilon}(\alpha_0)(1+d\cos{\theta}\cos{(\alpha_0-\alpha_{\mathrm{d}})}-d\sin{\theta}\cos{\varphi}\sin{(\alpha_0-\alpha_{\mathrm{d}})}),
\end{equation}
where $\tilde{\epsilon}(\alpha_0)=\int\dif t~\epsilon(t)\delta(\alpha_0,\alpha_0(t))$ is the detection efficiency expressed in terms of local sidereal time. To first order in $d$ and in $a$, the variations of both the intensity and the $\tilde{\epsilon}(\alpha_0)$ function do not contribute to the angular distribution since these variations are harmonic with $\alpha_0$ and thus cancel throughout the integration:
\begin{equation}
\label{eqn:dN_dnnl_bis}
\frac{\dif N}{\dif\nnl}\simeq A_0I_0\Delta t\cos{\theta}.
\end{equation}
In local angles, the angular distribution is thus, in this case, identical to the one that would have been obtained with a stable detector had the intensity been isotropic. Besides, the event rate $\dif N/\dif t$ can be obtained following the same reasoning. The result, still to first order in $d$ and in $a$, reads as
\begin{equation}
\label{eqn:dN_dt}
\frac{\dif N}{\dif t}\simeq \pi A_0I_0\epsilon(t)~\left(1+\frac{2}{3}d\cos{(\alpha_0(t)-\alpha_{\mathrm{d}})}\right),
\end{equation}
so that, when inserting equations~(\ref{eqn:dN_dnnl_bis}) and~(\ref{eqn:dN_dt}) into equation~(\ref{eqn:direxpo_bis}), two thirds of the variations of the event rate induced by the dipole are absorbed in the integrand. The remaining integration in equation~(\ref{eqn:direxpo_bis}), carried out through the shuffling of the events in equation~(\ref{eqn:direxpo_sh}), leads to 
\begin{eqnarray}
\label{eqn:direxpo_sh}
\mu^{(0)}(\alpha,\delta)=\mu(\alpha,\delta)\left(1+\frac{\pi}{6}d\cos{(\alpha-\alpha_{\mathrm{d}})}\right).
\end{eqnarray}
There is thus, in this example, a fraction $\pi/6$ of the dipole which is absorbed into the directional exposure, and consequently a fraction $1-\pi/6$ of the dipole which is reconstructed by the procedure. This is in agreement with figure~\ref{fig:shuff_mapintensity_iter1}. The intensity recovered through this procedure is hence a first guess of the underlying one, denoted as $\delta I_1^{(0)}$. This first iteration is generally sufficient to pick up small-scale structures out from the background noise. 

\begin{figure}[!h]
\centering\includegraphics[width=0.8\textwidth]{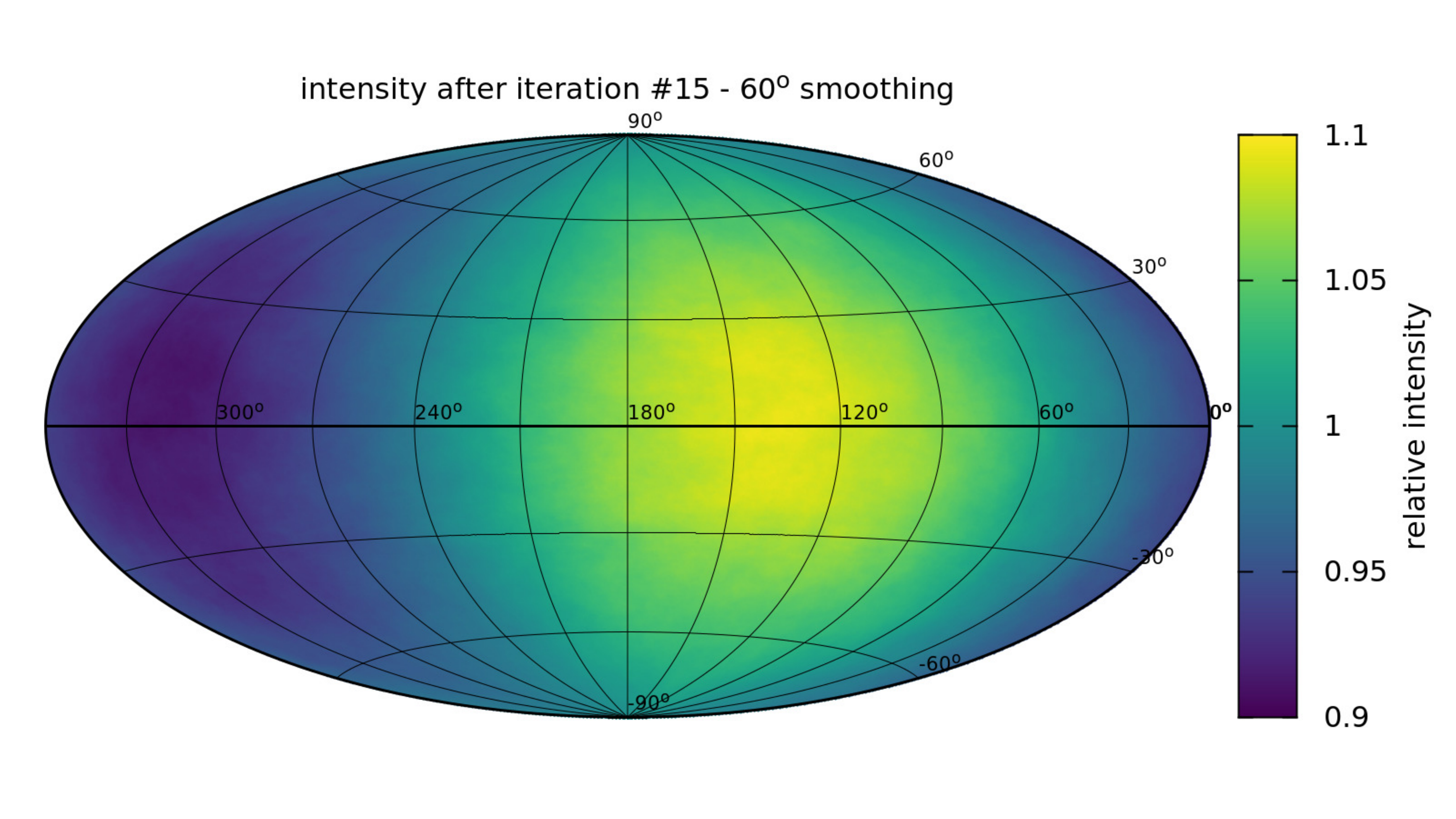}
\caption{Reconstructed intensity after the 15th iteration described in the text. The dipole amplitude is recovered.}
\label{fig:shuff_mapintensity_iter15}
\end{figure}
To improve the description of the underlying intensity, the substitutions operated in equations~(\ref{eqn:replacement1}) and~(\ref{eqn:replacement2}) need to be more accurate. Ideally, the proper estimates of $A_1(\nnl)\times\epsilon_1(\nnl)$ and $A_2(t)\times\epsilon_2(t)$ should be the event arrival directions and the event time rate that would be observed for an isotropic sky, say $\dif N_0/\dif\nnl$ and $\dif N_0/\dif t$. It is thus necessary to account for the impact of the anisotropies on the observed distributions to infer the best way these latter quantities. This can be done through an iterative procedure, on inserting the quantity $\delta I_1^{(0)}$ into the relationship between the observed distributions and expected ones for isotropy:
\begin{eqnarray}
\label{eqn:replacement1iter}
\dif N^{(1)}_0/\dif\nnl &\simeq& (1-c_1^{(0)}(\nnl))\times\dif N/\dif\nnl, \\
\label{eqn:replacement2iter}
\dif N^{(1)}_0/\dif t &\simeq& (1-c_2^{(0)}(t))\times\dif N/\dif t,
\end{eqnarray}
with the $c_1$ and $c_2$ functions defined as
\begin{eqnarray}
\label{eqn:c1}
c_1^{(0)}(\nnl)=\frac{\int\dif t~\delta I_1^{(0)}(\nnl,t)\frac{\dif^2N}{\dif t\dif\nnl}}{\int\dif t~\frac{\dif^2N}{\dif t\dif\nnl}}, \\
\label{eqn:c2}
c_2^{(0)}(t)=\frac{\int\dif\nnl~\delta I_1^{(0)}(\nnl,t)\frac{\dif^2N}{\dif t\dif\nnl}}{\int\dif\nnl~\frac{\dif^2N}{\dif t\dif\nnl}}.
\end{eqnarray}
With these new estimates of $\dif N_0/\dif\nnl$ and $\dif N_0/\dif t$, the estimate of the directional exposure is retuned as
\begin{eqnarray}
\label{eqn:direxpo_sh_final}
\mu^{(1)}(\nn)\simeq\frac{1}{N_{\mathrm{sh}}}\sum_{j=1}^{N_{\mathrm{sh}}}\sum_{i=1}^N(1-c_1^{(0)}(\nnl_i)-c_2^{(0)}(t_{\sigma_j(i)}))\times\delta(\nn(\nnl_i,t_{\sigma_j(i)}),\nn).
\end{eqnarray}
The fraction of the genuine dipole which is absorbed in this new estimation is now reduced. Repeating the same calculations as above yields in fact to the following estimate of the directional exposure after the $k$th iteration:
\begin{eqnarray}
\label{eqn:direxpo_sh_kth}
\mu^{(k)}(\alpha,\delta)=\mu(\alpha,\delta)\left(1+\left(\frac{\pi}{6}\right)^{k}d\cos{(\alpha-\alpha_{\mathrm{d}})}\right),
\end{eqnarray}
so that it is possible to iterate until convergence is reached, convergence which turns out to be attained after about ten iterations. An image of the genuine anisotropy is thus recovered as seen in figure~\ref{fig:shuff_mapintensity_iter15}, image which is no longer attenuated. 

Although exemplified here on a rather simple situation, this procedure (or its numerous variants~\cite{Cui2003,Amenomori2005,Ahlers2016a}) turns out to be extremely robust against spurious effects. The sensitivity to capture anisotropies is however reduced compared to a knowledge of the directional exposure independently of the data and requires Monte-Carlo studies. 

\section{First harmonic analysis in right ascension}
\label{sec:harmonic}

\subsection{The Rayleigh formalism}
\label{subsec:rayleigh}

Under stable detection conditions, the directional exposure as defined through equation~(\ref{eqn:direxpo}) is independent of the right ascension when integrating the local-angle-detection efficiency over full periods of sidereal revolution of the Earth. For that reason, directional data have most commonly been analysed through harmonic analysis in right ascension of the event counting rate within the declination band defined by the detector field of view. The analysis of the harmonics encoded in the right ascension distribution is indeed the most natural tool to reveal the large-scale anisotropy of CRs. The technique in itself is rather simple: the greatest difficulties are in the treatment of the data, that is, of the counting rates themselves. The description of this technique and of the related sanity checks to validate the analyses leading to large-scale anisotropy reports is the object of this section.

The main difficulty to measure the variations of the counting rate is that the uniformity of the detector responses in right ascension is only approximate: in addition to the challenging high continuity of operation required over long-term observations, detection systems are mostly located in outdoor ambients, generally at mountain altitude, being thus subjected to large atmospheric variations. The CR counting rate is then affected by these meteorological modulations, since the extensive air shower development depend on the air density through variations of the Moli\`ere radius and on pressure which impacts on the absorption of the electromagnetic component. Since the searched anisotropy amplitudes range down to $\simeq 10^{-4}$--$10^{-3}$, and since these sources of spurious variations of the CR counting rate can lead to larger modulation amplitudes, it is essential to account for them. In the following, the variations of the CR counting rate in right ascension expected from changes of experimental conditions are denoted by $\mu(\alpha)=\mu_0(1+\delta\mu(\alpha))$, where $\mu(\alpha)$ formally reads as $\mu(\alpha)=\int\dif\delta\cos{\delta}\mu(\alpha,\delta)$.

Denoting the declination-integrated directional intensity as $I(\alpha)=I_0(1+\delta I(\alpha))/2\pi$, the arrival direction distribution $\dif N/\dif\alpha$ considered as a function of the right ascension thus only results from genuine anisotropies $\delta I(\alpha)$ coupled to the directional exposure variations,
\begin{eqnarray}
\label{eqn:dNda0}
\frac{\dif N}{\dif\alpha}&=&\frac{I_0\mu_0}{2\pi} \left(1+\delta\mu(\alpha)\right)\left(1+\delta I(\alpha)\right) \nonumber \\
                                   &=&\frac{I_0\mu_0}{2\pi} \left(1+\delta\mu(\alpha)\right)\left(1+\sum_{n>0}~a_{\mathrm{c}n}\cos{n\alpha}+\sum_{n>0}~a_{\mathrm{s}n}\sin{n\alpha}\right).
\end{eqnarray}
where the anisotropies have been decomposed as an harmonic expansion in such a way that the Fourier coefficients $a_{\mathrm{c}n}$ and $a_{\mathrm{s}n}$ provide a fingerprint of the anisotropy in right ascension relative to the monopole (isotropic intensity). Considering the arrival direction distribution corrected for the directional exposure variations, the customary recipe to extract each harmonic coefficient makes use of the orthogonality of the trigonometric functions:
\begin{eqnarray}
\label{eqn:an-cs}
a_0&=&\frac{1}{2\pi}\int\frac{\dif\alpha}{1+\delta\mu(\alpha)}\frac{\dif N}{\dif\alpha}, \nonumber  \\
a_{\mathrm{c}n}&=&\frac{1}{\pi a_0}\int\frac{\dif\alpha}{1+\delta\mu(\alpha)}\frac{\dif N}{\dif\alpha}\cos{n\alpha}, \nonumber  \\
a_{\mathrm{s}n}&=&\frac{1}{\pi a_0}\int\frac{\dif\alpha}{1+\delta\mu(\alpha)}\frac{\dif N}{\dif\alpha}\sin{n\alpha}.
\end{eqnarray}
In practice, the event counting rate, $\dif N/\dif\alpha$, is a natural estimator of the arrival direction distribution. Modeling this counting rate as a sum of $N$ Dirac functions over the circle, $\dif N/\dif\alpha=\sum_i \delta(\alpha,\alpha_i)$, the coefficients can be estimated from the discrete sums running over the $N$ observed events\footnote{Throughout this review, over-lined symbols are used to indicate the \textit{estimator} of any quantity.}:
\begin{eqnarray}
\label{eqn:an-cs-est}
\bar{a}_{\mathrm{c}n}&=&\frac{2}{\tilde{N}} \sum_{1\leq i \leq N} \frac{\cos{n\alpha_i}}{1+\delta\mu(\alpha_i)},\\
\bar{a}_{\mathrm{s}n}&=&\frac{2}{\tilde{N}} \sum_{1\leq i \leq N} \frac{\sin{n\alpha_i}}{1+\delta\mu(\alpha_i)}.
\end{eqnarray}
In these expressions, the coefficient $a_0$ has been estimated by $\bar{a}_0=\tilde{N}/2\pi$, with the notation $\tilde{N}=\sum_i(1+\delta\mu(\alpha_i))^{-1}$ standing for the number of events that would have been observed had the directional exposure been uniform. 

The statistical properties of the estimators $\{\bar{a}_{\mathrm{c}n},\bar{a}_{\mathrm{s}n}\}$ can be derived from the Poissonian nature of the sampling of $N$ points over the circle distributed according to the underlying angular distribution, as presented in a short and punchy style by Linsley~\cite{Linsley1975}. From Poisson statistics indeed, the first and second moments of $\overline{\delta I}(\alpha)$ averaged over a large number of realisations of $N$ events read as
\begin{eqnarray}
\label{eqn:momentsphi}
\left\langle \overline{\delta I}(\alpha)\right\rangle_{\mathrm{P}}&=&\delta I(\alpha), \nonumber \\
\left\langle \overline{\delta I}(\alpha)\overline{\delta I}(\alpha^\prime)\right\rangle_{\mathrm{P}}&=&\delta I(\alpha)\delta I(\alpha^\prime)+\delta I(\alpha)\delta(\alpha,\alpha^\prime).
\end{eqnarray}
The mean and RMS of the estimators are then obtained by propagating these properties in equations~(\ref{eqn:an-cs}), and by considering $\bar{a}_0$ as a constant. This latter approximation is extremely accurate in most of practical cases. In this way, it is straightforward to see that the estimators are, on the one hand, unbiased:
\begin{eqnarray}
\label{eqn:moments1a}
\left\langle \bar{a}_{\mathrm{c}n}\right\rangle_{\mathrm{P}}&=&a_{\mathrm{c}n}, \nonumber \\
\left\langle \bar{a}_{\mathrm{s}n}\right\rangle_{\mathrm{P}}&=&a_{\mathrm{s}n},
\end{eqnarray}
and, on the other hand, obey the following covariance matrix coefficients:
\begin{eqnarray}
\label{eqn:moments2a}
\mathrm{cov}(\bar{a}_{\mathrm{c}m},\bar{a}_{\mathrm{c}n})&=& \frac{1}{\pi^2 a_0^2}\int\frac{\mathrm{d}\alpha}{1+\delta\mu(\alpha)}~\frac{I_0\mu_0}{2\pi}\delta I(\alpha)\cos{m\alpha}\cos{n\alpha}, \nonumber \\
\mathrm{cov}(\bar{a}_{\mathrm{s}m},\bar{a}_{\mathrm{s}n})&=& \frac{1}{\pi^2 a_0^2}\int\frac{\mathrm{d}\alpha}{1+\delta\mu(\alpha)}~\frac{I_0\mu_0}{2\pi}\delta I(\alpha)\sin{m\alpha}\sin{n\alpha}.
\end{eqnarray}
In the case of small anisotropies (\textit{i.e.} $|a_{\mathrm{c}n}|\ll 1$ and $|a_{\mathrm{s}n}|\ll 1$), the previous expressions allow the derivation of the uncertainties of the estimators as~:
\begin{eqnarray}
\label{eqn:moments3a}
\sigma_{\mathrm{c}n}(\bar{a}_{\mathrm{c}n})&=& \left[\frac{2}{\pi\tilde{N}}\int\frac{\mathrm{d}\alpha}{1+\delta\mu(\alpha)}~\cos^2{n\alpha}\right]^{1/2}, \nonumber \\
\sigma_{\mathrm{s}n}(\bar{a}_{\mathrm{s}n})&=& \left[\frac{2}{\pi\tilde{N}}\int\frac{\mathrm{d}\alpha}{1+\delta\mu(\alpha)}~\sin^2{n\alpha}\right]^{1/2}.
\end{eqnarray}
In practice, the variations of $\delta\mu(\alpha)$ are so small and so smooth that the integrals are accurately close to $\pi$, so that the uncertainties are $\sigma_{\mathrm{c}n}=\sigma_{\mathrm{s}n}=\sqrt{2/\tilde{N}}$. From the central limit theorem, the Gaussian probability density functions (p.d.f.) of the $\bar{a}_{\mathrm{c}n}$ and $\bar{a}_{\mathrm{s}n}$ coefficients, $p_{A_{\mathrm{c}n}}$ and $p_{A_{\mathrm{s}n}}$ respectively, are then entirely determined with the parameters $(\left\langle \bar{a}_{\mathrm{c}n}\right\rangle_{\mathrm{P}},\sigma^2)$ and $(\left\langle \bar{a}_{\mathrm{s}n}\right\rangle_{\mathrm{P}},\sigma^2)$, with $\sigma^2=2/\tilde{N}$.

Rather than in terms of Fourier coefficients $\bar{a}_{\mathrm{c}n}$ and $\bar{a}_{\mathrm{s}n}$, harmonic analyses are generally reported in terms of more intuitive geometrical parameters: the amplitude of the harmonic modulation, $\bar{r}_n=({\bar{a}_n}^2+{\bar{a}_n}^2)^{1/2}$, and the phase corresponding to the right ascension of the maximum intensity, $\bar{\phi}_n=\arctan{(\bar{a}_{\mathrm{s}n}/\bar{a}_{\mathrm{c}n})}$, defined modulo $2\pi/n$. It is thus of interest to infer the expected distributions of these variables under any distribution of arrival directions. For any realisation of $N$ events, $\bar{a}_{\mathrm{c}n}$ and $\bar{a}_{\mathrm{s}n}$ are random variables whose joint p.d.f., $p_{A_{\mathrm{c}n},A_{\mathrm{s}n}}$, can be factorized in the limit of large number of events in terms of the product $p_{A_{\mathrm{c}n}}p_{A_{\mathrm{s}n}}$. For any $n$, the joint p.d.f. of the estimated $\bar{r}_n$ and $\bar{\phi}_n$ is then obtained through the appropriate Jacobian transformation:
\begin{eqnarray}
\label{eqn:jointpdf1}
p_{R,\Phi}(\bar{r},\bar{\phi};r,\phi)=\frac{\bar{r}}{2\pi\sigma^2}~\exp{\left(-\frac{1}{2\sigma^2}\left(\bar{r}^2+r^2-2\bar{r}r\cos{\psi}\right)\right)},
\end{eqnarray}
with $\psi=\bar{\phi}-\phi$, and where the index $n$ has been dropped for simplicity. The p.d.f. of the amplitude (phase), $p_{R}$ ($p_{\Phi}$), is then the result of the marginalisation of $p_{R,\Phi}$ over the phase (amplitude):
\begin{equation}
\label{eqn:amp}
p_{R}(\bar{r};r)=\frac{\bar{r}}{\sigma^2} \exp{\left(-\frac{\bar{r}^2+r^2}{2\sigma^2}\right)}I_0\left(\frac{\bar{r}r}{\sigma^2}\right), 
\end{equation}
\begin{equation}
\label{eqn:phase}
p_{\Phi}(\bar{\phi};r,\phi)=\frac{\exp{\left(-\frac{r^2}{2\sigma^2}\right)}}{2\pi}\left[1+\sqrt{\frac{\pi}{2}}\frac{r}{\sigma}\cos{\psi}\exp{\left(\frac{r^2\cos^2{\psi}}{2\sigma^2}\right)}\left(1+\xi\text{erf}\left(\xi \frac{r\cos{\psi}}{\sqrt{2}\sigma}\right)\right) \right]
\end{equation}
with $I_n(\cdot)$ the modified Bessel function of first kind with order $n$, and $\xi=1$ if $|\psi|\leq\pi/2$ and $-1$ otherwise. Equations~(\ref{eqn:amp}) and~(\ref{eqn:phase}) correspond to the expressions derived by Linsley in the framework of his ``second alternative''~\cite{Linsley1975}. For an underlying isotropic distribution, $p_{\Phi}$ is uniform, and $p_{R}$ reduces to the Rayleigh distribution which allows a straightforward estimation of the probability (in fact, the $p$-value) that an observed amplitude $\bar{r}$ arises from pure statistical fluctuations as
\begin{equation}
\label{eqn:prob}
P(>\bar{r})=\int_{\bar{r}}^{\infty} \dif\bar{r}'p_{R}(\bar{r}';r=0)=\exp{\left(-\bar{r}^2/2\sigma^2\right)}.
\end{equation}
For a non-zero amplitude $r$, depending on the signal-to-noise ratio parameter $r/\sigma$, $p_{R}$ and $p_{\Phi}$ smoothly evolve from the Rayleigh and uniform distributions to bell curves well-defined about the values of $r$ and $\phi$. For $r/\sigma$ large enough, the bell curves are indistinguishable from Gaussian ones.

Uncertainties on any set of measured $\bar{r}$ and $\bar{\phi}$ values are generally estimated by propagating to first order the uncertainties $\sigma$ of the Fourier coefficients, leading to $\sigma_{\bar{r}}=\sigma$ and $\sigma_{\bar{\phi}}=\sigma/\bar{r}$. This empirical recipe is known, however, to overestimate in most of cases the real uncertainty when the relationship between new and old variables is not linear, as this is here the case  between $(r,\phi)$ and $(a_{\mathrm{c}},a_{\mathrm{s}})$. Alternatively to this propagation rule, uncertainties can be estimated from the two first moments of  $p_{R}$ and $p_{\Phi}$. This leads to a semi-analytical expression for $\sigma_{\bar{r}}$ and a numerical integration for $\sigma_{\bar{\phi}}$:
\begin{eqnarray}
\label{eqn:uncertainties-amp-ph}
\sigma_{\bar{r}}&=& \left[2\sigma^2+r^2-\frac{\pi\sigma^2}{2}L^2_{1/2}\left(\frac{-r^2}{2\sigma^2}\right)\right]^{1/2},\\
\sigma_{\bar{\phi}}&=&\left[ \int_{-\pi}^\pi\dif\psi\psi^2 p_{\Phi}(\bar{\phi};r,\phi)\right]^{1/2},
\end{eqnarray}
with $L^2_{1/2}(\cdot)$ the square of the Laguerre polynomial $L_{1/2}(\cdot)$. These expressions depend on the genuine amplitude $r$, which is searched for. A reasonable assumption can be to use $r=0$ in the background-dominated regime, leading to $\sigma_{\bar{r}}=\sqrt{(4-\pi)/\tilde{N}}$ and $\sigma_{\bar{\phi}}=2\pi/\sqrt{12}$, and the measured value $r=\bar{r}$ in the signal-dominated regime. 

The last essential tool of harmonic analysis is that providing the calculation of upper limits on the amplitude, which is particularly relevant in the noise-dominated regime. For a given confidence level CL, a first approach, frequentist, consists in searching for the amplitude $r_<$ such that the following relation is satisfied:
\begin{equation}
\label{eqn:ul0}
\int_0^{\bar{r}}\dif\bar{r}'p_{R}(\bar{r}';r=r_{<})=1-\text{CL}.
\end{equation}
According to this expression, any value $r\geq r_<$ should have led, with a probability greater than 1-CL, to a measured value greater than the observed one $\bar{r}$. This is exactly what is intended for when deriving an upper limit, but problems may arise in the case of a sub-fluctuating measurement $\bar{r}$, for there is nothing to prevent this relation from always being satisfied for $r_<\geq\bar{r}$. To circumvent this phenomenon, Linsley proposed, again in that same article~\cite{Linsley1975}, another way, to which he referred as his ``last alternative.'' This last alternative relies on the use of the Bayes theorem to infer the posterior p.d.f. $\tilde{p}_{R}(r;\bar{r})$ of the amplitude $r$ given the measured value $\bar{r}$. With this approach, the value of the underlying signal is no longer a constant to be surrounded, but becomes a variable. Thus, the roles of $\bar{r}$ and $r_<$ in equation~(\ref{eqn:ul0}) can be exchanged and the relation allowing the estimation of the upper limit $r_<$ becomes
\begin{equation}
\label{eqn:ul}
\int_0^{r_<}\dif r \tilde{p}_{R}(r;\bar{r})=\text{CL}.
\end{equation}
Compared to equation~(\ref{eqn:ul0}), this relationship always involves a solution for $r_<$ to be at least of the order of the (frequentist) upper bound for isotropy with the confidence level CL. In this way, and in contrast to the frequentist case, the derived upper limit can never take on values smaller than the sensitivity driven by the available number of events. Making use of the Bayes theorem implies obviously the choice of a prior distribution for the signal. Using a uniform distribution in the $(r,\theta)$ plane as the most general non-informative prior,\footnote{Note that Linsley used a different prior in his last alternative, considering a pure constant. The choice made here accounts for the cylindrical geometry.} the posterior joint p.d.f. $\tilde{p}_{R,\Phi}$ is obtained as
\begin{eqnarray}
\label{eqn:posterior-pdf}
\tilde{p}_{R,\Phi}(r,\phi;\bar{r},\bar{\phi})&=&\frac{r^2p_{R,\Phi}(\bar{r},\bar{\phi};r,\phi)}{\displaystyle \int\dif r\dif\phi~r^2p_{R,\Phi}(\bar{r},\bar{\phi};r,\phi)} \nonumber \\
&=&\sqrt{\frac{2}{\pi}}\frac{r^2}{\pi\sigma}\frac{\exp{\left(-\frac{r^2+\bar{r}^2/2-2r\bar{r}\cos\psi}{2\sigma^2}\right)}}{(\bar{r}^2+2\sigma^2)I_0\left(\frac{\bar{r}^2}{4\sigma^2}\right)+\bar{r}^2I_1\left(\frac{\bar{r}^2}{4\sigma^2}\right)},
\end{eqnarray}
where the upper bound for the integration in $r$ in the denominator is formally considered as infinite to allow for an analytical solution, which is an extremely accurate approximation in the regime of small $\bar{r}$ and small $\sigma$ values. This yields, after marginalisation over $\phi$, to the posterior p.d.f. $\tilde{p}_{R}$ to be used in equation~(\ref{eqn:ul}): 
\begin{equation}
\label{eqn:posterior-pdf}
\tilde{p}_{R}(r;\bar{r})=\sqrt{\frac{2}{\pi}}\frac{2r^2}{\sigma}\frac{\exp{\left(-\frac{r^2+\bar{r}^2/2}{2\sigma^2}\right)}I_0\left(\frac{r\bar{r}}{\sigma^2}\right)}{(\bar{r}^2+2\sigma^2)I_0\left(\frac{\bar{r}^2}{4\sigma^2}\right)+\bar{r}^2I_1\left(\frac{\bar{r}^2}{4\sigma^2}\right)}.
\end{equation}

\subsection{The solar time scale}
\label{subsec:solar}

As already stressed, an accurate knowledge of the directional exposure function $\mu(\alpha)$ is in general very challenging, since it requires an accurate knowledge of the on-time of the detection systems as well as an accurate control of the counting rate with atmospheric changes. In past studies, such subtle time variations of experimental origin have caused some anisotropy claims to be made which have not been confirmed by subsequent experiments. One famous example is that of Hess and Steinmaurer who claimed a $\lesssim 0.1\%$ amplitude of the first harmonic as a function of sidereal time for $\simeq 100$~TeV CRs with 10$\sigma$ confidence, pointing to $\simeq 20~$h~\cite{HessSteinmaurer1933}. This led Compton and Getting to interpret this signal as an apparent effect resulting from the rotation of the Galaxy, providing a ``strong presumption'' that all CRs have an extragalactic origin~\cite{ComptonGetting1935}. This interpretation was based on Doppler effect studies of the globular clusters and the extragalactic nebulae, implying a motion of the Earth of about 300~km~s$^{-1}$ due mainly to the rotation of the Milky Way. As a result of a snow-plow effect, an anisotropy pointing to $\simeq 20.6~$h and with an amplitude similar to the observed one was expected. Subsequent and modern measurements, however, did not confirm the Hess and Steinmaurer values. 

Here, the focus is given to the possible ways that can provide support that the estimation of $\mu$ is accurate enough to probe a sidereal anisotropy. The idea is to repeat the analysis for fictitious right ascension angles $\tilde{\alpha}(\omega)$ calculated with an angular frequency of the Earth's rotation $\omega$ different from the sidereal one $\omega_{\mathrm{sid}}$, 
\begin{equation}
\label{eqn:pseudoalpha}
\tilde{\alpha}(\omega)=\omega t+\alpha-\alpha_0(t)=\omega t-h(t). 
\end{equation}
The interest to consider different angular frequencies than the sidereal one relies on the fact that observations can then be confronted to known expectations. Results at various angular frequencies can be viewed as different measurements as soon as the separation $\Delta\omega$ between two considered values of $\omega$ is larger than the resolution on an individual $\omega$ with data collected during a period of time $\Delta t$, namely $\Delta\omega$ of the order of $2\pi/\Delta t$. 

The study of the solar angular frequency, $\omega_{\text{sol}}=2\pi/24~$rad~h$^{-1}$, is of particular interest. This is because changes of atmospheric conditions are known to modulate the event rate as a function of time due to the sensitivity of the development of an extensive air shower to the atmospheric pressure and air density, as studied for instance in~\cite{AugerAPP2009}. These effects have some influence on the measurement of the deposited energy at a fixed distance from the shower core, and consequently on the measurement of the energy: for some energy $E$ on top of the atmosphere, the energy reconstructed at ground level, $E_{\text{grd}}$, differs according to the atmospheric conditions at the moment when the event is recorded. Neglecting resolution effects, the modulations of the event rate above some fixed threshold energy $E_{\text{thr}}$ reads as
\begin{equation}
\label{eqn:rawrate}
\frac{\dif N}{\dif t}(E_{\text{grd}}>E_{\text{thr}})=A_0\int\dif\nn~\cos{\theta}\int_{>E_{\text{thr}}}\dif E'_{\text{grd}}~\epsilon(E'_{\text{grd}},\nn,t)I(E'_{\text{grd}},\nn,t).
\end{equation}
For a relationship such that $E_{\text{grd}}=(1+\eta(t))E$ to model the dependence of $E_{\text{grd}}$ with pressure and air density, and for $I(E_{\text{grd}})\dif E_{\text{grd}}=I(E)\dif E=KE^{-\gamma}\dif E$ to leading order (that is, neglecting here small anisotropies in the CR intensity, so that both dependences in local angles $\nn$ and time $t$ are neglected in $I$), the modulation factor induced by atmospheric changes thus behave to first order as $1+(\gamma-1)\eta(t)$. Since this factor follows the same dependences as do the pressure and air density with time, spurious variations of the event rate are thus amplified at the solar time scale. 

Different strategies can be followed to account for these undesired variations of the event rate. When estimating the directional exposure with the data themselves as explained in section~\ref{subsec:direxpo}, these variations are then encompassed in this effective directional exposure. Alternatively, by determining $\eta$, the observed energies can be converted to the ones that would have been measured at some fixed reference values of pressure and density~\cite{AugerAPP2009}. Conceptually, this latter approach is closer to the actual sequence, but provides a stable rate against atmospheric changes only in the energy range of full efficiency for triggering. The reason is that for a shower initiated by a CR with an energy $E$, the probability to trigger the detector array at ground is expected to be smaller (larger) under atmospheric conditions of pressure and density such that the observed energy $E_{\text{grd}}$ is smaller (larger). This is primarily due to the signal variations caused by atmospheric conditions and to the changes in the number of triggered detectors induced by these signal variations. The remaining variation of the event rate thus reads as
\begin{equation}
\label{eqn:rate_belowsat}
\frac{\dif N}{\dif t}(E>E_{\text{thr}})=A_0\int\dif\nn~\cos{\theta}\int_{>E_{\text{thr}}}\dif E'~\epsilon(E_{\text{grd}}(E',t),\nn,t)I(E',\nn,t),
\end{equation}
which requires additional corrections. 

Once a strategy has been identified and followed to correct for the spurious modulations, the recover of the expected pattern in the harmonic coefficients then provides confidence that the corrections are appropriate. At this angular frequency, the position of the mean Sun is fixed in the sky, conventionally to $\tilde{\alpha}_{\odot}=\pi$ so that the mean solar time $\omega_{\text{sol}}t$ corresponds to noon when the mean Sun culminates at the observer point, as expected. Given that, from the definition of the equatorial coordinate system, the right ascension of the mean Sun is $\alpha_{\odot}=\omega_{\text{orb}}t+\pi$ with $\omega_{\text{orb}}=2\pi/365.25~$rad~day$^{-1}$ the orbital angular frequency of the Earth around the mean Sun, and that the sidereal angular frequency $\omega_{\text{sid}}$ equals $\omega_{\text{sol}}+\omega_{\text{orb}}$, the fictitious right ascension $\tilde{\alpha}_{\text{sol}}\equiv\tilde{\alpha}(\omega_{\text{sol}})$ corresponds in fact to $\tilde{\alpha}_{\text{sol}}=\alpha-\alpha_{\odot}+\pi$, that is, to the right ascension of the mean Sun subtracted to the real right ascension and shifted by $\pi$. In the following, the unit vectors are denoted with a tilde superscript when referring to the solar reference frame.  

The motion of the Earth around the Sun causes a current of particles from the direction opposite to that of its motion. This leads to a small large-scale anisotropy, known as the solar Compton-Getting effect~\cite{ComptonGetting1935,GleesonAxford1968}. The anisotropy shaped by this effect is, to leading order, a dipole vector, the parameters of which can be determined as follows. Let $f(\mathbf{p})$ be the particle distribution function in momenta in the frame of the solar system, so that there are $f(\mathbf{p})\dif^3p$ particles per unit volume in the momentum interval $\dif^3p$ about the momentum $\mathbf{p}$. This distribution is considered to be homogeneous and isotropic in this frame, $f=f(p)$. On moving with velocity $\mathbf{v}_{\text{obs}}$ relative to this frame, an observer on Earth measures a mean intensity governed by the transformed particle distribution function $f^\prime$ in transformed momenta $\mathbf{p}^\prime$. Given that Lorentz invariance requires $f^\prime(\mathbf{p}^\prime)=f(\mathbf{p})$ and that $\mathbf{p}-\mathbf{p}^\prime\simeq p\mathbf{v}_{\text{obs}}/c$ for ultra-relativistic particles, $f^\prime$ can be explicitly expressed to first order as
\begin{equation}
\label{eqn:ff'}
f'(\mathbf{p}')\simeq f(\mathbf{p}')+\frac{p\mathbf{v}_{\text{obs}}}{c}\cdot\nabla_{\mathbf{p}'}f(p')=f(p')+p\frac{\hat{\mathbf{p}}'\cdot\mathbf{v}_{\text{obs}}}{c}\frac{\partial{f}}{\partial{p}'}.
\end{equation}
The differential intensity $I(p)$ is defined such that there are $I(p)\dif p$ particles in the momentum interval $\dif p$ about the momentum $p$ crossing a unit area perpendicular to $\mathbf{p}$ per unit time and per steradian. In the frame of the solar system, considering ultra-relativistic particles with speed $\simeq c$, the intensity is thus $I(p)=cp^2f(p)/4\pi$. To express the intensity $I'$ in the observer frame, it is convenient to express the $f'$ function in terms of its zeroth and first moment as 
\begin{equation}
\label{eqn:f}
f'(\mathbf{p}')\simeq \frac{1}{4\pi}\phi_0' +  \frac{3}{4\pi} \boldsymbol{\phi}_1'\cdot \hat{\mathbf{p}}'+...,
\end{equation}
with the mean density and current per momentum unit defined as
\begin{eqnarray}
\label{eqn:phi_0'1'}
\phi_0'&=&\int \dif\hat{\mathbf{p}}'f'(\mathbf{p}'),  \\
\boldsymbol{\phi}_1'&=&\int \dif\hat{\mathbf{p}}'\hat{\mathbf{p}}' f'(\mathbf{p}').
\end{eqnarray}
Similar quantities can be defined in the solar frame, with $\boldsymbol{\phi}_1=0$. On inserting the r.h.s. of equation~(\ref{eqn:ff'}) into $\phi_0'$ and $\boldsymbol{\phi}_1'$, it turns out that $\phi_0'\simeq\phi_0=4\pi f$ and, with the zenith angle oriented such that $\mathbf{v}_{\text{obs}}\cdot\mathbf{p}'=v_{\text{obs}}\cos{\theta}'$, that the only non-zero term for $\boldsymbol{\phi}_1'$ is
\begin{eqnarray}
\label{eqn:phi_1'}
\boldsymbol{\phi}_1'\simeq \frac{p\mathbf{v}_{\text{obs}}}{c}\int\dif\varphi'\dif\theta'\sin{\theta'}\cos{\theta'}\frac{\partial f}{\partial p'}\simeq\frac{4\pi}{3c}\frac{\partial f}{\partial \ln{p'}}\mathbf{v}_{\text{obs}}.
\end{eqnarray}
Hence, adopting the notation $\nn=-\hat{\mathbf{p}}$ standing for the arrival direction as observed on Earth, the intensity $I'$ reads as
\begin{eqnarray}
\label{eqn:I'}
I'(p,\nn)=I(p)\left(1-\frac{\dif\ln{\phi_0}}{\dif \ln{p}}\frac{\mathbf{v}_{\text{obs}}\cdot\nn}{c}\right).
\end{eqnarray}
The searched anisotropy thus takes the form $1+\mathbf{d}_{\text{CG}}\cdot\nn$, with $\mathbf{d}_{\text{CG}}$ a dipole vector. For $I(p)\propto p^{-\gamma}$, the amplitude of the dipole is $d_{\text{CG}}=(\gamma+2)v_{\text{obs}}/c$, while its direction is along the observer velocity $\mathbf{v}_{\text{obs}}$ relative to the rest frame of reference. 

In the case of the Earth motion around the mean Sun, the resulting Compton-Getting dipole is best revealed when analysing arrival directions on the solar angular frequency. If the rotation axis of the Earth were not tilted relative to the normal of the ecliptic plane, the Compton-Getting dipole vector would be simply aligned with orbital motion of the Earth at any time, the direction of which would be observed as fixed in this solar reference frame, $(\tilde{\alpha}_{\text{sol,CG}},\tilde{\delta}_{\text{sol,CG}})=(\pi/2,0)$, or $\mathbf{d}_{\text{sol,CG}}=d_{\text{CG}}~\tilde{\mathbf{u}}_2^0$. The time-dependent direction of the vector in the sidereal reference frame would then be obtained by means of a series of two rotations. A first time-dependent rotation with angle $-\omega_{\text{sol}}t$ around $\mathbf{u}_3^0$ would express the dipole vector in the local system, in co-rotation with the Earth. Then,  the coordinates of the vector in the sidereal frame would be obtained through a second time-dependent rotation with angle $+\omega_{\text{sid}}t$ around $\mathbf{u}_3^0$. Making use of the relation $\omega_{\text{sid}}=\omega_{\text{sol}}+\omega_{\text{orb}}$, the searched expression would be $\mathbf{d}_{\text{sid,CG}}(t)=d_{\text{CG}}\cos{\omega_{\text{orb}}t}~\mathbf{u}_2^0-d_{\text{CG}}\sin{\omega_{\text{orb}}t}~\mathbf{u}_1^0$. In the sidereal frame, the vector would therefore rotate and make a complete revolution in one year. While collecting data during several years, an observer would thus, as expected, detect the vector in the solar frame, but not in the sidereal frame due to the average time-integration of $\mathbf{d}_{\text{sid,CG}}(t)$ which oscillates around zero with an envelope amplitude decreasing as one over the integration time.

The tilt $\iota$ of the rotation axis of the Earth relative to the ecliptic plane makes things, however, somehow more tricky in the sense that, after several years of data taking, an observer is expected to detect an amplitude slightly reduced by a factor $\cos^2{\iota/2}$ in the solar reference frame~\cite{CutlerGroom1986}. Since the sidereal reference frame is the only one remaining fixed with time, it turns out that the most convenient approach to derive this factor is to express the solar Compton-Getting dipole first in this frame, and then to proceed with the change of frame in a similar way as described above, with inverted operations though. In the sidereal frame, the effect of the tilt is to induce a non-zero component along $\mathbf{u}_3^0$:
$$\mathbf{d}_{\text{sid,CG}}(t)=d_{\text{CG}}(\cos{\iota}\cos{\omega_{\text{orb}}t}~\mathbf{u}_2^0+\sin{\iota}\cos{\omega_{\text{orb}}t}~\mathbf{u}_3^0-\sin{\omega_{\text{orb}}t}~\mathbf{u}_1^0).$$
After a first rotation with angle $-\omega_{\text{sid}}t$ and a second one with angle $+\omega_{\text{sol}}t$ around $\mathbf{u}_3^0$, some elementary algebra leads to
$$\mathbf{d}_{\text{sol,CG}}(t)=d_{\text{CG}}\left((\cos^2{\iota/2}-\sin^2{\iota/2}\cos{2\omega_{\text{orb}}t})~\tilde{\mathbf{u}}_2^0+\sin{\iota}\cos{\omega_{\text{orb}}t}~\tilde{\mathbf{u}}_3^0-\sin^2{\iota/2}\sin{2\omega_{\text{orb}}t}~\tilde{\mathbf{u}}_1^0\right).$$
The instantaneous vector is thus experiencing a non-trivial motion in the solar reference frame. If an observer could access this instantaneous vector, this observer would detect an anisotropy with amplitude $d_{\text{CG}}$, but would see the dipole vector pointing to a different position with time. Observers, however, perform during long-term observations. While collecting data during several years, the average time-integration of $\mathbf{d}_{\text{sid,CG}}(t)$ then tends to the constant vector $\mathbf{d}_{\text{sol,CG}}\simeq d_{\text{CG}}\cos^2{\iota/2}~\tilde{\mathbf{u}}_2^0$. The expected solar amplitude is thus reduced by a factor $\cos^2{\iota/2}$, as used in~\cite{CutlerGroom1986}, and the right ascension of maximum of intensity is thus fixed to $\tilde{\alpha}_{\text{sol}}=\pi/2$. Note that a recent and similar demonstration can be found in~\cite{AhlersMertsch2016}. 

With $v_{\text{obs}}=29.8~$km~s$^{-1}$, $\iota=23.4^\circ$, and $\gamma=2.7$ typical of the spectral index below the knee energy, the amplitude of the average $\mathbf{d}_{\text{sol,CG}}$ vector turns out to be very small, $d_{\text{sol,CG}}=4.5\times10^{-4}$. When measuring this effect by means of the harmonic analysis in (fictitious) right ascension only, an additional projection factor enters into play since the measurement is folded into the declination response of the experiment. For a pure dipolar intensity, the relation between $d_{\text{sol,CG}}$ and $r_{\text{sol,CG}}$ follows $ r_{\text{sol,CG}}\simeq\langle \cos{\tilde{\delta}}\rangle d_{\text{sol,CG}}$, where $\langle\cdot\rangle $ stands for the average over declination of the response of the considered detector to an isotropic intensity. This factor encompasses the latitude of the site on Earth and the efficiency of the detector as a function of the zenithal angle. A significant measurement of this effect requires a number of events as large as several hundreds of millions to beat the statistical fluctuations, which can only be met for energies below $\simeq 100~$TeV. It also requires an extremely accurate control of $\mu(\tilde{\alpha};\omega_{\text{sol}})$. This effect has been successfully measured in several places, first by Cutler and Groom~\cite{CutlerGroom1986}.

\subsection{The anti- and extended-sidereal time scales}
\label{subsec:antisid}

Although results examined at different angular frequencies separated by $\Delta\omega$ are fully decoupled from each other as soon as the collection time $\Delta t$ is long enough so that $2\pi/\Delta t<\Delta\omega$, subtle correlations between some angular frequencies can occur due to long-term variations of the daily counting rate. Such effects were understood by Farley and Storey to lead to sidebands which can show up at the sidereal time scale~\cite{FarleyStorey1954}. A simple example is a counting rate related to temperature, the daily variations of which would be greater in winter than in summer. 

To see the sideband effect at work, let us follow the original argument put forward by Farley and Storey. Let us suppose that the rate undergoes, in addition to a solar diurnal modulation and a sidereal modulation, some variation of the diurnal amplitude along the year with amplitude $2B$. The observed counting rate then takes the form
\begin{eqnarray}
\label{eqn:n_t_exp}
\frac{\dif N(t)}{\dif t}=\frac{1}{\Delta t}\left(N_0+\left[A+2B\cos{(\omega_{\text{orb}}t+\varphi_2)}\right]\cos{(\omega_{\text{sol}}t+\varphi_1)}+C\cos{(\omega_{\text{sid}}t+\varphi_3)}\right).
\end{eqnarray}
Some algebra allows this rate to be expressed as a sum of cosines at different angular frequencies:
\begin{eqnarray}
\label{eqn:n_t_exp}
\frac{\dif N(t)}{\dif t}&=&\frac{1}{\Delta t}\big(N_0+A\cos{(\omega_{\text{sol}}t+\varphi_1)}+B\cos{((\omega_{\text{sol}}-\omega_{\text{orb}})t+\varphi_1-\varphi_2)} \nonumber\\
&&+~B\cos{((\omega_{\text{sol}}+\omega_{\text{orb}})t+\varphi_1+\varphi_2)}+C\cos{(\omega_{\text{sid}}t+\varphi_3)}\big).
\end{eqnarray}
There are therefore two terms with amplitude $B$ appearing at angular frequencies $\omega_{\text{sol}}+\omega_{\text{orb}}$ and $\omega_{\text{sol}}-\omega_{\text{orb}}$. The first one is nothing else but the sidereal angular frequency, and the second one has been called the anti-sidereal angular frequency. The net result of a sidereal search is a term behaving as $D\cos{(\omega_{\text{sid}}t+\varphi_4)}$, with $D$ and $\varphi_4$ polluted by $B$ and $\varphi_2$. Hence, an amplitude $B$ standing out from the background noise at the unphysical anti-sidereal time scale, be it in terms of the counting rate and/or of the fictitious right ascension $\tilde{\alpha}_{\text{a-s}}=(\omega_{\text{sol}}-\omega_{\text{orb}})t-h(t)$, is thus indicative of important spurious effects of instrumental origin in the measurement of the first harmonic coefficients at the sidereal time scale through a sideband mechanism.

Formally, similar correlations can occur between the annual and the sidereal time scales. In this case, a non-zero sidereal modulation would pollute the solar modulation and falsify the measurement, when the number of events is large enough to probe it, of the solar Compton-Getting effect. The unphysical angular frequency to probe is then called extended-sidereal. The fictitious right ascension is then obtained through $\tilde{\alpha}_{\text{o-s}}=(\omega_{\text{sol}}+2\omega_{\text{orb}})t-h(t)$.

\subsection{A directional-exposure independent method: the East/West method}
\label{subsec:ew}

To circumvent the difficulty of estimating in practice the directional exposure, alternative methods have been designed to measure the harmonic coefficients while avoiding to introduce any corrections of the counting rate. One of them is the East-West method. The original idea was proposed in the early 1940s to study asymmetries in the intensity of solar CRs (see, e.g.,~\cite{Kolhorster1941,AlfvenMalmfors1943}. The approach presented here follows from~\cite{Bonino2011}.

Let us separate the field of view of an observatory into two sectors, the geographic Eastern and Western ones in terms of local coordinates. Both the Eastern and the Western field of views experience different kind of variations during a sidereal day, of experimental and/or astrophysical origin. The effects of experimental origin, being largely independent of the incoming direction, are expected to affect the same way both sectors and thus to cancel when considering, conventionally, the Western counting rate subtracted to the Eastern one. On the flip side, in the presence of a genuine dipolar distribution of CRs, the difference between the Eastern and the Western counting rates is not expected to cancel. Indeed, as the Earth rotates eastward, the Eastern sky is closer to the dipole excess region for half a sidereal day each sidereal day; then, after the field of view has traversed the excess region, the Western sky becomes closer to the excess region and thus bears higher counting rates than the Eastern sky. The East/West differential counting rate is thus subject to oscillations, the amplitude and phase of which are expected to be related to those of the genuine large-scale anisotropy.

These qualitative arguments can be quantified formally, with some approximations and restrictions. The Eastern and Western counting rate definitions follow from equation~(\ref{eqn:rawrate}), restricting the integration over azimuth to each sector. Dropping the energy dependence for simplicity, they read as
\begin{eqnarray}
\label{eqn:erate}
\tau_E(t)&=&A_0\int_{-\pi/2}^{\pi/2}\dif\varphi\int_0^{\theta_{\text{max}}}\dif\theta~\sin{\theta}\cos{\theta}\epsilon(\nn,t)I(\nn,t),\\
\label{eqn:wrate}
\tau_W(t)&=&A_0\int_{\pi/2}^{3\pi/2}\dif\varphi\int_0^{\theta_{\text{max}}}\dif\theta~\sin{\theta}\cos{\theta}\epsilon(\nn,t)I(\nn,t).
\end{eqnarray}
The aim is to relate the difference $\tau_E-\tau_W$ to the counting rate $\tau$ that would be obtained with a perfectly stable detector, that is, without any time dependence in the efficiency function:
\begin{equation}
\label{eqn:idealrate}
\tau(t)=A_0\int_{0}^{2\pi}\dif\varphi\int_0^{\theta_{\text{max}}}\dif\theta~\sin{\theta}\cos{\theta}\epsilon_0(\nn)I(\nn,t).
\end{equation}
The time dependence of the ideal rate $\tau$ comes exclusively  from the variation of the intensity in local coordinates. One immediate limitation is that $\tau$ keeps track of a genuine anisotropy in $I$ only for limited instantaneous fields of view in terms of right ascension. In contrast, for observatories located close to one pole of the Earth such as Ice Cube/Ice Top, the whole range of right ascension is constantly in the field of view. In this case, this definition of the counting rate leads to a constant value, no matter the anisotropy amplitude in $I$. 

To guarantee that the Eastern and Western sectors are equivalent in terms of counting rates, any dependence of $\epsilon$ in azimuth needs to be symmetrical. For simplicity, a uniform detection efficiency is assumed hereafter in azimuth; but similar conclusions still hold as long as the symmetry between the sectors is respected, which is a reasonable assumption in practice. It is also reasonable to assume that the temporal variations $\zeta$ of the efficiency are small, and that those variations decouple from the zenith angle dependent ones, $\epsilon(\nn,t)=\epsilon_0(\theta)(1+\zeta(t))$. On inserting, to fix the ideas, a dipolar intensity into equations~(\ref{eqn:erate}) and~(\ref{eqn:wrate}), that is $I(\nn(\alpha,\delta))=I_0(1+\mathbf{d}\cdot\nn)/4\pi$ with $\mathbf{d}=d\cos{\delta_{\text{d}}}\cos{\alpha_{\text{d}}}~\uu_1^0+d\cos{\delta_{\text{d}}}\sin{\alpha_{\text{d}}}~\uu_2^0+d\sin{\delta_{\text{d}}}~\uu_3^0$ expressed in the $(\uu_x,\uu_y,\uu_z)$ basis, the integrations in local angles lead to
\begin{eqnarray}
\label{eqn:ewrates}
\tau_E(t)&=&\frac{A_0I_0}{4\pi}\left(\pi g_{11}(t)+2d_x(t)g_{12}(t)+\pi d_zg_{21}(t)\right),\\
\tau_W(t)&=&\frac{A_0I_0}{4\pi}\left(\pi g_{11}(t)-2d_x(t)g_{12}(t)+\pi d_zg_{21}(t)\right).
\end{eqnarray}
where $g_{ij}(t)=(1+\zeta(t))f_{ij}$ and $f_{ij}=\int\dif\theta~\epsilon_0(\theta)\cos^i{\theta}\sin^j{\theta}$. To leading order, that is, neglecting second-order terms in $\zeta d$, the $\tau_E(t)-\tau_W(t)$ difference is thus
\begin{eqnarray}
\tau_E(t)-\tau_W(t)\simeq\frac{-A_0I_0}{\pi}f_{12}d_x(t).
\end{eqnarray}
Meanwhile, the derivative of the ideal rate $\tau$ is $\dif\tau/\dif t=-A_0I_0\cos{\lambda}f_{21}d_x(t)/2$, so that the searched relation reads as
\begin{eqnarray}
\label{eqn:ewrate}
\tau_E(t)-\tau_W(t)\simeq\frac{2}{\pi\cos\lambda}\frac{f_{12}}{f_{21}}\frac{\dif\tau}{\dif t}.
\end{eqnarray}
Noticing that the ratio $f_{12}/f_{21}$ can be estimated from any data set through the empirical average ratio $\langle\sin\theta\rangle/\langle\cos\theta\rangle$, the Eastern and Western counting rates can thus be related to the ideal counting rate through a simple integration with a proportionality factor which can be entirely determined from any data set. Hence, from the estimate of the first harmonic coefficients defined as $\bar{a}^{\text{c}}_{\text{EW}}=\frac{2}{N}\sum_i\cos{(\alpha_{0i}+\zeta_i)}$ and $\bar{a}^{\text{s}}_{\text{EW}}=\frac{2}{N}\sum_i\sin{(\alpha_{0i}+\zeta_i)}$ with $\zeta_i=0$ if the event is coming from the East or $\zeta_i=\pi$ if coming from the West, unbiased estimates of the amplitude and phase can be obtained as
\begin{eqnarray}
\label{eqn:ew_amp_phase}
\bar{r}_{\text{EW}}=\frac{\pi\cos\lambda}{2}\frac{f_{21}}{f_{12}}\sqrt{(\bar{a}^{\text{c}}_{\text{EW}})^2+(\bar{a}^{\text{s}}_{\text{EW}})^2}, \hspace{1cm} \bar{\phi}_{\text{EW}}=\arctan{(\bar{a}^{\text{s}}_{\text{EW}}/\bar{a}^{\text{c}}_{\text{EW}})}+\pi/2,
\end{eqnarray}
where the factor $\pi/2$ is to account for the passage from $\dif\tau/\dif t$ to $\tau$. 

\begin{wrapfigure}{L}{8. cm}
{\includegraphics[width=0.45\textwidth]{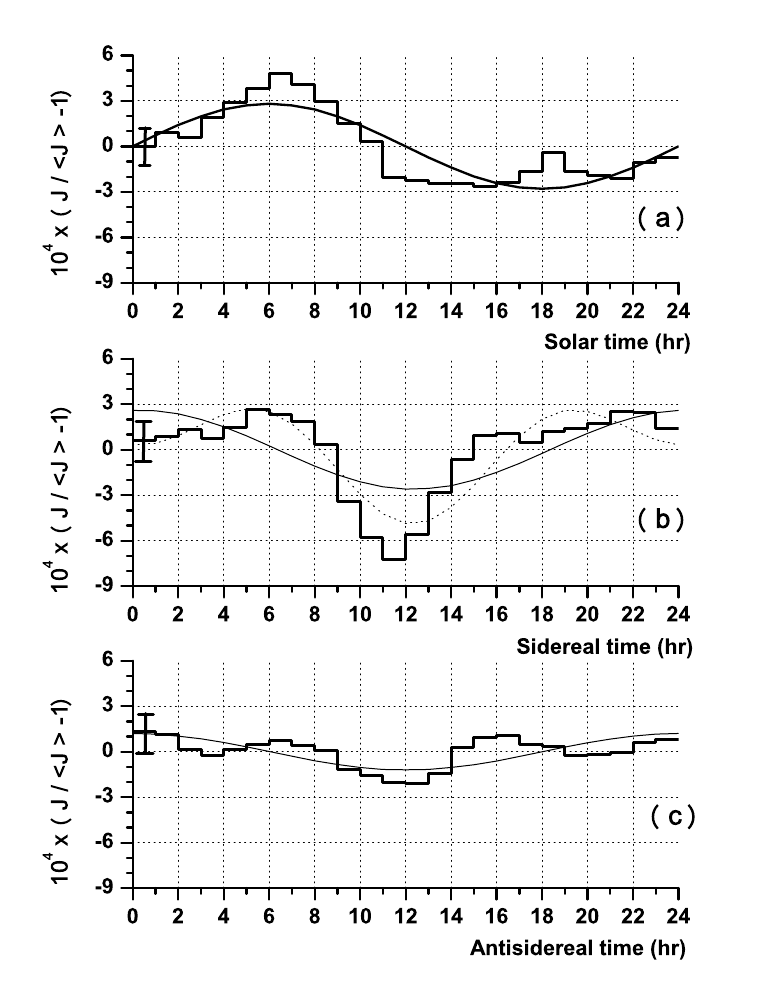}}
\caption{Counting rate curves in solar (a), sidereal (b) and anti-sidereal (c) time as measured at the EAS-TOP observatory~\cite{EASTOP2009}.}
\label{fig:eastop_3timescales}
\end{wrapfigure}

This method has thus the advantage of avoiding the need to correct the total counting rate for instrumental and atmospheric effects. But this can be applied for moderate latitudes of observation only (the local and equatorial coordinate systems need to be ``well separated''), and this is with the cost of a reduced sensitivity to pick up an anisotropy, since, compared to the optimal uncertainties $\sigma_{\bar{r}}$ previously discussed, the uncertainties scale as
\begin{eqnarray}
\label{eqn:ew_resolution}
\sigma_{\bar{r}_{\text{EW}}}=\frac{\pi\cos\lambda}{2}\frac{f_{21}}{f_{12}}\sigma_{\bar{r}}.
\end{eqnarray}
For most of the cases, the scaling factor is about 2.5, so that about 6 times more data are needed to detect a non-zero amplitude with this method compared to the optimal one. 

An example of application of this method is shown in figure~\ref{fig:eastop_3timescales}, resulting from the EAS-TOP experiment~\cite{EASTOP2009}. About 1.5$\times 10^9$ events with energies above $1.1\times10^{14}$~eV were analysed in this study. The East/West counting rates shown on this plot are emblematic of the use of the harmonic analysis technique for interpreting CR counting rates in a robust way: on the one hand, a null result at the anti-sidereal time scale; and on the other hand, an amplitude and a phase compatible with that expected from the Compton-Getting effect at the solar time scale.  The statistical uncertainty for each time bin is given in the first one. The signal reported as a function of local sidereal time has a first harmonic (thin line) with about 3$\sigma$ confidence, while the East/West counting rate is improved by the addition of a second harmonic (dotted line).

\subsection{Review of observations}
\label{subsec:review}

In this subsection, some attention is given on the observational results of the first harmonic in right ascension as a function of energy, together with some basic interpretation of these results. More educated interpretations will be given in next sections. It should be noted that the amplitude measurement in right ascension is a quantity that depends on the characteristics of each experiment, through the latitude of the site and the particular shape of the directional exposure in the considered declination band. There is no way to ``deconvolve'' these distortion effects without knowing the shape of the underlying intensity\footnote{This is because any family of anisotropies described by $Y_{\ell 1}$ and $Y_{\ell -1}$ spherical harmonic functions contribute to the first harmonic in right ascension.}. Furthermore, the energy resolutions are generally different as well as the detection efficiencies relative to different primaries, so that the all-particle anisotropy could be slightly different from the one reported by an experiment having better sensitivity to some primaries. Energy scales are also affected by systematic uncertainties. The comparison of the results between several experiments is therefore not easy, and it is preferable to look at the evolution of the coefficients as a function of the energy in a global way, without seeking in the dispersion of the points any big disagreement between all the existing measurements.

\begin{figure}[!h]
\centering\includegraphics[width=1.0\textwidth]{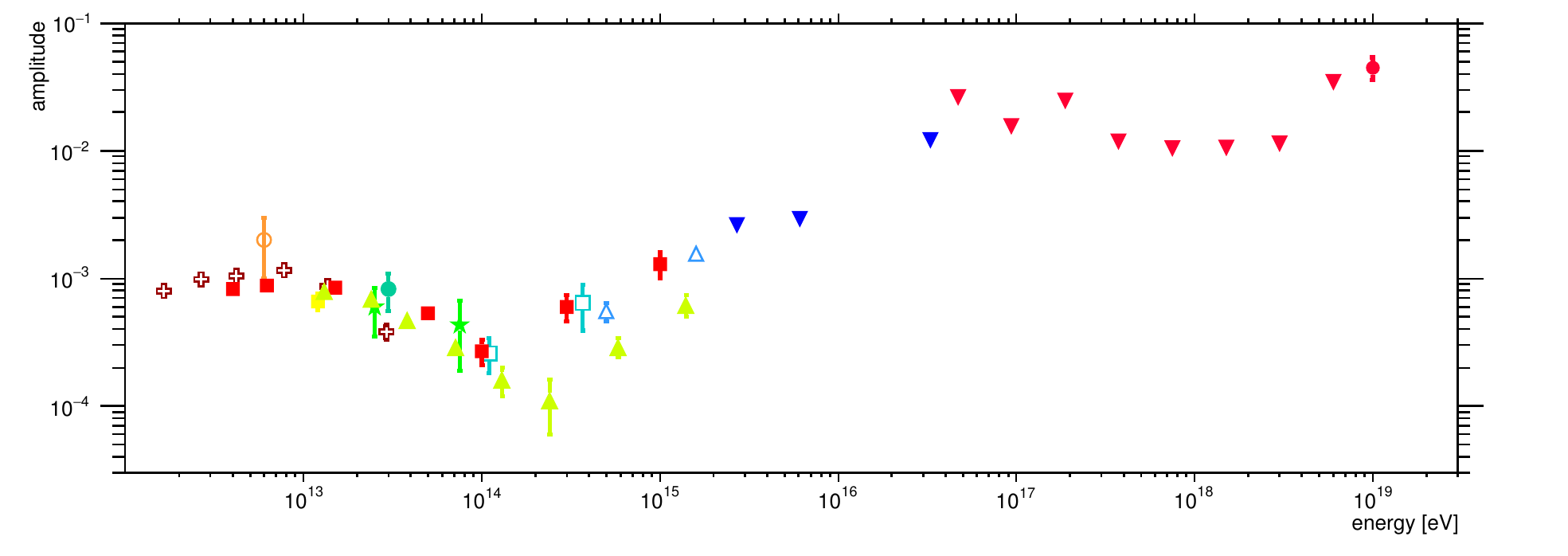}
\centering\includegraphics[width=1.0\textwidth]{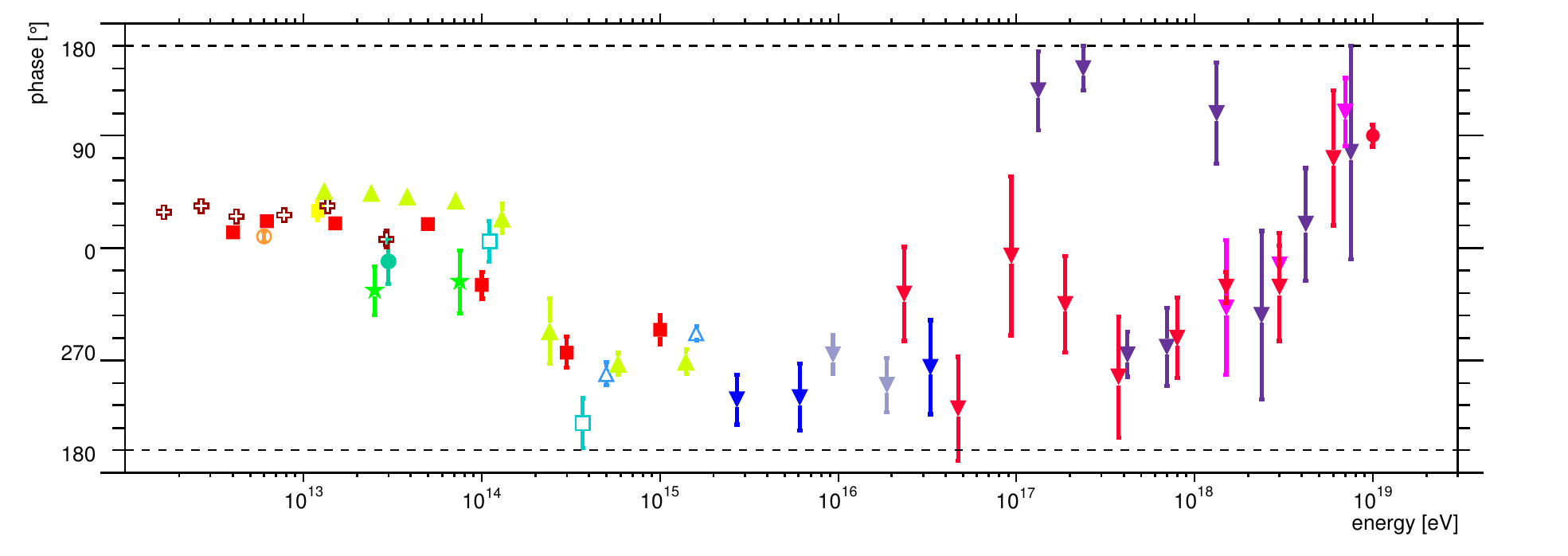}
\centering\includegraphics[width=1.0\textwidth]{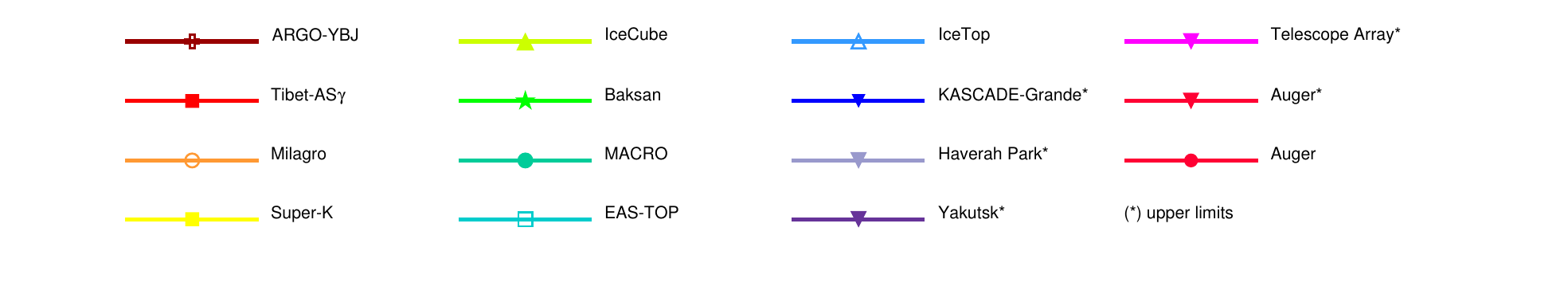}
\caption{Amplitude (top) and phase (bottom) measurements of the first harmonic in right ascension as a function of energy, from various reports. Amplitudes drawn as triangles with apex pointing down are the most stringent upper limits up to date in the considered energy ranges.}
\label{fig:ampphase}
\end{figure}

In the TeV--PeV energy range, complex patterns have been revealed in the arrival directions of CRs thanks to the large statistics collected in the last decade by several experiments, and anisotropy contrats at the $10^{-4}-10^{-3}$ level are now established at large scales. Consistent measurements from experiments located in both hemispheres were reported: MACRO~\cite{MACRO2003}, Tibet AS$\gamma$~\cite{Tibet2005}, Super-Kamiokande~\cite{SK2007}, Milagro~\cite{Milagro2008}, EAS-TOP~\cite{EASTOP2009}, Baksan~\cite{Baksan2009} and ARGO-YBJ~\cite{ARGO2013} in the Northern hemisphere, and IceCube~\cite{IceCube2010} and IceTop~\cite{IceTop2013} in the Southern hemisphere. A collection of amplitude and phase measurements is shown in figure~\ref{fig:ampphase}. The amplitude is observed to increase with energy up to $\simeq 10~$TeV before flattening, and the phase is observed to be smoothly evolving before undergoing a sudden flip at $\simeq 0.3~$PeV. 

Assuming to first order that a first harmonic modulation in right ascension is entirely due to a pure dipolar pattern on the sphere, the first harmonic parameters as derived in figure~\ref{fig:ampphase} are generally considered with special interest in the context of spatial diffusion. The dipole moment is then naively expected to provide a way to probe the particle density gradient shaped by the diffusion in interstellar magnetic fields on scales of the scattering diffusion length. In this picture, for stationary sources smoothly distributed in the Galaxy, the dipole vector should align roughly with the direction of the Galactic center with an amplitude increasing with energy in the same way as the diffusion coefficient, typically $E^{0.3-0.6}$. However, this simple picture is not confirmed by the measurements, showing that the dipole amplitude is not described by a single power law and that the dipole phase does not align with the Galactic center and undergoes a rapid flip at an energy of 0.1-0.3 PeV. Recent studies have put this (too) simple picture into question, and will be discussed farther in \S~\ref{subsec:tevpev}.

At higher energy, in the PeV--EeV energy range, the expected increase of anisotropy contrasts does not compensate, yet, the decrease in the collected statistics with increasing energy. The most constraining data are provided by the IceTop~\cite{IceTop2013}, KASCADE-Grande~\cite{KG2015} and Auger~\cite{AlSamarai2015} experiments. Except for the IceTop amplitude, only upper limits are currently provided in this energy range, drawn as triangles with apex pointing down. Meanwhile, it is interesting to note that an apparent constancy of phase, even though the significances of the amplitudes are relatively small, has been pointed out previously in surveys of measurements with ion chambers and counter telescope made in the range $\simeq 0.1<E/\mathrm{PeV}<100$~\cite{Greisen1962}. A clear tendency for maxima to occur around 20 hours in local sidereal time was observed, not far from the Galactic center. Interestingly, the phases reported by more contemporary experiments (adding data from Haverah Park~\cite{Edge1978} and Yakutsk~\cite{Yakutsk2012} with respect to the collection of amplitudes, where only the most stringent upper limit data appear) also show a consistent tendency to align in the general right ascension of the Galactic center ($\simeq 268^\circ$ in right ascension). 

\begin{wrapfigure}{L}{8. cm}
{\includegraphics[width=0.5\textwidth]{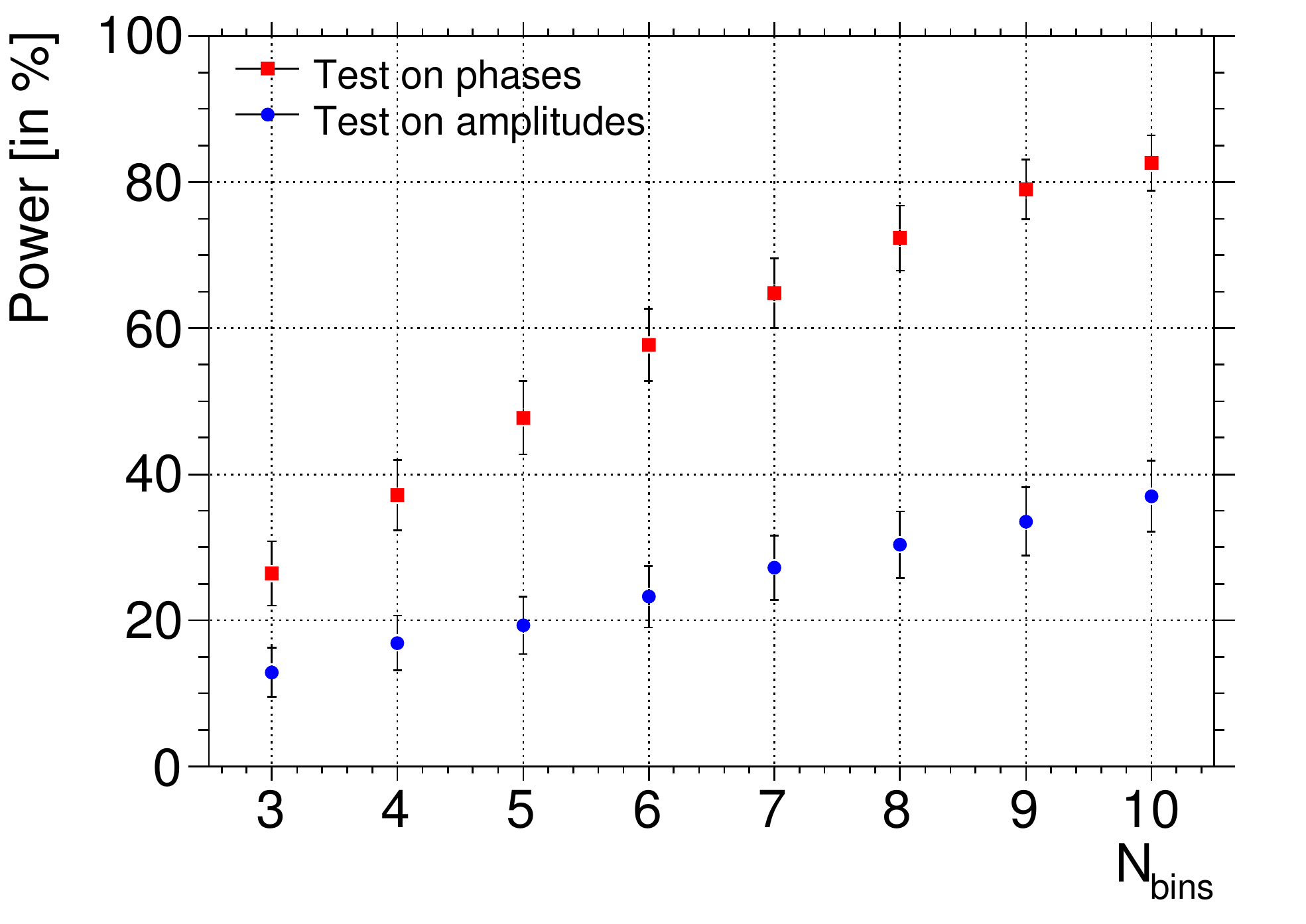}}
\caption{Power of the tests on amplitudes (in blue) and on phases (in red) as a function of the number of bins $N_b$ entering in each test,  in the case of a genuine signal $s=1\%$ and with $N=30,000$ events in each bin.}
\label{fig:power}
\end{wrapfigure}
As already pointed out by Linsley long time ago, the consistency of the phase measurements in ordered energy intervals is potentially indicative of a real anisotropy, because such a consistency is expected to be revealed with a smaller number of events than needed to detect the amplitude with high statistical significance~\cite{Edge1978}. For an anisotropy evolving smoothly with the energy, testing the consistency of independent phase measurements, such as a constancy or a smooth evolution in adjacent energy bins, is indeed a powerful tool for detecting anisotropy. Let us exemplify this based on a simple working example. A likelihood ratio test can be designed to quantify whether or not a parent random distribution of arrival directions better reproduces the phase measurements in different energy intervals than an alternative dipolar parent distribution~\cite{AugerAPP2011}. The likelihood ratio $\Lambda$ is built from the p.d.f. of each independent measurement (let us suppose here $N_b$ independent bins ordered in energy, with $N=30,000$ events in each bin) under the hypothesis of isotropy ($p^{\rm{iso}}_\Phi(\phi)$, uniform distribution) and under the hypothesis of a signal ($p^{\rm{sig}}_\Phi(\phi)$ calculated from equation~(\ref{eqn:phase})). The expected amplitude values $r$ entering into each $p^{\rm{sig}}_\Phi(\phi_i)$ can be estimated by the expected mean noise given the available statistics. The statistic of the variable $-2\ln{\Lambda}$ is then built under the hypothesis of isotropy by means of a large number of Monte-Carlo isotropic samples. The probability that the hypothesis of isotropy better reproduces the measurements compared to the alternative hypothesis is then calculated by integrating the normalised distribution of $-2\ln{\Lambda}$ above the value found in the data set. The power of this test is illustrated in figure~\ref{fig:power}, where its efficiency (shown as the red squares) to detect a genuine anisotropy for a threshold value of 1\% is shown as a function of the number of bins $N_b$. Compared to the efficiency of the "$2K$" test on amplitudes (see~\cite{Edge1978} for details about the "$2K$" test), it is apparent that the test on the consistency of the phase measurements leads to a better power by a factor greater than 2.

Above $1~$EeV, a change of phase is observed in Yakutsk, Telescope Array~\cite{TA2012} and Auger data towards, roughly, the opposite of the one at energies below $1~$EeV. The percent limits to the amplitude of the anisotropy exclude the presence of a large fraction of Galactic protons at EeV energies~\cite{AugerApJS2012,TA2017}. Accounting for the reports from both the Pierre Auger Observatory and the Telescope Array that protons are in fact abundant at those energies, this might indicate that this sub-component is already extragalactic, gradually taking over a Galactic one. The low level of anisotropy would then be the sum of two vectors with opposite directions, naturally reducing the amplitudes. This scenario is further detailed in~\S~\ref{subsec:tevpeveev} and~\S~\ref{subsec:transeev}. 

At higher energies, the Auger collaboration recently reported the existence of a non-zero first harmonic above $8~$EeV by studying the distribution of the arrival directions of more than 30,000 UHECRs~\cite{AugerScience2017}. Although the actual sources of UHECRs are still to be identified, the direction of maximum of intensity, being in opposition of phase to that of the Galactic center, is suggestive that these sources are of extragalactic origin. Extragalactic sources are then expected to produce some anisotropy given the high inhomogeneity at large scale of the matter in the nearby Universe. The observed anisotropy is in line with these general expectations; more will be presented in \S~\ref{subsec:transeev}.

\section{Astrophysical consequences of first harmonic measurements}
\label{sec:astro}

Although first harmonic coefficients in right ascension are only a limited description of the anisotropy function $\delta I(\nn)$, several astrophysical consequences can be drawn from these measurements, consequences which are reviewed in this section. Note that generally, little distinction is made at this stage between the first harmonic coefficients and the 3D dipole moment that will be discussed in~\S~\ref{sec:3dreco}.

\subsection{Diffusion and dipole anisotropy}
\label{subsec:diff}

The strength of the Galactic magnetic field is estimated to be of order several microgauss within a disk of $\simeq 300~$pc thickness in the Galaxy, as inferred from optical and synchrotron polarization measurements and Faraday rotation measures of pulsars and extragalactic sources. It can be roughly described by a structure with arms with reversing field direction between the arms and displaying variations of strength within them~\cite{JanssonFarrar2012,Tinyakov2011}. Meanwhile, there are uncertainties in the way the field falls off along the direction perpendicular to the disk and in the Galactic halo. The main features are a northerly directed poloidal component falling off slowly with the distance from the disk, and oppositely directed toroidal fields in the halo~\cite{JanssonFarrar2012,Tinyakov2011}. Additional turbulent fields with significant variations from arm to arm are also present on correlation lengths of $\simeq 10-100~$pc. Such field values are largely sufficient for particle trajectories to resemble random walks except at very high energies, given that the gyroradius of a CR nucleus with charge $Z$ is $r_{\mathrm{g}}\simeq 1.1\times10^{-3}~(R/\mathrm{TV})/(B/\mu G)~$pc, with $R=pc/(Ze)$ the rigidity of the particle. Over a few scattering times, a diffusion regime develops and the momenta of the particles are ``isotropised'' to some extent. What is meant through ``isotropised to some extent'' is the object of this section. Given the high conductivity of the interstellar medium, electric fields are neglected. 

In the same way as in \S~\ref{subsec:solar}, let us denote the phase-space density function giving the number of CRs at time $t$ in the volume $\mathrm{d}^3x$ about the position $\mathbf{x}$ and in the momentum interval $\mathrm{d}^3p$ about $\mathbf{p}$. In the absence of source terms and of energy losses, the Liouville theorem states that this phase-space density function is conserved, $f(t,\mathbf{x},\mathbf{p})=f(t_0,\mathbf{x}_0,\mathbf{p}_0)$. For relativistic particles, this condition implies for $f$ to satisfy the following transport (Vlasov) equation,
\begin{eqnarray}
\label{eqn:vlasov}
\frac{\partial{f}}{\partial{t}}+c\hat{p}_i\frac{\partial{f}}{\partial{x_i}}+\epsilon_{ijk}p_j\omega_k\frac{\partial{f}}{\partial{p_i}}=0,
\end{eqnarray}
with $\epsilon_{ijk}$ the $ijk$ component of the anti-symmetric tensor $\underline{\epsilon}$ and $\omega_i=ecB_i/E$ the component $i$ of the rotation vector imprinted on the particle trajectories by the magnetic field. Because of the fluctuations in the field leading to fluctuations in the rotation vector components $\omega_i=\langle \omega_i\rangle+\delta \omega_i$, individual test particles undergo irregular motions that can be characterised in a statistical way by introducing fluctuations in the function, $f=\langle f\rangle+\delta f$. Hereafter, the $\langle \cdot\rangle$ symbol stands for the average quantities, taken over several space and time correlation scales of the turbulent field. The fluctuations are considered as ergodic, in the sense that averaging over an ensemble of systems would lead to the same average quantities as through the operation $\langle \cdot\rangle$. Using these ans\"atze for $f$ and $\mathbf{\omega}$ in equation~(\ref{eqn:vlasov}) and carrying out the ensemble averaging yield to the following (Boltzmann) equation for $\langle f\rangle$~\cite{Jokipii1972}:
\begin{eqnarray}
\label{eqn:boltzmann}
\frac{\partial{\langle f\rangle}}{\partial{t}}+c\hat{p}_i\frac{\partial{\langle f\rangle}}{\partial{x_i}}+\epsilon_{ijk}p_j\langle\omega_k\rangle\frac{\partial{\langle f\rangle}}{\partial{p_i}}=-\epsilon_{ijk}p_j\left\langle\delta\omega_k\frac{\partial(\delta f)}{\partial p_i}\right\rangle.
\end{eqnarray}
It thus appears that the effect of the fluctuations in the field is to induce an effective collision term in the r.h.s. of the transport equation governing the evolution of $\langle f\rangle$. This collision term invokes the unknown function $\delta f$. Different approximations can then be adopted to deal with this term.

\subsection{TeV--PeV dipole anisotropies as a probe of the local magnetic field}
\label{subsec:tevpev}

The first approximation, called decay-time approximation~\cite{BKG1954,Welander1954}, consists in replacing the collision term by a relaxation term with rate $\nu$:
\begin{eqnarray}
\label{eqn:approxbkg}
-\epsilon_{ijk}p_j\left\langle\delta\omega_k\frac{\partial(\delta f)}{\partial p_i}\right\rangle \rightarrow -\nu\left(\langle f\rangle-\frac{\phi_0}{4\pi}\right).
\end{eqnarray}
Here, $\phi_0$ is the zeroth moment of the $f$ distribution, as defined in equation~(\ref{eqn:f}). The approximation therefore consists in considering that the effective collisions tend to bring the ensemble-average $f$ function to its isotropic mean. Therefore, by introducing this term into equation~(\ref{eqn:boltzmann}) and taking as in~\cite{Jones1990} the zeroth and first moments with respect to $\hat{\mathbf{p}}$, the following coupled partial differential equations are obtained (after integrating by parts several terms and noting that $p_i\partial(\cdot)/\partial p_i=\hat{p}_i\partial(\cdot)/\partial \hat{p}_i$):
\begin{eqnarray}
\label{eqn:diffusion_coupled}
\frac{\partial{\langle \phi_0\rangle}}{\partial{t}}+c\frac{\partial{\langle \phi_{1j}\rangle}}{\partial{x_i}}\delta_{ij}&=&0,\\
\label{eqn:diffusion_coupled_bis}
\frac{\partial{\langle \phi_{1i}\rangle}}{\partial{t}}+\frac{c}{3}\frac{\partial{\langle \phi_{0}\rangle}}{\partial{x_i}}+\epsilon_{ijk}\langle\omega_j\rangle\langle \phi_{1k}\rangle&=&-\nu\langle \phi_{1i}\rangle.
\end{eqnarray}
Limiting the intensity to the dipolar term, it follows from the definition of $f$ that $\delta I(\nn)=\mathbf{d}\cdot\nn=-3\boldsymbol{\phi}_{1}\cdot\hat{\mathbf{p}}/\phi_0$; solving the set of equations~(\ref{eqn:diffusion_coupled}) and~(\ref{eqn:diffusion_coupled_bis}) hence allows for estimating the average dipole vector expected from the diffusion regime approximation. In this approximation, the dipolar component is assumed to be slowly varying over the relaxation time $1/\nu$, so that $\partial\langle\phi_{1i}\rangle/\partial t$ is about to be null. In these conditions, the way to solve this system of coupled equations is to express the components of the vector $\boldsymbol{\phi}_{1}$ as a function of the scalar $\phi_{0}$, which requires the inversion of the system 
\begin{eqnarray}
\label{eqn:diffusion_inverse}
-\frac{\partial{\langle \phi_{0}\rangle}}{\partial{x_i}}=\frac{3\nu}{c}\langle \phi_{1j}\rangle\delta_{ij}+\frac{3}{c}\epsilon_{ikj}\langle\omega_k\rangle\langle \phi_{1j}\rangle.
\end{eqnarray}
The r.h.s. can formally be written as $cD^{-1}_{ij}\langle \phi_{1j}\rangle$ with $\underline{D}$ the diffusion tensor. To satisfy the identity $D_{ij}D^{-1}_{jk}=\delta_{ik}$, the most general decomposition of the components of $\underline{D}$ have to obey
\begin{eqnarray}
\label{eqn:difftensor_inverse}
\left(\frac{3\nu}{c^2}\delta_{ij}+\frac{3}{c^2}\epsilon_{imj}\langle\omega_m\rangle\right)\left(A\delta_{jk}+B\langle\omega_j\rangle\langle\omega_k\rangle+C\epsilon_{jkn}\langle\omega_n\rangle\right)=\delta_{ik}.
\end{eqnarray}
By identifying the coefficients $A, B, C$, and noting that $\epsilon_{ijk}\epsilon_{lmn}=\delta_{jm}\delta_{kn}-\delta_{jn}\delta_{km}$, the diffusion tensor is conventionally written as
\begin{eqnarray}
\label{eqn:difftensor}
D_{ij}=\frac{c^2}{3\nu_{\parallel}}\hat{b}_i\hat{b}_j+\frac{c^2}{3\nu_{\perp}}(\delta_{ij}-\hat{b}_i\hat{b}_j)+\frac{c^2}{3\nu_A}\epsilon_{ijk}\hat{b}_k,
\end{eqnarray}
with $\langle\omega_i\rangle=\langle\omega\rangle \hat{b}_i$ ($\hat{\mathbf{b}}$ being the unit vector of the coherent field), $\nu_{\parallel}=\nu$ the effective scattering rate along the field, $\nu_{\perp}=(\nu^2+\langle\omega\rangle^2)/\nu$ the perpendicular one, and $\nu_{A}=(\nu^2+\langle\omega\rangle^2)/\langle\omega\rangle$ the axial one. 

\begin{wrapfigure}{L}{8. cm}
{\includegraphics[width=0.45\textwidth]{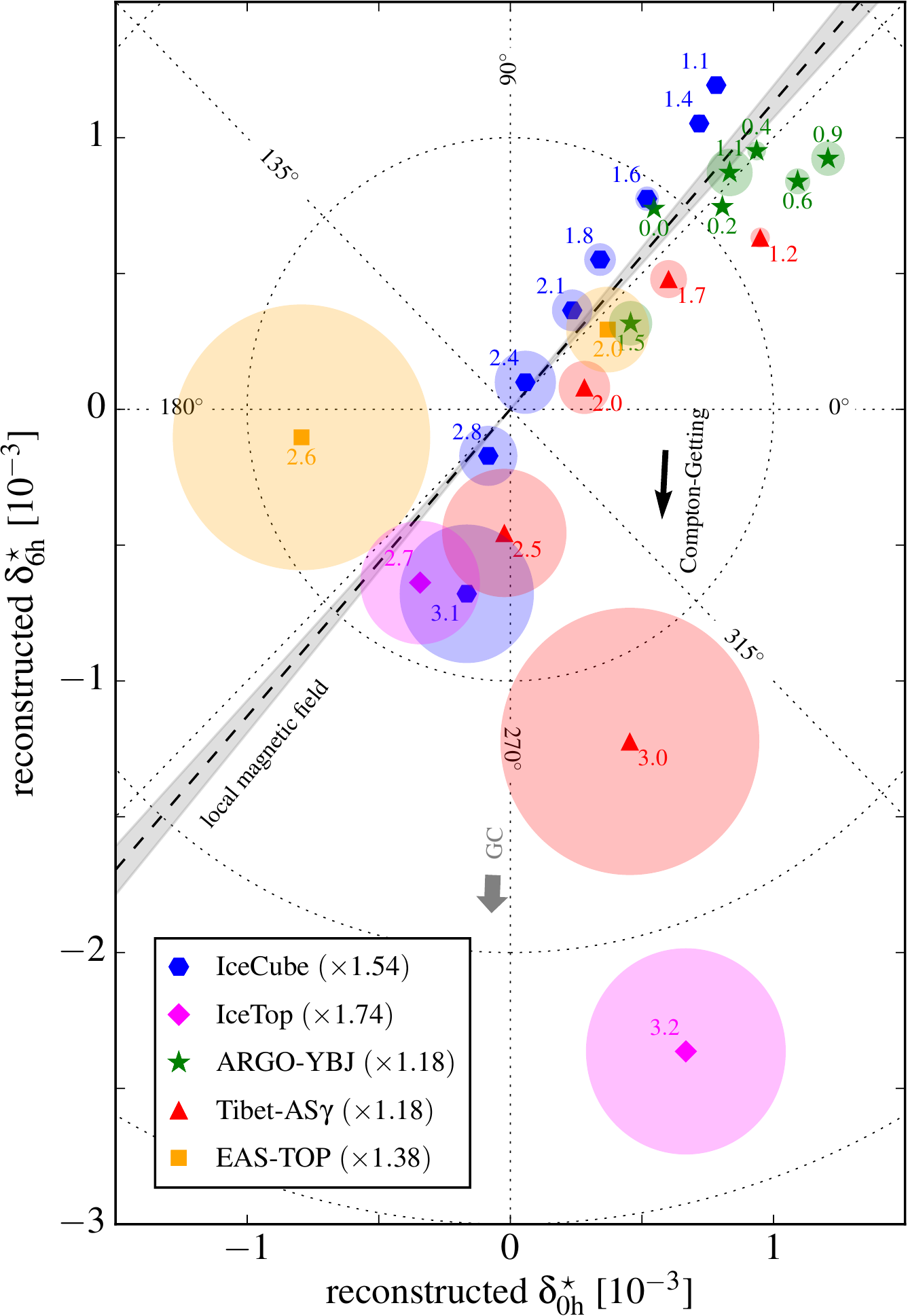}}
\caption{Reconstructed equatorial components of the dipole as a function of energy (decimal logarithm of the median energy in TeV indicated next to each data point), after subtraction of the dipole induced by the Compton-Getting effect~\cite{Ahlers2016}.}
\label{fig:reconstructed_delta}
\end{wrapfigure}
Having the expression of $\underline{D}$, it is now straightforward to invert equation~(\ref{eqn:diffusion_inverse}). The result is a Fick equation for $\langle\boldsymbol{\phi}_{1}\rangle$,
\begin{eqnarray}
\label{eqn:fick}
\langle\phi_{1i}\rangle=-\frac{1}{c}D_{ij}\frac{\partial{\langle \phi_{0}\rangle}}{\partial{x_j}},
\end{eqnarray}
which allows for estimating the dipole anisotropy of the intensity at any time and any position as
\begin{eqnarray}
\label{eqn:diffdipole}
d_{i}=-3\frac{\langle \phi_{1i}\rangle}{\langle \phi_{0}\rangle}=\frac{3}{c}D_{ij}\frac{\partial{(\ln{n})}}{\partial{x_j}},
\end{eqnarray}
where $n=p^2\phi_0$ is the spectral density of CRs. This is the searched expression for interpreting the results presented in \S~\ref{subsec:review} in the TeV--PeV energy range. In our local environment, the magnetic field lines, as deduced both from the IBEX ribbon~\cite{IBEX} and from polarisation of optical starlight from stars within a few tens of parsecs~\cite{Frisch2012,Frisch2015}, are thought to extend in a coherent tube. When the coherent field is much stronger than the turbulent one, the anisotropic diffusion induced by the local ordered magnetic field is such that circular motions predominate over random scattering. In this case, $\nu_{\parallel}\gg\nu_\perp$ and $\nu_\parallel\gg\nu_A$, and from equation~(\ref{eqn:diffdipole}), the dipole anisotropy is expected to result from the projection of the density gradient of CRs onto the direction of the ordered magnetic field, and thus to be aligned along the field lines. This is indeed what is observed~\cite{Schwadron2014}, aside from some distortions attributable to the draping of the magnetic field lines around the heliosphere on the one hand, and from the small Compton-Getting effect arising from the motion of the solar system relative to the frame of the local plasma. 

The summary plot of the reconstructed TeV--PeV dipole components in the equatorial plane, as obtained in~\cite{Ahlers2016}, is shown in figure~\ref{fig:reconstructed_delta}. The small Compton-Getting effect is here subtracted by assuming that the local standard of rest corresponds to the local plasma frame. The dipole vector is here characterised by its components in the equatorial plane associated to the unit vectors pointing towards local sidereal times 0~h and 6~h. The numbers attached to the data indicate the median energy of the bins as $\log_{10}{(E_\mathrm{med}/\mathrm{TeV})}$. The dashed line and gray-shaded area indicate the magnetic field direction and its uncertainty (projected onto the equatorial plane) inferred from IBEX observations. A close alignment of the inferred dipole components with the local magnetic field direction is quite clearly depicted. Overall, within this framework of anisotropic diffusion, although the projection of the dipole onto the magnetic field axis does not allow one to reconstruct the CR gradient, a range of possible directions for this CR gradient can be determined by the dipole phase and aligns with Galactic longitudes between 120$^\circ$ and 300$^\circ$ below 0.1-0.3~PeV. In this regard, and among other requirements, the Vela SNR is shown to be a plausible source candidate for explaining the anisotropy in~\cite{Ahlers2016}. \\

So far, only the fact that the solar system is immersed in a tube of field lines oriented coherently in a certain volume has been used to interpret the TeV--PeV large-scale anisotropy in right ascension, anisotropy characterized by its first harmonic only. Recently, it has been put forward that the shape of the right ascension distribution, beyond the first harmonic, can potentially constrain the power spectrum of turbulence in the local interstellar environment~\cite{Giacinti2017}. Assuming that locally, the turbulence scale is about a few tens of parsecs, this turbulence would indeed have the characteristics of a coherent field for any CR penetrating into this zone. Some attention is now given to this promising study.

The approach to derive these constraints differs from the previous one in that it relies on a Fokker-Planck-type treatment of the equation~(\ref{eqn:boltzmann}). The primary goal this time is to obtain this time an equation governing $\delta f$, and to introduce the corresponding solution in the collision term of equation~(\ref{eqn:boltzmann}) without resorting to the decay-time approximation. Following~\cite{Jokipii1972}, the starting point is the ensemble-averaged equation~(\ref{eqn:vlasov}) from which equation~(\ref{eqn:boltzmann}) is subtracted. This gives to first order the following equation
\begin{eqnarray}
\label{eqn:boltzmann_df}
\frac{\partial{(\delta f)}}{\partial{t}}+c\hat{p}_i\frac{\partial{(\delta f)}}{\partial{x_i}}+\epsilon_{ijk}p_j\langle\omega_k\rangle\frac{\partial{(\delta f)}}{\partial{p_i}}=-\epsilon_{ijk}p_j\delta\omega_k\frac{\partial\langle f\rangle}{\partial p_i},
\end{eqnarray}
whose solution is
\begin{eqnarray}
\label{eqn:df}
\delta f(\mathbf{x},\mathbf{p},t)=\delta f(\mathbf{x}_0,\mathbf{p}_0,t_0)-\epsilon_{ijk}\int_{t_0}^t\dif t'\left[p_j\delta\omega_k\frac{\partial\langle f\rangle}{\partial p_i}\right]_{U(t')},
\end{eqnarray}
where the integration is carried out along the unperturbed trajectory $U(t')$ (that is, considering only the coherent field). On inserting this solution into the collision term in the r.h.s. of equation~(\ref{eqn:boltzmann_df}), one finds that 
\begin{eqnarray}
\label{eqn:collision_df}
-\epsilon_{ijk}p_j\left\langle\delta\omega_k\frac{\partial{(\delta f)}}{\partial{p_i}}\right\rangle&=&\epsilon_{imn}\epsilon_{jkl}p_m\left\langle\delta\omega_n\frac{\partial}{\partial p_i}\int_{t_0}^t\dif t'\left[p_k\delta\omega_l\frac{\partial\langle f\rangle}{\partial p_j}\right]_{U(t')}\right\rangle\nonumber \\
&-&\epsilon_{ijk}p_j\left\langle\delta\omega_k\frac{\partial{(\delta f_0)}}{\partial{p_i}}\right\rangle.
\end{eqnarray}
For an homogeneous random field and $c\mu(t-t_0)$ much greater than the coherence length of the field, the integration over $t'$ can be taken from $-\infty$ to $+\infty$ so the integrand becomes independent of $t_0$, and so is the second term since the l.h.s. is itself independent of $t_0$. In this case, the second term is actually zero since the correlation vanishes if one extends $t_0$ sufficiently back to a time such that $\mathbf{x}-\mathbf{x}_0$ is much greater than the coherence length of the field~\cite{Jokipii1972}. Considering in addition the special case in which the constant average field is in the $z-$direction, $\langle f\rangle$ depends only on $p_z, z,$ and $t$, and only the pitch-angle scattering $\mu=p_z/p$ is then relevant. This is the typical ansatz of a quasi-linear theory~\cite{Jokipii1966,Hall1967}, in which a particle travelling along the $z-$axis at a velocity $c\mu$ is scattered in pitch angle as it interacts with the magnetic irregularities. For small changes of orbits in a correlation length, and on inserting the non-zero part of the r.h.s. of equation~(\ref{eqn:boltzmann_df}) into equation~(\ref{eqn:boltzmann}), an average over gyrophase and an integration along $U(t')$ yields to the Fokker-Planck equation for $\langle f\rangle$:
\begin{eqnarray}
\label{eqn:fokker_planck}
\frac{\partial{\langle f\rangle}}{\partial{t}}+c\mu\frac{\partial{\langle f\rangle}}{\partial{z}}=\frac{\partial}{\partial{\mu}}\left(D_{\mu\mu}\frac{\partial{\langle f\rangle}}{\partial{\mu}}\right).
\end{eqnarray}
The pitch-angle diffusion coefficient $D_{\mu\mu}$ depends on the way the statistical properties of the random field are modeled. In the quasi-linear approach, for an outer scale of the turbulence $l$, this coefficient is the result of the following sum~\cite{Kulsrud1969}:
\begin{eqnarray}
\label{eqn:Dmumu}
D_{\mu\mu}=\langle\omega\rangle^2(1-\mu^2)\int\dif\mathbf{k}\sum_{n=-\infty}^\infty\left(\frac{n^2J_n^2(u)}{u^2}\mathcal{I}_{\mathrm{A}}(\mathbf{k})+\frac{k_\parallel^2J_n^{'2}(u)}{k^2}\mathcal{I}_{\mathrm{S,F}}(\mathbf{k})\right) R_n(k_\parallel v_\parallel-\Omega+n\langle\omega\rangle),
\end{eqnarray}
with $k_\parallel$ and $k_\perp$ the components of $\mathbf{k}$ parallel and perpendicular to the local field lines, $u=k_\perp r_{\mathrm{g}}\sqrt{1-\mu^2}$, $J_n(u)$ the Bessel function of the first kind ($'$ denotes a derivative with respect to $u$), $\mathcal{I}_{\mathrm{A,S,F}}$ the normalised energy spectra of the Alfv\'en, slow and fast modes of the field, and $R_n$ is a resonance function describing the scattering between particles with gyrofrequency $\langle\omega\rangle$ and the angular frequency of the waves $\Omega$. In the limit of the quasi-linear theory, this function is a Dirac one. 

\begin{figure}[!h]
\centering\includegraphics[width=0.49\textwidth]{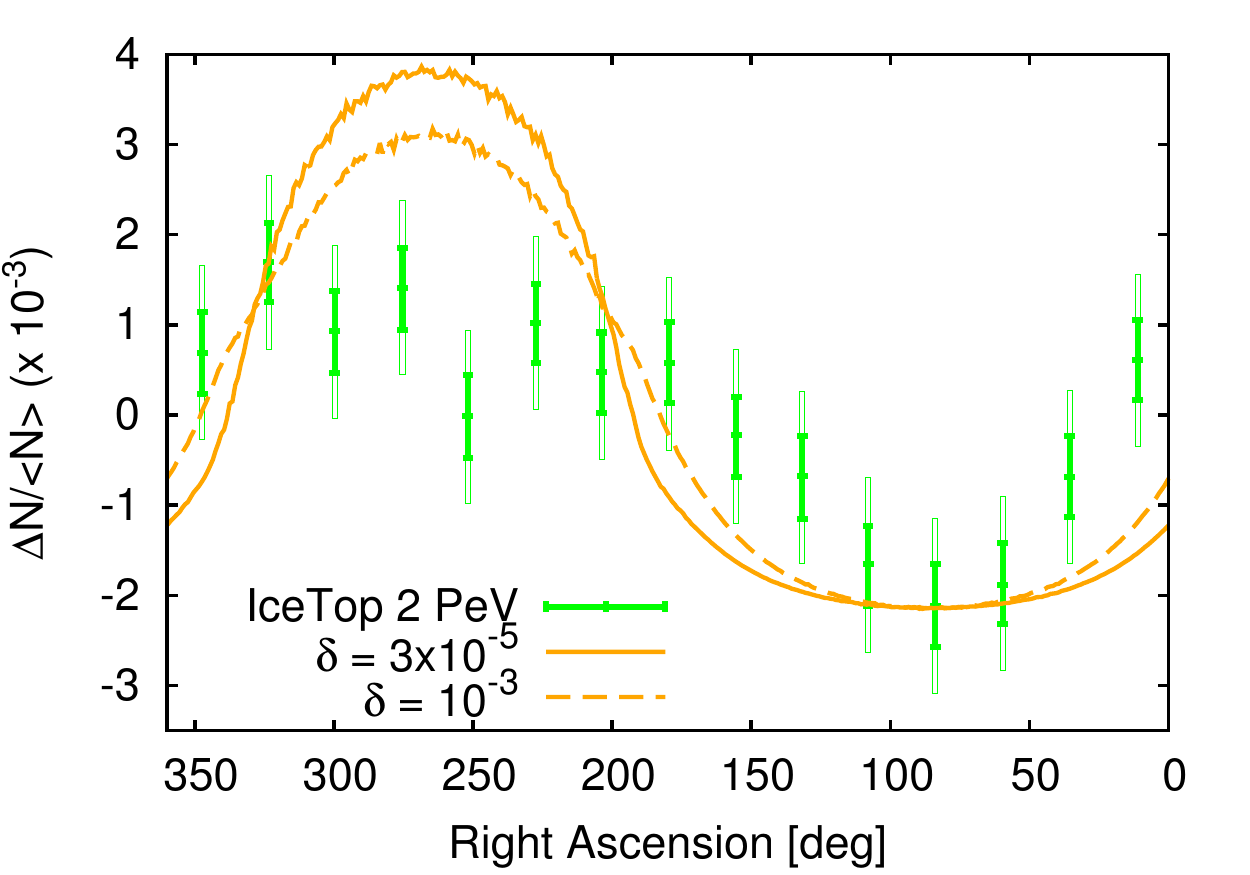}
\centering\includegraphics[width=0.49\textwidth]{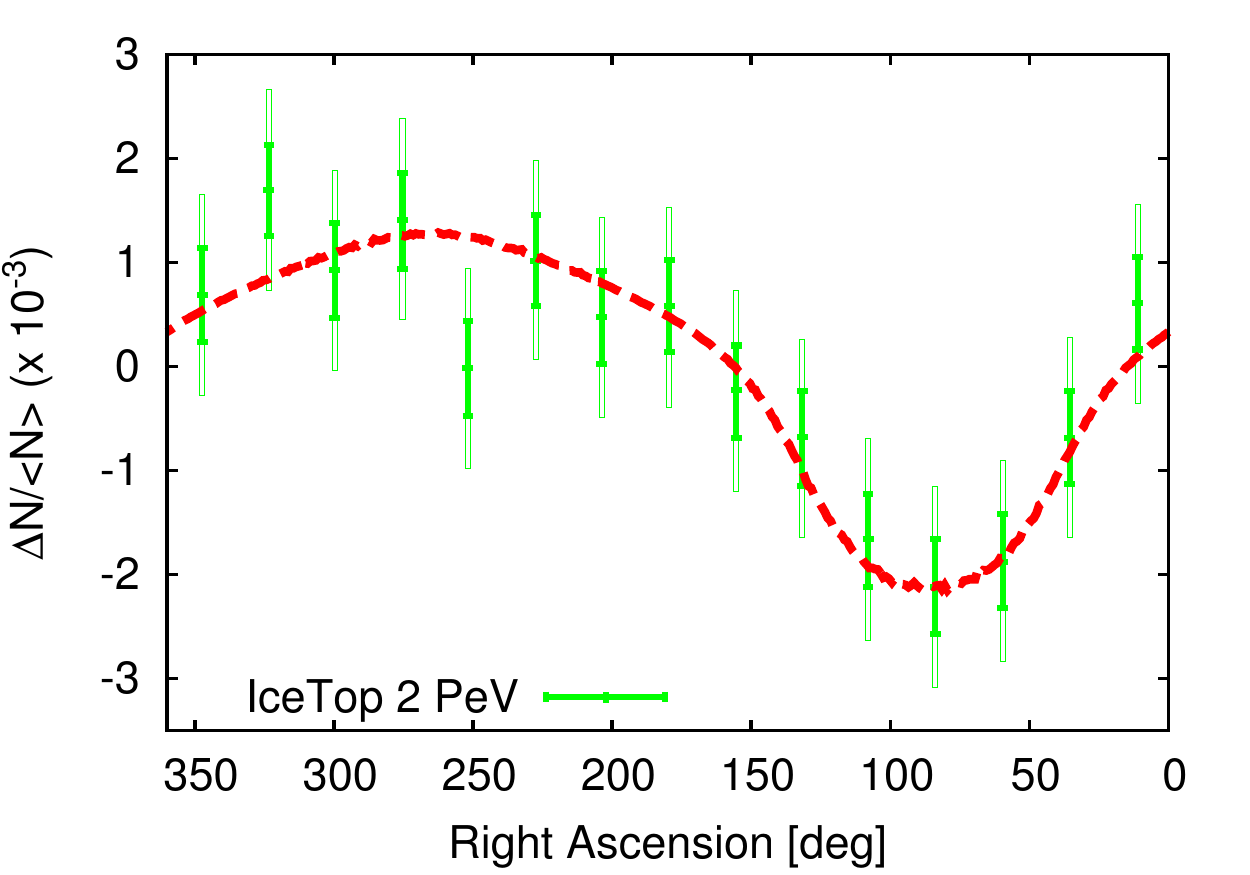}
\caption{Relative CR intensity as a function of right ascension, compared to IceTop data at 2~PeV, for two different models of energy spectrum and resonance function of the local turbulence (see text)~\cite{Giacinti2017}. In the left panel, $\delta$ is the ratio between the interstellar Alfv\'en speed and the CR velocities.}
\label{fig:generalised_dipole}
\end{figure}
From this formalism, investigations of how the shape of the large-scale anisotropy of TeV--PeV CRs depends on the properties of the turbulence can be made. Indeed, in a stationary situation, the solution of equation~(\ref{eqn:fokker_planck}) can be written as $f\propto (1+ag(\mu))$ with $a$ some constant and $g(\mu)$ given by
\begin{eqnarray}
\label{eqn:g}
g(\mu)=-\frac{c}{2}\int_0^\mu\dif\mu'\frac{1-\mu'^2}{D_{\mu'\mu'}}.
\end{eqnarray}
The function $g$ is thus entirely shaping the anisotropy. The dipolar shape is recovered in the case $g(\mu)=\mu$, but a flattening of the anisotropy is generally expected in directions perpendicular to the field lines, something which is compatible with data showing narrower excesses and deficits than those expected from a pure dipole. To estimate $g$, a series of phenomenological inputs needs to be done concerning the considered energy spectra and resonance functions. Different scenarios are investigated in~\cite{Giacinti2017}. Narrow and broad resonance functions are tried for each model of turbulence. The narrow resonance function is modeled as a Breit-Wigner distribution shaped  by the correlation time of the turbulence~\cite{Chandran2000}. In this case, only waves with angular frequency $\Omega$ satisfying closely the condition $k_\parallel v_\parallel-\Omega+n\langle\omega\rangle\simeq 0$ contribute to scatter particles with gyrofrequency $\langle\omega\rangle$. The broad resonance function, following from~\cite{Yan2008}, accounts for the fluctuations of the large-scale magnetic field and thus for the focussing and defocussing of magnetic field lines. This leads to a broadening of the resonance through the variations of the pitch angle. Generic compressible turbulences are modeled with fast modes dominating over Alfv\'en and slow ones and having an isotropic energy spectrum following $\mathcal{I}_{\mathrm{F}}\propto k^{-3/2}$. Incompressible turbulences, following from~\cite{Goldreich1995}, are on the other hand modeled with Alfv\'en and slow modes dominating over fast ones, and having an anisotropic power spectrum such that $\mathcal{I}_{\mathrm{A,S}}\propto k_\perp^{-10/3}\exp{(-k_\parallel l^{1/3}/k_\perp^{2/3})}$~\cite{Cho2002,Yan2002}. Smaller eddies are thus much more elongated. Fluctuations with $|k_\parallel|\geq|k_\perp|^{2/3}l^{-1/3}$ are strongly suppressed in this case.  

Among these different scenarios, let us focus on two distinct ones for exemplify purpose: fast modes with narrow resonance on the one hand, and Alfv\'en and slow modes with broad resonance on the other hand. Results in terms of large-scale anisotropies as reported around 2~PeV using data recorded at IceTop~\cite{IceTop2013} are shown in figure~\ref{fig:generalised_dipole}. It is seen that the first scenario of turbulence (left panel) provides a poor fit of the data. In this case, it turns out that the $n=0$ term dominates the CR transport for pitch angles perpendicular to the magnetic field, resulting in hot/clod spots with a rather flat intensity. In contrast, the second scenario (right panel) provides a satisfactory fit of the data: scattering is now dominated by $n=0$ term of the slow modes for pitch angles perpendicular to the field lines, while the $n=0$ term of the Alfv\'en modes dominates for pitch angles parallel to the field lines, which results in excesses and deficits narrower than in the case of a dipole. 

Hence, CR anisotropy data might be used to constrain at the same time the turbulence in the local interstellar medium and the transport of the particles (through the resonance function). Current data seem to favour moderately broad resonance functions. More systematic studies based on all existing TeV--PeV data in both hemispheres could provide more stringent constraints. This is a technique to be explored further to probe the nature and anisotropy of the local magnetic turbulence.

\subsection{Galactic cosmic-ray density gradients}
\label{subsec:tevpeveev}

Let us focus now on the interpretation of the observed amplitude as a function of energy. As already mentioned in~\S~\ref{subsec:review}, simple arguments in the context of isotropic diffusion predict an increasing amplitude with energy, with an increase expected to scale as the (isotropic) diffusion coefficient (typically as $E^{0.3-0.6}$). Such an increase is not observed in the data shown in figure~\ref{fig:ampphase}. These arguments have often been put forward in the literature to stress the tension between the paradigm of CR sources densely distributed in the disk of the Galaxy such as supernova remnants for instance and the observed anisotropies. This tension has however been alleviated in several studies. 

One important feature to consider within the paradigm of supernova remnants is the stochasticity in the spatial and temporal distribution of these objects. This stochasticity impacts the energy spectrum observed at Earth today~\cite{Busching2005,Blasi2012a}, and the anisotropy as well~\cite{Erlykin2006,Blasi2012b,Sveshnikova2013}. Considering the diffusion formalism in presence of a source term and neglecting for simplicity the energy losses and the rate of spallation of nuclei, the (diffusion) equation governing the transport of CRs from a point source located at a position $\mathbf{x}_{\mathrm{s}}$ and injecting a spectrum $N(E,M)$ of particles of species $M$ at time $t_{\mathrm{s}}$ is obtained by plugging equation~(\ref{eqn:fick}) into equation~(\ref{eqn:diffusion_coupled}):
\begin{eqnarray}
\label{eqn:diffusion}
\frac{\partial{n(E,\mathbf{x},t,M)}}{\partial{t}}=\frac{\partial}{\partial{x_i}}\left(D_{ij}\frac{\partial n(E,\mathbf{x},t,M)}{\partial{x_j}}\right)+N(E,M)\delta(t,t_{\mathrm{s}})\delta(\mathbf{x},\mathbf{x}_{\mathrm{s}}).
\end{eqnarray}
In~\cite{Blasi2012b}, the source term is randomly probed to reproduce the rate of supernova remnants along the spiral arms of the Galaxy. The boundary conditions for the diffusion region are chosen to be a cylinder with infinite radius and half height $H$ (mimicking a turbulent magnetic field concentrated within the Galactic disk), and the escape of the particles is considered to occur through the upper and lower boundaries of the cylinder. The escape is thus modeled here by imposing $n(z=\pm H)=0$. For a spatially constant diffusion coefficient, the Green function that satisfies these conditions for each species $M$ is obtained through the image charge method\footnote{Note that the solution is here simplified, since the rate of spallation of nuclei was also considered in the reference~\cite{Blasi2012b} driving the discussion.}:
\begin{eqnarray}
\label{eqn:green}
G(\mathbf{x},t,\mathbf{x}_{\mathrm{s}},t_{\mathrm{s}})=\frac{N(E)}{(4\pi D(t-t_{\mathrm{s}}))^{3/2}}\exp{\left(-\frac{(\mathbf{x}_{\mathrm{s}}-\mathbf{x})^2}{4D(t-t_{\mathrm{s}})}\right)}\sum_{k=-\infty}^{+\infty}(-1)^k\exp{\left(-\frac{(z-z'_k)^2}{4D(t-t_{\mathrm{s}})}\right)},
\end{eqnarray}
where $z'_k=(-1)^kz_{\mathrm{s}}+2kH$ are the $z-$coordinates of the image sources. 

From this Green function, the gradient of CRs at Earth can be determined for any random realisation of the underlying distribution of point-like and bursting sources. This gradient includes both the contribution from the large-scale distribution of (old) sources, and from the random configuration of nearby and recent sources. However, it turns out that for a halo size below $\simeq 5~$kpc, the contribution of the nearby sources dominates over the ``background'' diffuse emission. So, even if their contribution to the total intensity is only subdominant, local and young sources are found to shape the anisotropy at Earth. Because the most contributing sources, located at different places, can have different ``ages'', they do not contribute the same way as a function of energy. As a result of the summation of competing gradient vectors associated to each source, the amplitude of the total vector can undergo bumps and dips as a function of energy. In particular, the observed behaviour of the evolution of the amplitude with energy in the TeV--PeV range is not unlikely to occur for a diffusion coefficient scaling as $E^{1/3}$, a halo size $H=2~$kpc, and a rate of supernova remnants of 1/100 per year~\cite{Blasi2012b}.

As mentioned in~\S~\ref{subsec:tevpev}, the anisotropy observed on Earth is fueled by the density gradient, but it is ultimately shaped by the projection of this density gradient along the local field. This projection itself can generate non-trivial effects depending on the energy, resulting from the fact that particles of different rigidities have different effective scattering lengths and thus probe different regions of the magnetic field. These effects are most amplified for a gradient almost perpendicular to the local field direction~\cite{Mertsch2015}: large fluctuations can then be observed in the evolution of the amplitude with energy for different realisations of the random field; and in particular, the observed behaviour is likely to be reproduced through this mechanism. \\

From a few PeV to a few hundred of PeV energies, results presented in~\cite{Blasi2012b} suggest as a relevant possibility an increasing amplitude that could reach the percent level. An additional difficulty to frame in a relevant way the density gradients in this energy range is that, for coherence lengths of turbulences between 10~pc and 100~pc and gyration radius of particles lying roughly within 1~pc and 1~kpc, the spiral geometry of the coherent field is not negligible any longer to describe the diffusive transport. This translates into the use of the full diffusion tensor (cf. expression~(\ref{eqn:difftensor})). In a spiral geometry, the gradient density is expected to be predominantly shaped by the axial term of $\underline{D}$ inducing drift motions once the gyration radius of the particles becomes larger than the coherence length of the random fields~\cite{Ptuskin1993}. For stationary sources distributed in the disk, the gradient is then perpendicular to the field and anisotropies are in average amplified, growing linearly with energy~\cite{Candia2002}. However, the effect of the fluctuations due to the random distribution of the sources in space and time has not yet been studied in the context of anisotropic diffusion. Overall, and as already stressed in~\S~\ref{subsec:review}, increased statistics is necessary to probe the anisotropy contrast levels that may exist in this energy range, as suggested by the phase almost-constancy. \\

\begin{wrapfigure}{L}{7.8 cm}
{\includegraphics[width=0.48\textwidth]{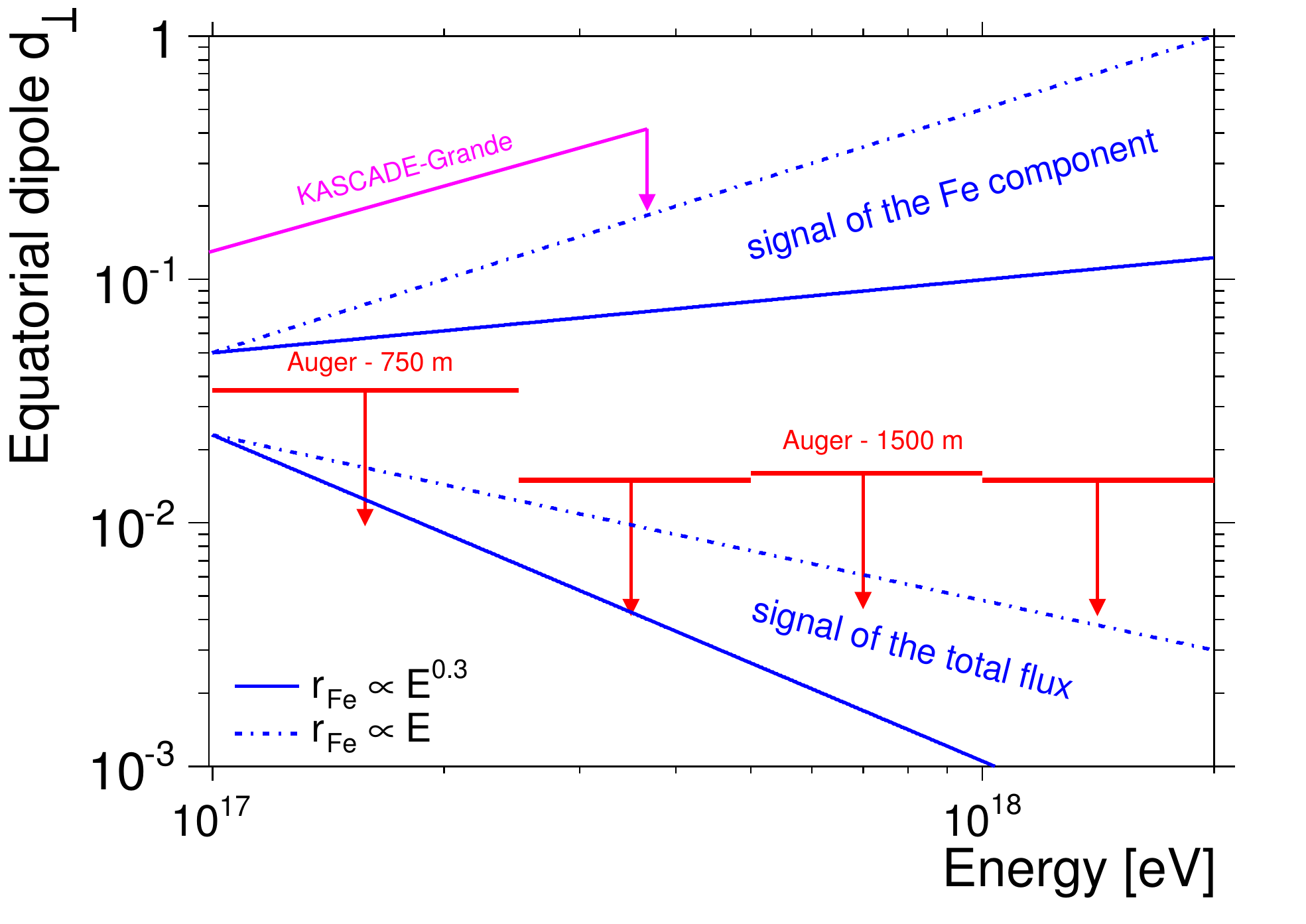}}
\caption{Illustration of a mechanism authorising a large first-harmonic amplitude for the end of the Galactic
  component -- see text~\cite{Deligny2014}.}
\label{fig:toyanis}
\end{wrapfigure}

Above a few hundred of PeV, the putative dipole-like signal responsible for the phase alignment is required to have a rather low amplitude, being below the percent level to meet the conditions fixed by the upper limits obtained from the whole population of CRs detected between $\simeq 100~$PeV and $\simeq 1~$EeV. Although hardly predictable in a quantitative way given the unknown source distributions in space and time and the difficulty to model the propagation of particles around $\simeq 1~$EeV in the intervening magnetic fields known only approximately, such an amplitude seems too low to be naturally explained by a Galactic scenario only. A relevant possibility to explain such a low global level of anisotropy could be that the component of heavy elements marking the end of the Galactic CRs is steadily extinguishing from $\simeq 100~$PeV up to $\simeq 1~$EeV, and does show important large-scale anisotropies in its distribution of arrival directions. The anisotropy of this subdominant component of heavy nuclei could be diluted in an almost isotropic component of extragalactic origin -- likely protons in that energy range, in agreement with the observational results on composition -- so that the anisotropy of the all-particle component could satisfy the upper limits obtained between $\simeq 100~$PeV and $\simeq 1~$EeV. To illustrate this scenario, let us consider the total intensity $I(E,\nn)$ as the sum of a dominant isotropic component $I_X(E)$ and of a subdominant anisotropic component $I_{\mathrm{Fe}}(E,\nn)$ of iron-nuclei elements:
\begin{equation}
I(E,\mathbf{n})=I_X^0E^{-\gamma_X}+I_{\mathrm{Fe}}^0E^{-\gamma_{\mathrm{Fe}}}(1+d_{\mathrm{Fe}}(E)~\hat{\mathbf{d}}\cdot\mathbf{n}).
\end{equation}
The anisotropy of the iron component is here characterised by an energy-dependent amplitude $d_{\mathrm{Fe}}(E)$
and the directions of the unit vector $\hat{\mathbf{d}}$. The anisotropy amplitude of the total intensity is then diluted as follows:
\begin{equation}
d(E)=\frac{I_{\mathrm{Fe}}^0E^{-\gamma_{\mathrm{Fe}}}}{I_X^0E^{-\gamma_X}+I_{\mathrm{Fe}}^0E^{-\gamma_{\mathrm{Fe}}}}d_{\mathrm{Fe}}(E).
\end{equation}
For reasonable choices of the normalisation parameters and spectral indexes such that $\gamma_{\mathrm{Fe}}$ is larger than $\gamma_{\mathrm{X}}$, a \textit{decreasing} amplitude $d(E)$ with energy can be naturally obtained in this scenario even for an anisotropy of the iron component increasing with energy and lying above the all-particle upper limits. This is illustrated in figure~\ref{fig:toyanis} for two \textit{toy} energy evolutions of the anisotropy inspired from diffusion-dominated ($d_{\mathrm{Fe}} \propto E^{0.3}$) or drift-dominated ($d_{\mathrm{Fe}} \propto E$) propagation of CRs in the Galactic magnetic field. Although lots of fine-tunings are involved in this simplistic illustration, the global picture depicted in this plot may not be too unrealistic. 

Another possibility to mention is that the anisotropy of the iron component is not diluted, but almost compensated by an anisotropy of the same order imprinted in the arrival directions of the component of light elements. Since the anisotropies considered here are mainly described in terms of dipoles, this condition can be met by considering that the vectors characterising the dipoles of each component have an amplitude of the same order but an almost \textit{opposite} direction. This scenario provides a mechanism to reduce significantly the amplitude of the vector describing the arrival directions of the whole population of CRs -- which is observationally the only one at reach so far --, and to induce a change of phase in a narrow energy interval. Along these lines, it is interesting to note the change of phase observed around $1~$EeV
in figure~\ref{fig:ampphase} between $\simeq 260^\circ$ and $\simeq 100^\circ$ in right ascension. 

Overall, the current limitation of the measurements to provide further insights in this energy range is that neither spectra nor anisotropies can yet be studied as a function of the mass of the particles with adequate statistical precision, measurements that could allow a distinction between Galactic and extragalactic angular distributions above a few hundred of PeV. This requires investing many experimental efforts to identify the primary mass on an event-by-event basis, which represents a fair amount of remaining work.

\subsection{Trans-EeV anisotropies and extragalactic cosmic rays}
\label{subsec:transeev}

\begin{wrapfigure}{R}{8. cm}
{\includegraphics[width=0.5\textwidth]{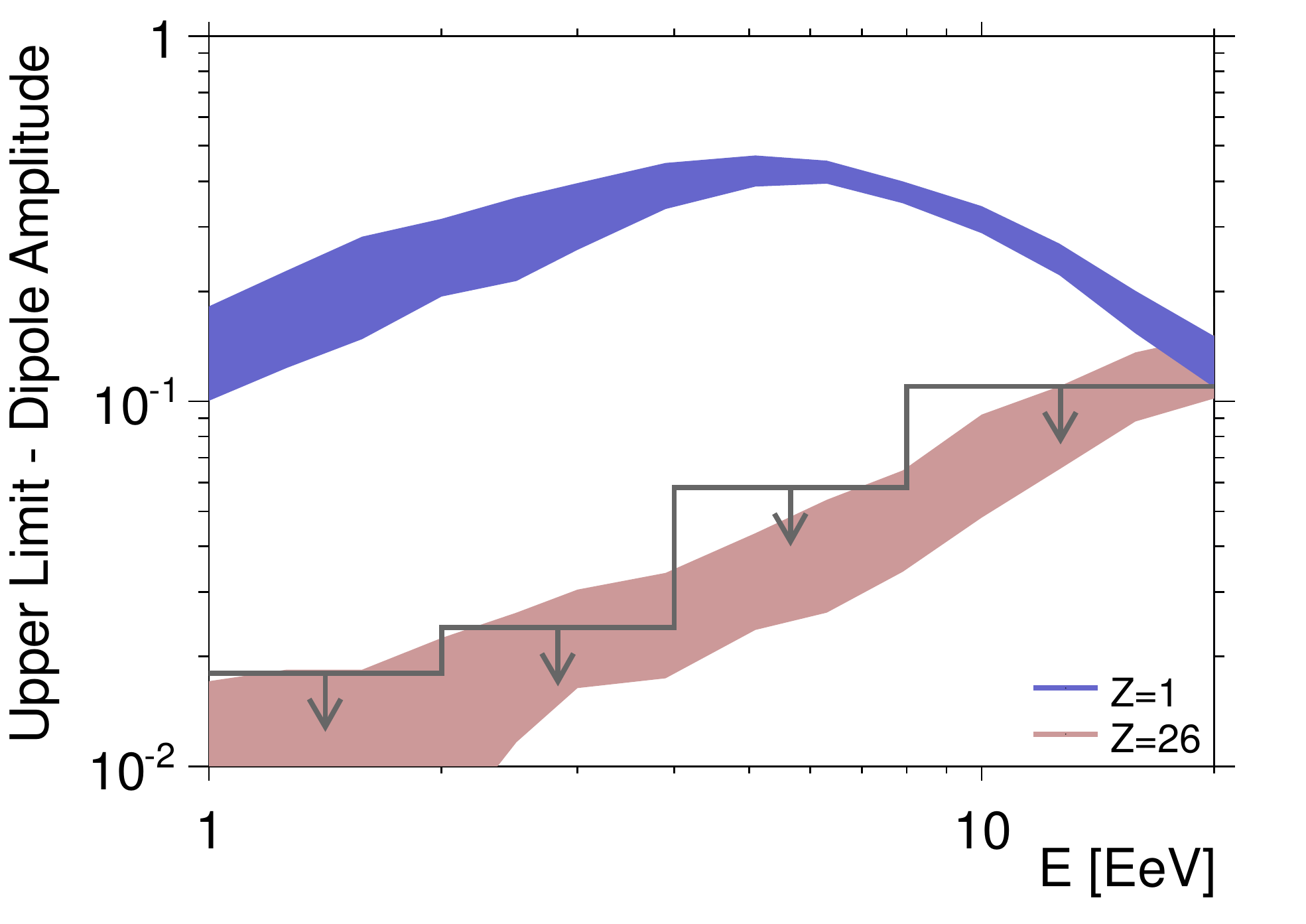}}
\caption{99\% C.L. upper limits on dipole amplitude as a function of energy, as obtained from Auger data~\cite{Almeida2013}. Some generic anisotropy expectations from stationary Galactic sources distributed in the disk are also shown, for two distinct assumptions on the CR composition.}
\label{fig:ul_dipole3d}
\end{wrapfigure}
At EeV energies, there are lots of debates as to whether such particles are produced in the Galaxy or in distant extragalactic objects. Establishing the energy at which the intensity of CRs starts to dominate the intensity of Galactic ones is still, as of today, a fundamental question in astroparticle physics. Although the existence of the ankle -- a hardening of the energy spectrum around 5~EeV first discovered by Linsley at the Volcano Ranch experiment~\cite{Linsley1963} -- is beyond controversy, its interpretation is still unclear. A time-honoured picture is that it may be the spectral feature marking the transition between Galactic and extragalactic CRs. 

Models of expected sky maps from hypothetical Galactic sources of EeV protons, based on numerical integrations of trajectories, show that the arrival directions should reflect to some extent the geometry of the Galaxy~\cite{Ptuskin1998,Giacinti2012,AugerApJS2012,AugerApJL2013,TA2017}, in contrast to what is expected from extragalactic sources. Such signatures have been extensively searched for but not found in the data provided by the Pierre Auger Observatory and the Telescope Array. The most stringent up-to-date upper limits on dipole amplitudes, as reported by the Auger collaboration~\cite{Almeida2013}, are shown in figure~\ref{fig:ul_dipole3d} along with generic estimates of the dipole amplitudes expected from stationary Galactic sources distributed in the disk considering two extreme cases of single primaries: protons and iron nuclei. The percent limits on the amplitude of the anisotropy exclude the presence of a large fraction of Galactic protons at EeV energies. Accounting for the inference from both the Pierre Auger Observatory and the Telescope Array that protons are in fact abundant at these energies~\cite{AugerPRD2014}, the lack of strong anisotropies provides some indication that this component of protons is extragalactic, gradually taking over a Galactic one. Increased statistics is however necessary to probe the anisotropy contrast levels that may exist in this energy range. \\

\begin{figure}[!h]
\centering\includegraphics[width=0.8\textwidth]{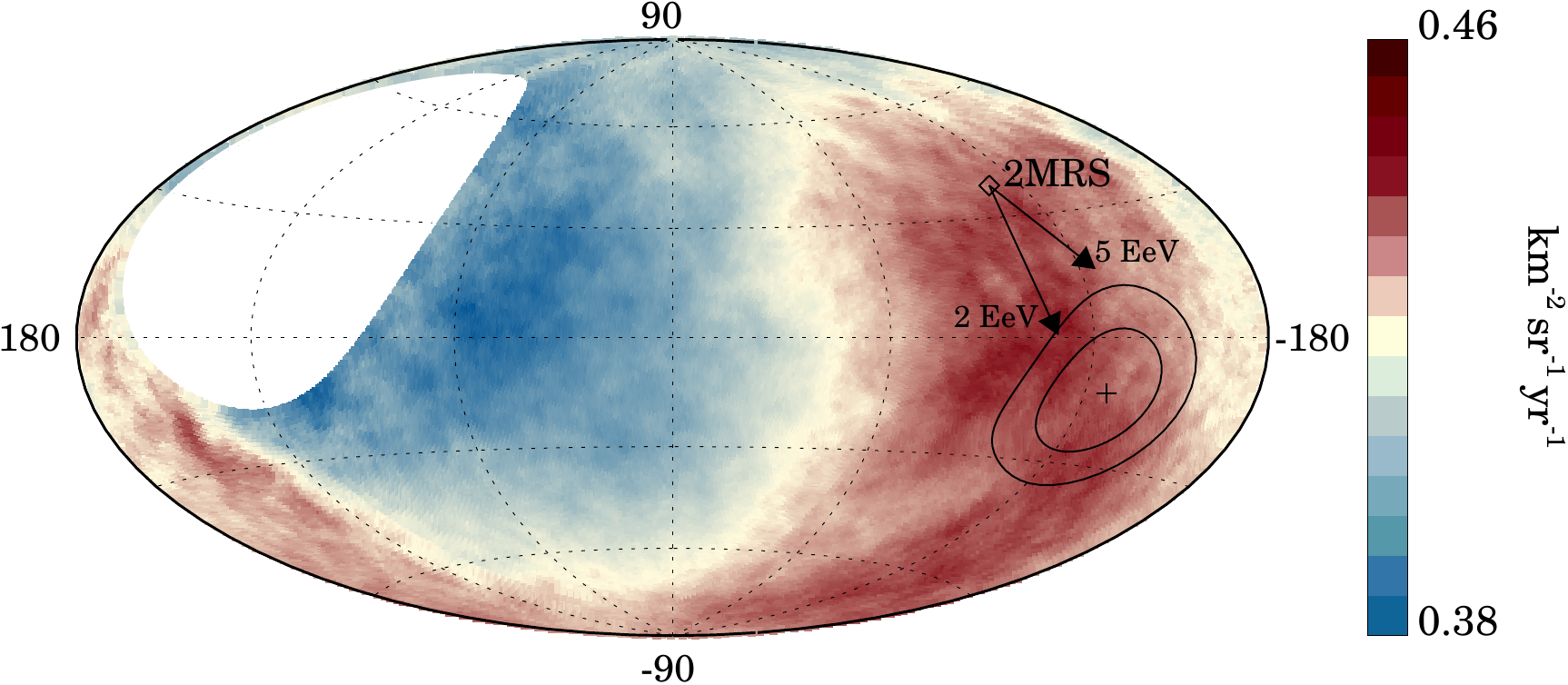}
\caption{Intensity integrated above 8~EeV as a function of the direction in the sky, in Galactic coordinates~\cite{AugerScience2017}. The intensity is smoothed out at a 45$^\circ$ angular scale, to exhibit better the large-scale structures. The measured dipole direction is indicated as the cross, and the contours denote the 68\% and 95\% confidence level regions. The dipole in the 2MRS galaxy distribution~\cite{Erdoglu2006} is indicated, and the deflections expected for a particular model of the Galactic magnetic field on particles with $E/Z=5$ or 2~EeV are shown by the arrows.}
\label{fig:augerdipole}
\end{figure}
At higher energies, above 8~EeV, the anisotropy reported by the Auger collaboration does not reflect, to any extent, the geometry of the Galaxy. The direction of maximum intensity, $(l,b)=(233^\circ,-13^\circ)$, shown as the cross in figure~\ref{fig:augerdipole} within the 68\% and 95\% confidence level regions, cannot be associated with putative sources in the plane or center of our Galaxy for any realistic configuration of the Galactic magnetic field. The natural alternative to match the observed pattern with some prediction is to look at the geometry of extragalactic sources. Such  sources being still unknown, a plausible and conservative hypothesis about their distribution in space is that, regardless of their nature, they follow at large scales the distribution of the baryonic matter. Within a scale of 100~Mpc, the matter distribution in the Universe is inhomogeneous, with large over-densities of matter corresponding to clusters of galaxies, sheets and filaments, and under-densities corresponding to voids. For a density of acceleration sites large enough, the distribution of galaxies could thus be a good tracer of the distribution of UHECR sources.  \\

\begin{wrapfigure}{L}{8. cm}
{\includegraphics[width=0.55\textwidth]{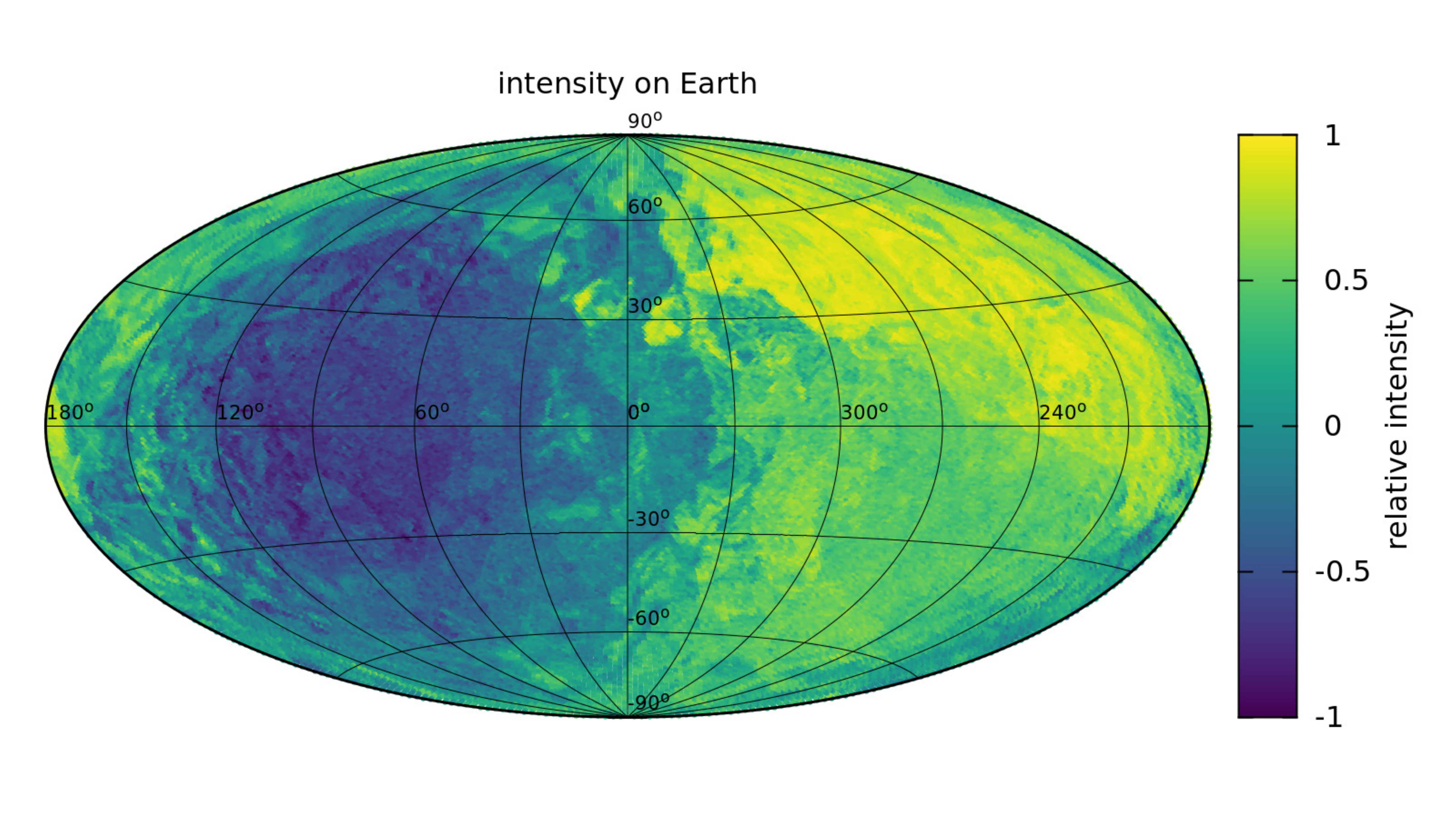}}
\caption{Sky map that would be observed on Earth once the magnetic deflections following from the model~\cite{JanssonFarrar2012} are accounted for, if the distribution of cosmic rays with a rigidity of 3~EV was drawn from a dipole outside from the Galaxy pointing towards the one of the 2MASS Redshift Survey galaxies.}
\label{fig:gmfdipole}
\end{wrapfigure}
The 3D positions of the closest of these structures are known from complete galaxy catalogs. For mean rigidities ranging from 2 to 5~EV as suggested by results on mass composition above 8~EeV~\cite{AugerPRD2014}, scattering of UHECRs in the extragalactic magnetic fields can still give rise to large deflections for field amplitudes ranging in few nanogauss and extended over coherence lengths of the order of one megaparsec. The angular distribution of UHECRs is thus expected to be influenced by the contribution of sources in the neighborhood of the Milky Way. The contribution of nearby sources is even expected to become dominant as the energy of CRs increases due to the reduction of the horizon of UHECRs induced by energy losses more important at higher energies. For the infrared-detected galaxies in the 2MASS Redshift Survey catalog~\cite{Erdoglu2006} mapping the distribution of galaxies out to $\simeq 115~$Mpc, the flux-weighted dipole points in Galactic coordinates in the direction $(l, b) = (251^\circ, 38^\circ)$, $55^\circ$ away from the best-fit position of the cross in figure~\ref{fig:augerdipole}. \\

Considering to first order that arrival directions of UHECRs should follow outside from the Galaxy the same dipolar pattern as the one of the 2MASS Redshift Survey galaxies, the observed pattern on Earth is expected to be further modified by the deflections imprinted by the Galactic magnetic field. The technique to ``fold'' the extragalactic dipole through the magnetic field consists in back-tracking antiparticles (distributed isotropically) from the Earth to outside the Galaxy, and in weighting each back-tracked antiparticle by an amount proportional to the density gradient in the retro-propagated direction outside the Galaxy. This is a consequence of the Liouville theorem. Note that the contribution of the electric field experienced by UHECRs travelling through far away regions due to the rotation of the Galaxy relative to the system in which the field is purely magnetic is negligible for the anisotropy~\cite{Harari2010}. For rigidities of 3~EV, and using the Galactic magnetic field model described in~\cite{JanssonFarrar2012}, the intensity that would be observed on Earth in that case is shown in figure~\ref{fig:gmfdipole}, in arbitrary units. Although a complex structure appears since the field connects the phase-space density at Earth to the one outside from the Galaxy in a non-trivial way, the pattern remains to first order the injected dipole slightly attenuated and rotated. For rigidities of 2 and 5~EV for illustration, the rotation of the dipole direction outside the Galaxy once folded through the magnetic field is indicated by the two arrows in figure~\ref{fig:augerdipole}. The agreement between the directions of the dipoles is then clearly improved. \\

These findings constitute the first firm observational evidence that the origin of UHECRs is indeed extragalactic. Searches for identifying the acceleration sites will be presented in~\S~\ref{sec:xgal}.

\section{Reconstruction of anisotropies beyond the dipole}
\label{sec:3dreco}

\begin{figure}[!h]
\centering\includegraphics[width=0.7\textwidth]{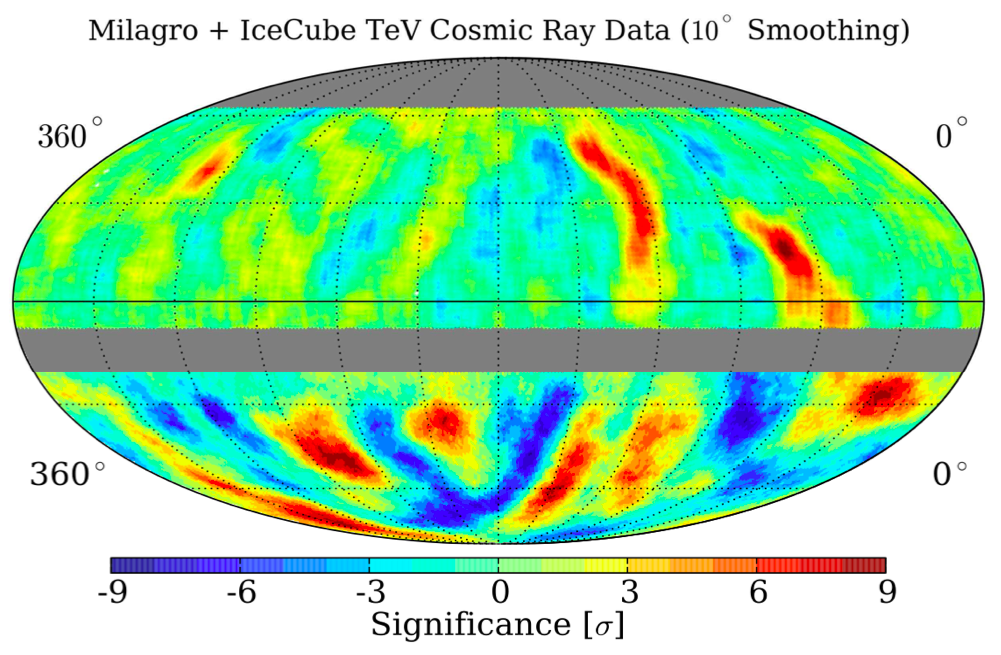}
\caption{Intensity in equatorial coordinates at median energies of 1~TeV in the Northern hemisphere (from Milagro~\cite{Milagro2008}) and of 20~TeV in the Southern one (from IceCube~\cite{IceCube2011}). The intensity is smoothed out at an angular scale of 10$^\circ$. The meta-analysis of Milagro and IceCube data to produce this single plot is from~\cite{Desiati2013}.}
\label{fig:skymap_icecube_milagro}
\end{figure}

Beyond the large-scale anisotropies captured by the first harmonic in right ascension, smaller significant anisotropy contrasts at the $10^{-4}$ level at intermediate and even small scales have also been reported in the TeV--PeV energy range in the last decade~\cite{Milagro2008,IceCube2011,ARGO2013,HAWC2014,IceTop2016}, and searches for such effects have been conducted at higher energies as well~\cite{AugerApJS2012,AugerTA2014,AugerJCAP2017}. An example of such anisotropies is displayed in figure~\ref{fig:skymap_icecube_milagro} at TeV energies and above, from Milagro and IceCube data~\cite{Desiati2013}. The dipole anisotropy thus appears as the leading order effect, but a description of the intensity beyond the dipole is required. A spherical harmonic decomposition of the intensity is then the natural tool to perform such a description; but technical difficulties to extract the multipolar moments appear due to the partial-sky coverage of ground-based experiments. The techniques to infer these moments and the underlying power spectrum are now reviewed. 

\subsection{Intensity reconstruction with partial-sky coverage}
\label{subsec:alm}

In general, and in contrast to the simplified approach presented in~\S~\ref{subsec:rayleigh}, the intensity function $I(\nn)$ depends on both the right ascension and the declination and thus must be decomposed in terms of a multipolar expansion in the spherical harmonics $Y_{\ell m}(\nn)$:
\begin{equation}
\label{eqn:almexpansion}
I(\nn)=\sum_{\ell\geq0}\sum_{m=-\ell}^\ell a_{\ell m}Y_{\ell m}(\nn).
\end{equation}
Non-zero amplitudes in the $\ell$ modes arise from variations of the intensity on an angular scale ${\simeq}1/\ell$ radians. With full-sky but non-uniform coverage, the customary recipe for decoupling directional exposure effects from anisotropy ones consists in defining the recovered coefficients as~\cite{Sommers2001}
\begin{eqnarray}
\label{eqn:est_alm}
\bar{a}_{\ell m}=\int_{4\pi}\frac{\dif\nn}{\mu(\nn)}\frac{\dif N(\nn)}{\dif\nn}Y_{\ell m}(\nn).
\end{eqnarray}
Introducing for convenience $\mu_\mathrm{r}(\nn)$ as the (dimensionless) relative directional exposure function normalized here to unity at its maximum and $f_1=\int\dif\nn~\mu_{\mathrm{r}}/4\pi$ as the covered fraction of the sky, and using $\dif N/\dif\nn=\sum_i\delta(\nn,\nn_i)$, the estimation reads as
\begin{eqnarray}
\label{eqn:est_almbis}
\bar{a}_{\ell m}=\frac{4\pi f_1}{\Omega_0}\sum_{i=1}^N\frac{Y_{\ell m}(\nn_i)}{\mu_\mathrm{r}(\nn_i)},
\end{eqnarray}
with $\Omega_0$ the total exposure of the experiment. Repeating the same operations as in~\S~\ref{subsec:rayleigh} but substituting $\alpha$ by $\nn$ in equations~(\ref{eqn:momentsphi}), it is seen that the estimator is unbiased, $\langle \bar{a}_{\ell m}\rangle_{\mathrm{P}} = a_{\ell m}$, and, in the case of small anisotropies $|a_{\ell m}/a_{00}|\ll 1$, that the uncertainty $\sigma_{\ell m}$ on each $a_{\ell m}$ multipole reflects the Poisson fluctuations induced by the finite number of events:
\begin{equation}
\label{eqn:rms_alm}
\sigma_{\ell m}=\left[\frac{4\pi f_1N}{\Omega_0^2}\int_{4\pi}\frac{\dif\nn}{\mu_{\mathrm{r}}(\nn)}~Y_{\ell m}^2(\nn)\right]^{1/2}.
\end{equation}

However, with ground-based observatories, coverage of the full sky is not possible with a single experiment. The partial-sky coverage of ground-based observatories prevents the multipolar moments $a_{\ell m}$ to be recovered in the direct way just presented. This is because the solid angle on the sky where the exposure is zero prevents one from making use of the completeness relation of the spherical harmonics. Indirect procedures have to be used, one of them consisting in considering first the ``pseudo-multipolar'' moments\footnote{Note the change of dimension between the $a_{\ell m}$ (in km$^{-2}$yr$^{-1}$sr$^{-1/2}$) and $\tilde{a}_{\ell m}$ (in sr$^{-1/2}$) coefficients.}
\begin{equation}
\label{eqn:alm_tilde}
\tilde{a}_{\ell m} = \int \dif\mathbf{n}~\mu(\mathbf{n})I(\mathbf{n})Y_{\ell m}(\mathbf{n}),
\end{equation}
and then the system of linear equations relating these pseudo moments to the real ones:
\begin{equation}
\label{eqn:kernel}
\tilde{a}_{\ell m} = \sum_{\ell'\geq0}\sum_{m'=-\ell}^\ell a_{\ell' m'}\int \dif\mathbf{n}~\mu(\mathbf{n})Y_{\ell m}(\mathbf{n})Y_{\ell' m'}(\mathbf{n})\equiv \sum_{\ell'\geq0}\sum_{m'=-\ell}^\ell [K]_{\ell m\ell' m'}~a_{\ell' m'}.
\end{equation}
Formally, the coefficients $a_{\ell m}$ are related to the pseudo-ones $\tilde{a}_{\ell m}$  through a convolution. The kernel $\underline{K}$, which imprints the interferences between modes induced by the non-uniform and partial coverage of the sky, is entirely determined by the directional exposure function $\mu(\nn)$ and is dimensioned like this function. Assuming a bound $\ell_{\mathrm{max}}$ beyond which $a_{\ell m}=0$, the first $\tilde{a}_{\ell m}$ pseudo-coefficients with $\ell\leq\ell_{\mathrm{max}}$ can be related to the non-vanishing $a_{\ell m}$ coefficients by the square matrix $\underline{K}$ truncated to $\ell_{\mathrm{max}}$. The truncated matrix can then be inverted, allowing recovery of the moments $a_{\ell m}$. However, this truncation induces large changes with $\ell_{\mathrm{max}}$ in the coefficients of the inverse kernel $\underline{K}^{-1}_{[\ell_{\mathrm{max}}]}$. Repeating again the operations leading to expression~(\ref{eqn:rms_alm}), the net effect is that the obtained uncertainty on each moment depends on $\underline{K}^{-1}_{[\ell_{\mathrm{max}}]}$~\cite{Billoir2008}:
\begin{equation}
\label{eqn:rms_alm_bis}
\sigma_{\ell m} \simeq \left[\frac{N}{\Omega_0}[K^{-1}_{[\ell_{\mathrm{max}}]}]_{\ell m\ell m}\right]^{1/2}.
\end{equation}

In numbers, it turns out that $\sigma_{\ell m}$ increases exponentially with $\ell_{\mathrm{max}}$. This dependence is nothing else but the mathematical translation of it being impossible to know the angular distribution of CRs in the uncovered region of the sky. In most of the practical cases, the small values of the energy-dependent $a_{\ell m}$ coefficients combined with the available statistics in the different energy ranges do not allow for an estimation of the individual coefficients with a relevant resolution as soon as $\ell_{\mathrm{max}}>2$. Nonetheless, although the use of a large value of $\ell_{\mathrm{max}}$ forbids giving for any $\bar{a}_{\ell m}$ coefficient an interpretation of an individual multipolar moment, the full set of coefficients $\{\bar{a}_{\ell m}\}$ still provides a sensible description of the intensity \textit{in the covered region of the sky}. This property allows for reconstructing the intensity as a multipolar expansion for any individual ground-based observatory, and can also be used to obtain a ``likely minimum value'' of $\ell_{\mathrm{max}}$ from the data themselves~\cite{Billoir2008}.

\subsection{Special cases: geometric representations of the dipole and quadrupole moments}
\label{subsec:dip_quad}

Some attention is given here to the two first moments of the multipolar expansion, and in particular to the dipole one given its special interest as exemplified in~\S~\ref{sec:astro}. From the estimation of the spherical harmonic coefficients, a more geometric and more intuitive representation of the dipole and quadrupole moments is generally used for an expansion truncated at $\ell_{\mathrm{max}}=2$:
\begin{equation}
\label{eqn:phidipquad}
I(\nn)=\frac{I_0}{4\pi}\left(1+d\,\hat{\mathbf{d}}\cdot\nn+\lambda_+(\hat{\mathbf{q}}_+\cdot\nn)^2+\lambda_0(\hat{\mathbf{q}}_0\cdot\nn)^2+\lambda_-(\hat{\mathbf{q}}_-\cdot\nn)^2+\ldots\right).
\end{equation}
In this picture, the dipole moment is thus characterized by a vector, whose amplitude $d$ and two angles of the unit vector $\hat{\mathbf{d}}$ are related to the $a_{1m}$ coefficients through, using here the equatorial coordinate system:
\begin{equation}
\label{eq:dip_param}
d=\frac{\sqrt{3}}{a_{00}}\left[a_{10}^2+a_{11}^2+a_{1-1}^2\right]^{1/2},\quad \delta_d=\arcsin{(\sqrt{3}a_{10}/a_{00})},\quad \alpha_d=\arctan{(a_{1-1}/a_{11})}.
\end{equation}
The amplitude $d$ corresponds to the maximal anisotropy contrast of a pure dipolar flux. The quadrupole, on the other hand, is characterized by the amplitudes $(\lambda_+,\lambda_0,\lambda_-)$ and unit vectors $(\hat{\mathbf{q}}_+,\hat{\mathbf{q}}_0,\hat{\mathbf{q}}_-)$ which are the eigenvalues and eigenvectors of a second order, traceless and symmetric tensor $1/2~\underline{Q}$ whose five independent components are related to the $a_{2m}$ coefficients through
\begin{equation}
Q_{xx} =  \frac{\sqrt{5}}{a_{00}}\,\left(\sqrt{3}a_{22}-a_{20}\right),\quad
Q_{xy} =  \frac{\sqrt{15}}{a_{00}}\,a_{2-2},\quad
Q_{xz} = -\frac{\sqrt{15}}{a_{00}}\,a_{21},\nonumber
\end{equation}
\begin{equation}
\label{eq:quad_tensor}
Q_{yy} = -\frac{\sqrt{5}}{a_{00}}\,\left(\sqrt{3}a_{22}+a_{20}\right),\quad
Q_{yz} = -\frac{\sqrt{15}}{a_{00}}\,a_{2-1}.
\end{equation}
The eigenvalues are ranked from the largest to the smallest one and assigned to the vectors $(\hat{\mathbf{q}}_+,\hat{\mathbf{q}}_0,\hat{\mathbf{q}}_-)$ that form the principal axes coordinate system. The traceless condition of the quadrupole tensor $\underline{Q}$ forces the relation $\lambda_++\lambda_0+\lambda_-=0$ to be satisfied, so that only two of these amplitudes are independent. Hence, two diagnostic parameters are used to characterize a quadrupole anisotropy: the quadrupole magnitude that takes on the value $\lambda_+$, and the anisotropy contrast of a quadrupolar flux $\beta=(\lambda_+-\lambda_-)/(2+\lambda_++\lambda_-)$. The orientation of the quadrupole is then described by the three Euler angles determined from the eigenvectors corresponding to each of the principal axes and characterizing the orientation of these principal axes with respect to some reference coordinate system.

Characterising the dipole vector in the same way as the first harmonic in right ascension is in fact possible, by mimicking the formalism of Linsley detailed in~\S~\ref{subsec:rayleigh} to the sphere instead of the circle. Denoting $\bar{d}_x, \bar{d}_y, \bar{d}_z$ the estimates of the Cartesian coordinates of $\mathbf{d}$, the joint p.d.f. $p_{\mathbf{d}}(\bar{d}_x,\bar{d}_y,\bar{d}_z)$ can be factorised in terms of three Gaussian distributions thanks to the central limit theorem:
\begin{equation}
\label{eqn:Pxyz}
p_{\mathbf{d}\mathrm{C}}(\bar{d}_x,\bar{d}_y,\bar{d}_z)=p_{D_x}(\bar{d}_x,\sigma_x)p_{D_y}(\bar{d}_y,\sigma_y)p_{D_z}(\bar{d}_z,\sigma_z),
\end{equation}
with the $\sigma_i$ parameters determined by propagating the change of variables from the $a_{1m}$ to the $d_i$ coefficients (with the correspondence $i=(x,y,z) \rightleftarrows m=(1,-1,0)$) into expression~(\ref{eqn:rms_alm_bis}):
\begin{equation}
\label{eqn:rms_d}
\sigma_{i} \simeq \left[\frac{3\Omega_0}{4\pi N}[K^{-1}_{[\ell_{\mathrm{max}}]}]_{1i1i}\right]^{1/2}.
\end{equation}
Under an underlying isotropic distribution ($d_i=0$), and restricting the discussion for convenience to the case of an axisymmetric directional exposure around the axis defined by the North and South equatorial poles (so that $\sigma_x=\sigma_y=\sigma$), the joint p.d.f. for $\bar{d}, \bar{\delta}_d, \bar{\alpha}_d$ (spherical coordinates of $\mathbf{d}$) is obtained from equation~(\ref{eqn:Pxyz}) by making the proper jacobian transformation:
\begin{eqnarray}
\label{eqn:jointPdda}
p_{\mathbf{d}\mathrm{S}}(\bar{d},\bar{\delta}_d,\bar{\alpha}_d)=\frac{\bar{d}^2\cos\bar{\delta}_d}{(2\pi)^{3/2}\sigma^2\sigma_z}\exp{\left[-\frac{\bar{d}^2\cos^2\bar{\delta}_d}{2\sigma^2}-\frac{\bar{d}^2\sin^2\bar{\delta}_d}{2\sigma_z^2}\right]}.
\end{eqnarray}
From this joint p.d.f., the p.d.f. of the dipole amplitude (declination and right ascension) is finally obtained by marginalising over the other variables, yielding to\footnote{Note that for $\sigma_z=\sigma$ (uniform and full-sky coverage), the amplitude distribution is the Maxwell-Boltzmann distribution.}:
\begin{eqnarray}
\label{eqn:Pdda}
p_{d}(\bar{d})&=& \frac{\bar{d}}{\sigma\sqrt{\sigma_z^2-\sigma^2}}\,\mathrm{erfi}\left(\frac{\sqrt{\sigma_z^2-\sigma^2}}{\sigma\sigma_z}\frac{\bar{d}}{\sqrt{2}}\right) \exp\left(-\frac{\bar{d}^2}{2\sigma^2}\right), \\
p_{\Delta}(\bar{\delta}_d)&=&\frac{\sigma\sigma_z^2}{2}\frac{\cos\bar{\delta}_d}{(\sigma_z^2\cos^2\bar{\delta}_d+\sigma^2\sin^2\bar{\delta}_d)^{3/2}},\\
p_{A}(\bar{\alpha}_d)&=&\frac{1}{2\pi},
\end{eqnarray}
with $\mathrm{erfi}(z)$ the imaginary error function $\mathrm{erfi}(z)=\mathrm{erf}(iz)/i$. Finally, from $p_d$, quantities of interest such as the expected mean noise $\langle\bar{d}\rangle_{\mathrm{P}}$, the RMS $\sigma_d$ and the probability of obtaining an amplitude greater than $\bar{d}$ can be derived:
\begin{eqnarray}
\label{eqn:stat_dipole}
\langle\bar{d}\rangle_{\mathrm{P}}&=&\sqrt{\frac{2}{\pi}}\,\left(\sigma_z+\frac{\sigma^2\mathrm{arctanh}(\sqrt{1-\sigma^2/\sigma_z^2}}{\sqrt{\sigma_z^2-\sigma^2}}\right), \\
\sigma_d&=&\sqrt{2\sigma^2+\sigma_z^2-\left\langle \bar{d} \right\rangle_{\mathrm{P}}^2}, \\
P(\geq\bar{d})&=&\mathrm{erfc}\left(\frac{\bar{d}}{\sqrt{2}\sigma_z}\right)+\frac{\sigma}{\sqrt{\sigma_z^2-\sigma^2}}\mathrm{erfi}\left(\frac{\sqrt{\sigma_z^2-\sigma^2}}{\sqrt{2}\sigma\sigma_z}\bar{d}\right)\mathrm{exp}\left(-\frac{\bar{d}^2}{2\sigma^2}\right),
\end{eqnarray}
which are here the searched expressions for characterising any set of dipole measurements under the null hypothesis of isotropy. For $d_i\neq0$, expressions get more complex. Using non-centered Gaussian functions in equation~(\ref{eqn:Pxyz}) with $d_x=d\cos\delta_d\cos\alpha_d$, $d_y=d\cos\delta_d\sin\alpha_d$ and $d_z=d\sin\delta_d$, the same methodology leads to the following semi-analytical p.d.f. for instance for $\bar{d}$, which is independent of $\alpha_d$:
\begin{equation}
\label{eqn:Pdda_bis}
p_{d}(\bar{d};d,\delta_d)=\frac{\bar{d}^2\exp\left(-\frac{\bar{d}^2}{2\sigma^2}-\frac{d^2}{2}\left(\frac{\cos^2{\delta_d}}{\sigma^2}+\frac{\sin^2{\delta_d}}{\sigma_z^2}\right)\right)}{\sigma\sqrt{2\pi\sigma_z^2}}\,G\left(\frac{d\bar{d}\sin{\delta_d}}{\sigma_z^2},\frac{\bar{d}^2}{2}\frac{\sigma_z^2-\sigma^2}{\sigma^2\sigma_z^2},\frac{d\bar{d}\cos{\delta_d}}{\sigma^2}\right),
\end{equation}
with $G(x,y,z)$ defined as
\begin{equation}
\label{eqn:Pdda_bis}
G(x,y,z)=\int_{-1}^{1}\dif u~\exp\left(xu+yu^2\right)I_0\left(z\sqrt{1-u^2}\right).
\end{equation}

\begin{wrapfigure}{L}{8. cm}
{\includegraphics[width=0.5\textwidth]{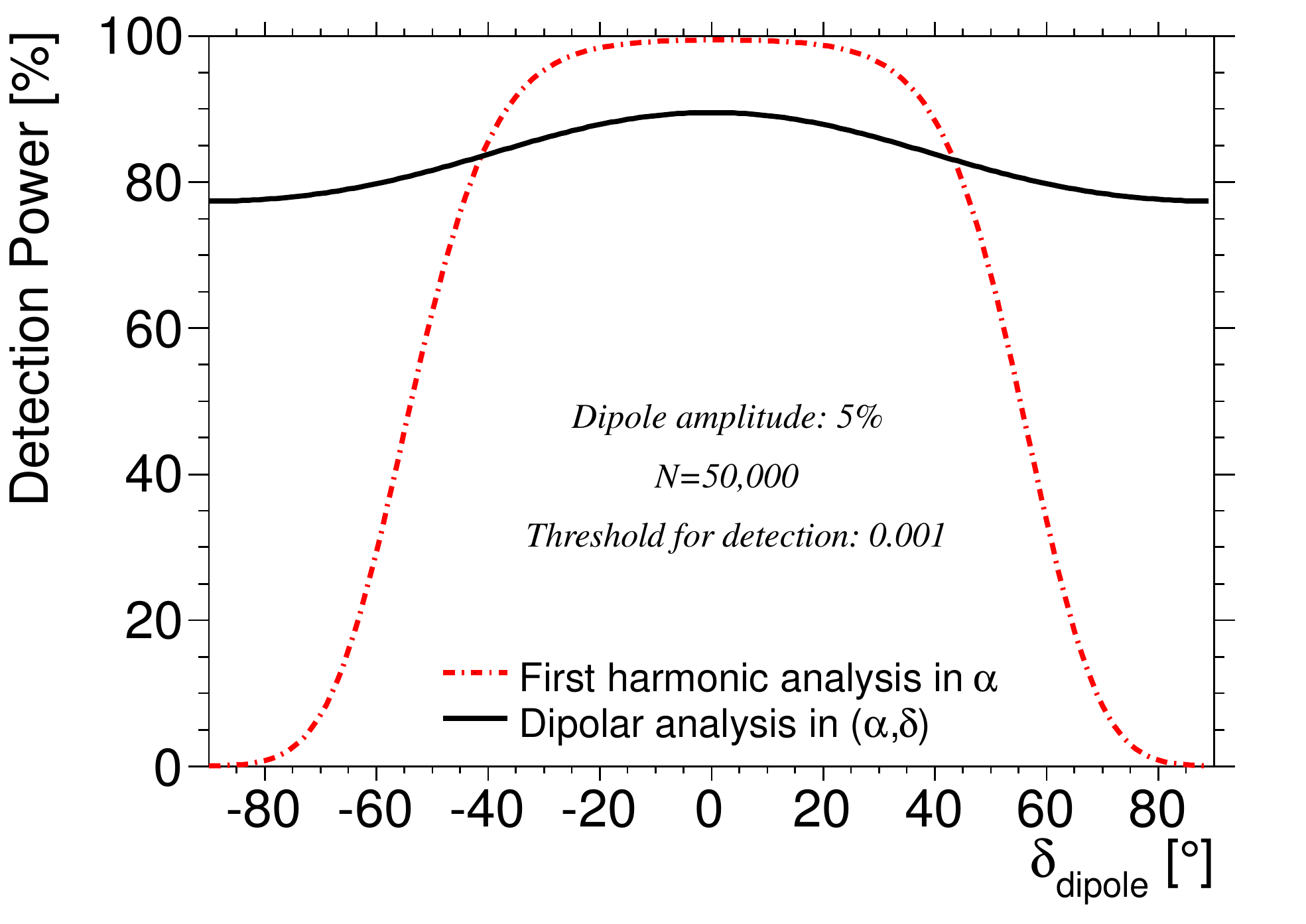}}
\caption{Comparison of the detection power of the 2D/3D methods vs the declination of the dipole.}
\label{fig:power_2dvs3d}
\end{wrapfigure}
In contrast to the first harmonic analysis presented in~\S~\ref{subsec:rayleigh} (``2D analysis''), the performances of this ``3D analysis'' are almost the same whatever the direction of the dipole is on the sphere. A comparison of the detection power of the two methods is displayed in figure~\ref{fig:power_2dvs3d} as a function of the declination for a threshold of $10^{-3}$, and for a pure dipole with an amplitude of 5\% with a statistics of 50,000 events. The reconstruction of the dipole parameters is performed with $\ell_{\mathrm{max}}=1$. While the 2D analysis performs slightly better for $\delta_{\mathrm{d}}$ close to the equatorial plane, the 3D analysis performs much better as soon as the orientation of the dipole is far from this plane.  

On the other hand, there is no analytical description for characterising the amplitudes and angles of the quadrupole. The corresponding distributions are thus estimated from Monte-Carlo simulations.

\subsection{Angular power spectrum}
\label{subsec:cl}

As discussed in~\S~\ref{subsec:alm}, the small values of the energy-dependent $a_{\ell m}$ coefficients combined to the available statistics in the different energy ranges does not allow, in general, for an estimation of the individual coefficients with a relevant resolution as soon as $\ell_{\mathrm{max}}>2$. However, based on analysis techniques previously developed in the CMB community~\cite{Hivon2002}, it is possible, under some restrictions detailed below, to reconstruct the angular power spectrum coefficients defined as
\begin{eqnarray}
\label{eqn:Cl}
{C}_{\ell}=\sum^{\ell}_{m=-\ell}\frac{|a_{\ell m}|^{2}}{2\ell +1},
\end{eqnarray}
and to do so within a statistical resolution independent of the bound $\ell_{\mathrm{max}}$~\cite{Deligny2004}. The starting point is to consider that the anisotropic part of the intensity, $\delta I(\nn)$, is a particular realisation of an underlying stochastic process, process which can be assumed Gaussian in a conservative way, and which is thus entirely characterised by its first two moments $\langle\delta I(\mathbf{n})\rangle$ and $\langle\delta I(\mathbf{n})\delta I(\mathbf{n'})\rangle$. In the absence of knowledge of this stochastic process, the simplest non-trivial configuration is to consider that the anisotropies cancel in ensemble average and produce a second order moment that does not depend on the position on the sphere but only on the angular separation $\gamma$ between $\mathbf{n}$ and $\mathbf{n'}$. These conditions, characterising an homogeneous and isotropic random process, are known as the conditions of ``stationarity.'' Expanding then $\langle\delta I(\mathbf{n})\delta I(\mathbf{n}')\rangle$ onto the Legendre polynomials $P_\ell$ allows for translating stationarity conditions in space into stationarity conditions in the reciprocal $\ell-$space:
\begin{eqnarray}
\label{eqn:stationarity_space}
\langle\delta I(\mathbf{n})\delta I(\mathbf{n}')\rangle=\sum_{\ell\geq 0}\frac{2\ell+1}{4\pi}C_\ell P_\ell(\nn\cdot\nn') \equiv \sum_{\ell\geq 0}\sum_{m=-\ell}^{\ell}C_\ell Y_{\ell m}(\nn)Y_{\ell m}(\nn'),
\end{eqnarray}
where spherical harmonics addition theorem is used. It is thus seen by identification that, under stationarity, the underlying $a_{\ell m}$ coefficients are not correlated to each other:
\begin{eqnarray}
\label{eqn:stationarity}
\langle a_{\ell m}a_{\ell' m'}\rangle=C_\ell\delta_{\ell\ell'}\delta_{mm'}.
\end{eqnarray}
The ${C}_{\ell}$ coefficients can thus be viewed as a measure of the variance of the $a_{\ell m}$ coefficients.  

The stationarity conditions are well suited for CMB studies where the power spectrum results from primordial and fundamental fluctuations. They are however questionable in the context of CR physics, where invoking a random nature for $\delta I(\nn)$ follows ``only'' from the stochastic modeling of unknown source positions and non-well known propagation mediums and regimes. Nonetheless, despite their limited range of applications, there are several benchmark scenarios of interest that the stationarity conditions can describe or approximate. One of these relies on sources drawn at random with a granularity in accordance with some density. The ensemble-average angular distribution is then isotropic by construction, and the particular anisotropies observed in a given realisation are dominantly due to the fluctuations of the positions of the most contributing local sources. More generally, even in the case of local sources not drawn at random but following some structure, the random localisation of the observer allows a description of the process through the stationarity, to first order at least. On the other hand, since the stationarity conditions are obviously not comprehensive of all stochastic processes, there are scenarios preventing the power spectrum to be fairly captured with the type of method described here. Caution should thus be kept in mind when interpreting angular power spectra measured with partial-sky coverages, since the angular distributions in the uncovered regions of the sky could show patterns that would lead to different power spectra.

Under the stationarity hypotheses, a quadratic estimation of the $a_{\ell m}$ coefficients allows for estimating the power spectrum even with partial-sky coverage. This is because in ensemble average, the correlation coefficient between the pseudo-multipolar moments is related to the power spectrum coefficients through
\begin{eqnarray}
\label{eqn:pseudoalm_correlation}
\langle \tilde{a}_{\ell m}\tilde{a}_{\ell' m'}\rangle=\sum_{i\geq0}\sum_{j=-i}^{i} K_{\ell mij}K_{\ell'm'ij}C_i.
\end{eqnarray}
In this situation, the pseudo-power spectrum $\tilde{C}_{\ell}=\sum^{\ell}_{m=-\ell}|\tilde{a}_{\ell m}|^{2}/(2\ell +1)$, which is directly measurable, is related to the real power spectrum through
\begin{eqnarray}
\label{eqn:Cl_tilde}
\tilde{C}_{\ell} = \sum_{\ell'}M_{\ell\ell'}C_{\ell'},
\end{eqnarray}
where the matrix $\underline{M}$ describing the cross-talk induced by the non-uniform exposure between genuine modes is entirely determined by the knowledge of the exposure function~\cite{Hivon2002}:
\begin{eqnarray}
\label{eqn:M}
M_{\ell\ell'} = \frac{1}{2\ell+1}\sum_{m=-\ell}^{\ell}\sum_{m'=-\ell'}^{\ell'}\left|K_{\ell m\ell'm'}\right|^2.
\end{eqnarray}
In some cases of interest, the inversion of $\underline{M}$ is unambiguously defined, independently of the bound $\ell_{\mathrm{max}}$. From the expression of $\underline{K}$ indeed, and making use of the addition theorem of the spherical harmonics, the matrix elements $M_{\ell\ell'}$ can be written as
\begin{eqnarray}
\label{eqn:Mbis}
M_{\ell\ell'} = \frac{2\ell'+1}{2}\int_{-1}^1\dif(\cos{\gamma})~\mathcal{W}(\cos\gamma)P_\ell(\cos{\gamma})P_{\ell'}(\cos{\gamma}),
\end{eqnarray}
where the average over position and orientation on the sphere of $\mu(\nn)$ at angular separation $\gamma$ is defined as 
\begin{eqnarray}
\label{eqn:W}
\mathcal{W}(\cos\gamma)=\iint\frac{\dif\mathbf{n}\dif\mathbf{n'}}{8\pi^2}\mu(\mathbf{n})\mu(\mathbf{n'})\delta(\mathbf{n}\cdot\mathbf{n'}-\cos\gamma).
\end{eqnarray}
On multiplying both sides of the identity $\sum_{\ell'}M_{\ell\ell'}M^{-1}_{\ell'\ell''}=\delta_{\ell\ell''}$ by $(2\ell+1)P_{\ell}(\cos\gamma')/2$, on summing over $\ell$, and on using the completeness relation of the Legendre polynomials, the following set of relations satisfied by the elements of $\underline{M}^{-1}$ is then obtained:
\begin{eqnarray} 
\label{mll2}
\mathcal{W}(\cos\gamma')\sum_{\ell'}^{\ell_{\mathrm{max}}}\frac{2\ell'+1}{2}P_{\ell'}(\cos\gamma')M^{-1}_{\ell'\ell''}=\frac{2\ell''+1}{2}P_{\ell''}(\cos\gamma').
\end{eqnarray}
For a non-vanishing $\mathcal{W}(\cos\gamma')$ function, a condition which is generally satisfied or close-to-satisfied by CR ground-based observatories, the inversion of $\underline{M}$ is then straightforward by multiplying both sides of these equations by $P_\ell(\cos\gamma')$, by integrating over $\cos\gamma'$, and by making use of the orthogonality of the Legendre polynomials:
\begin{eqnarray} \label{mll3}
M^{-1}_{\ell\ell''}=\frac{2\ell''+1}{2}\int_{-1}^1\frac{\dif(\cos\gamma')}{\mathcal{W}(\cos\gamma')}~P_{\ell}(\cos\gamma')P_{\ell''}(\cos\gamma').
\end{eqnarray}
The recovering of the power spectrum then proceeds from this inversion. 

In practice, the finite sampling of $\overline{\delta I}(\nn)$ induces Poisson fluctuations that lead to a predictable bias for the recovered $\bar{C}_{\ell}$ coefficients. Adopting here a different convention for the pseudo-multipoles such that they provide a direct measurement of the relative intensity in the reciprocal space,
\begin{equation}
\bar{\tilde{a}}_{\ell m} = \int_{4\pi} \dif\mathbf{n}~Y_{\ell m}(\mathbf{n})\frac{\dif N/\dif\mathbf{n}-(N/4\pi f_1)\mu(\mathbf{n})}{N/4\pi f_1},
\end{equation}
the statistical properties for the $\bar{C}_{\ell}$ coefficients are found by propagating the one- and two-point functions of $\delta I(\nn)$~\cite{Deligny2004}. On average, the power spectrum is recovered modulo a bias term such that
\begin{equation}
\label{eqn:bias_cl}
\langle\langle\bar{C}_{\ell}\rangle\rangle_{\mathrm{P}} = C_\ell+\frac{4\pi}{N}\frac{f_1^2}{f_2},
\end{equation}
with $f_2=\int\dif\mathbf{n}~\mu^2(\mathbf{n})/4\pi$. For isotropic samples, the resolution obtained on each recovered power for each mode $\ell$ behaves as
\begin{equation}
\label{eqn:variance_cl}
\sigma(C_{\ell}) = \left(\frac{4 \pi f_1}{N}\right) \left(\frac{2}{2\ell+1} M^{-1}_{\ell \ell}\right)^{1/2}.
\end{equation}

It is worth noting that this formalism holds in the case of the reconstruction of the power spectrum of $\delta I(\nn)$, that is, when the directional exposure function is known as a function of both the right ascension and the declination. When reconstructing $\delta I_1(\nn)$ only, the subtraction of the $|a_{\ell 0}|^2$ term must be imposed in the definition of the $C_\ell$ coefficients. This translates into a modified definition of the $\underline{M}$ matrix as~\cite{Ahlers2016a}:
\begin{eqnarray}
\label{eqn:Mter}
\tilde{M}_{\ell\ell'} = M_{\ell\ell'}-\frac{1}{2\ell+1}\sum_{m=-\ell}^\ell|K_{\ell 0\ell'm}|^2.
\end{eqnarray}

\section{Some implications of high-order multipoles}
\label{sec:highorder}

\subsection{TeV--PeV power spectrum}
\label{subsec:origin_cl}

\begin{figure}[!h]
\centering\includegraphics[width=0.6\textwidth]{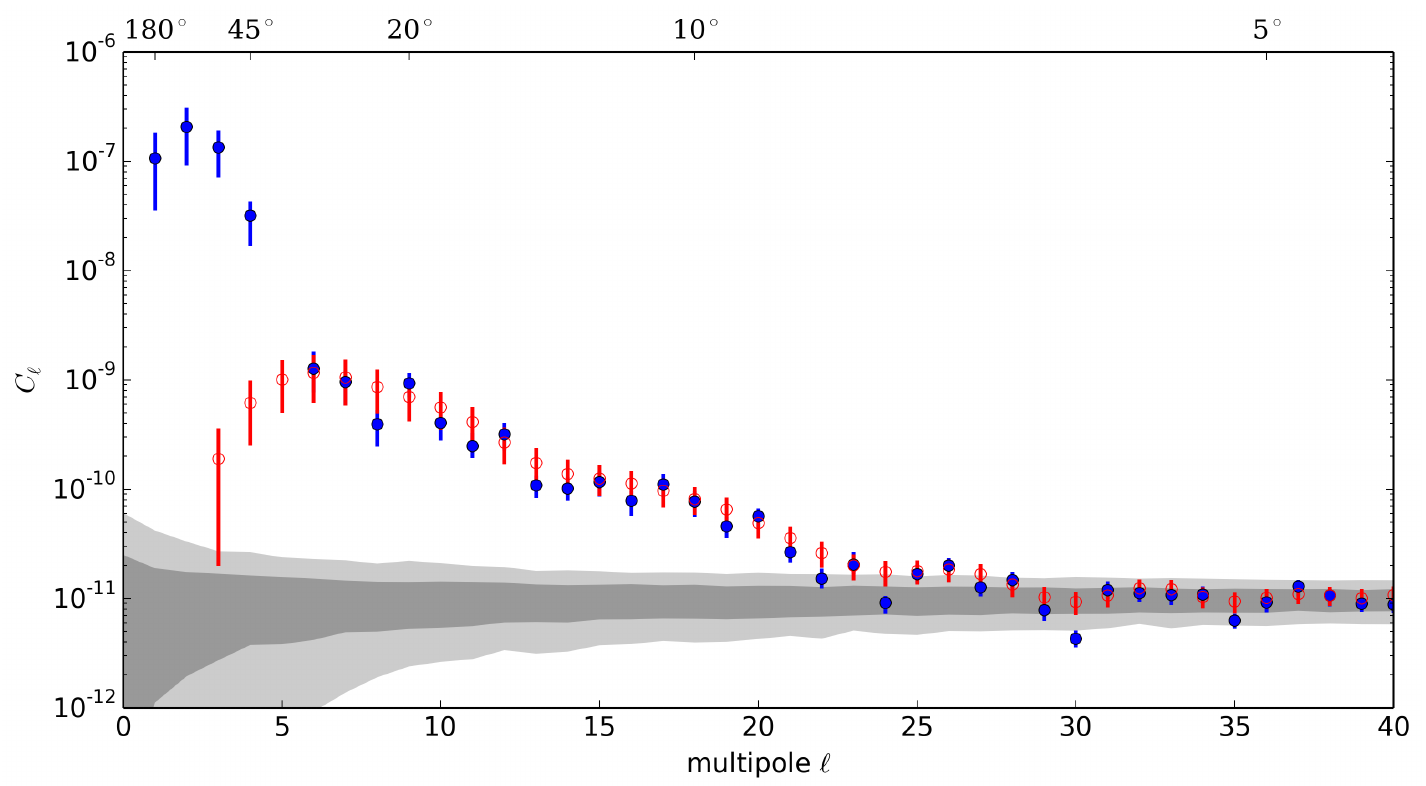}
\caption{Angular power spectrum as derived from IceCube data for a median energy of 20~TeV~\cite{IceCube2011} (blue points). The 68\% and 95\% confidence regions for isotropy (cf. equations~(\ref{eqn:bias_cl}) and~(\ref{eqn:variance_cl})) are indicated as the gray bands. The red points are the resulting spectrum after the subtraction of some large-scale contribution.}
\label{fig:powerspectrum_icecube}
\end{figure}

The power spectrum as derived from the IceCube data for a median energy of 20~TeV is shown in figure~\ref{fig:powerspectrum_icecube}~\cite{IceCube2011}. Significant power is observed up to multipoles of the order of 20, multipole above which the Poisson noise from equation~(\ref{eqn:bias_cl}) does not allow anymore to resolve a genuine signal given the available statistics. From the various caveats stressed in~\S~\ref{subsec:cl} related to the estimation of an underlying power spectrum with a limited coverage of the sky, the results on each mode should not be taken at face value. However, it is clear that there is a non-zero power from large scales down to small ones, and that the power spectrum roughly follows a decreasing power law. These results have challenged the long-standing picture that the CR propagation in the Galactic magnetic field should lead predominantly to a dipole moment only. Several interpretations have been put forward. Some attention is here given to the most conservative one first put forward in~\cite{GiacintiSigl2012}, being based on the fact that the observed complex patterns, whose positions appear randomly distributed, are mainly the consequence of the particular structure of the turbulent magnetic field within the last sphere of diffusion encountered by CRs. 

From equation~(\ref{eqn:stationarity_space}), it is straightforward to see that, within a convenient unit system, the power spectrum is related to the second order moments of $f$:
\begin{equation}
\label{eqn:defcl}
\langle C_{\ell}\rangle = \frac{1}{4\pi}\iint\dif\hat{\mathbf{p}}_1\dif\hat{\mathbf{p}}_2~P_\ell(\hat{\mathbf{p}}_1\cdot\hat{\mathbf{p}}_2)\langle f(\mathbf{p}_1)f(\mathbf{p}_2)\rangle.
\end{equation}
The estimation of this second order moment was first performed through Monte-Carlo simulations for PeV energies through numerical integration of test particle trajectories in~\cite{GiacintiSigl2012}, where CRs experience and probe a single realisation of the magnetic field up to a few tens of parsecs. With a density gradient of CRs at the entrance of this local turbulence considered as the last ``sphere of diffusion'', the turbulent magnetic field is frozen on the time scale of CR propagation and thus plays the same role as a structured one. The field is then expected to connect regions of higher density outside from the last sphere of diffusion to regions of lower density as seen from Earth, and \textit{vice versa}. This mechanism does produce intermediate- and small-scale anisotropies only from a gradient density of CRs encoded in $f$ at a time $-T$ large enough to describe the phase-space density outside from the local turbulence. The same argument holds at TeV energies, where CRs may start to experience magnetic fields in the heliosphere~\cite{DesiatiLazarian2012,Zhang2014}.  CR anisotropies at TeV--PeV energies may thus be a powerful tool to probe local magnetic fields, as also recently studied in~\cite{LopezBarquero2016} where the turbulence was modeled as a low compressible magnetohydrodynamic system with the external mean magnetic field as the only free parameter: the power spectrum can then be reproduced reasonably well for a sensible range of mean fields.

From a gradient density outside the last sphere of diffusion (thus a dipolar state for $f$), a universal expression of the power spectrum observed within the sphere has also been derived in the context of diffusion~\cite{AhlersMertsch2015}. An analytical expression can be obtained, providing insights on the mechanisms at work to produce the non-zero power in each mode $\ell$ despite some simplifying assumptions. The main steps go as follows. The aim is to derive an equation governing the evolution of $\langle f(\mathbf{x}_1(t),\mathbf{p}_1(t),t)f(\mathbf{x}_2(t),\mathbf{p}_2(t),t)\rangle$ and to solve it in the stationary case. Note that for convenience, the time and space dependences are dropped in the following unless necessary for the sake of clarity. Starting from the Boltzmann equation~(\ref{eqn:boltzmann}), it is straightforward to get 
\begin{eqnarray}
\label{eqn:boltzmann_f1f2}
-\frac{\partial}{\partial t}\langle f(\mathbf{p}_1)f(\mathbf{p}_2)\rangle &=&\left\langle f(\mathbf{p}_1)\left(c\hat{p}_{2i}\frac{\partial}{\partial x_{2i}}+\epsilon_{ijk}p_{2j}\langle\omega_k\rangle \frac{\partial}{\partial p_{2i}} +\epsilon_{ijk}p_{2j}\delta\omega_k \frac{\partial}{\partial p_{2i}}\right)f(\mathbf{p}_2)\right\rangle\nonumber \\
&+& (1\leftrightarrow 2),
\end{eqnarray}
where $(1\leftrightarrow 2)$ stands for substituting in the first term of the r.h.s. $\mathbf{p}_1$ for $\mathbf{p}_2$ and \textit{vice versa}. Considering the expansion of $f$ in terms of $\phi_0$ and $\boldsymbol{\phi}_1$ and the Fick equation~(\ref{eqn:fick}), the first term of the r.h.s. can be approximated to leading order as
\begin{equation}
\label{eqn:gradient_term_f1f2}
\left\langle f(\mathbf{p}_1)c\hat{p}_{2i}\frac{\partial f(\mathbf{p}_2)}{\partial x_{2i}}\right\rangle \simeq -\frac{3D_{ij}}{(4\pi)^2} \frac{\partial\langle\phi_0\rangle}{\partial x_{1j}}\frac{\partial\langle\phi_0\rangle}{\partial x_{2k}}\hat{p}_{1i}\hat{p}_{2k}.
\end{equation}
The second term describes a global rotation which does not contribute to the power spectrum. The last term needs to be simplified to be workable. Inspired by the Brownian diffusion of momenta on the sphere where an initial configuration of momenta in the same direction $\mathbf{p}_0$ relaxes at later times to $\sum_{\ell m}Y_{\ell m}(\mathbf{p})Y_{\ell m}(\mathbf{p}_0)\exp{(-\ell(\ell+1)\nu t/2)}$~\cite{Yosida1949}, the decay-time approximation can be thought as applying to $f$ an angular momentum operator $\mathbf{L}^2$. Analogously, the last term can be thought as resulting from the composition of two angular momenta $\mathbf{J}=\mathbf{L}_1+\mathbf{L}_2$:
\begin{eqnarray}
\label{eqn:bkg_f1f2}
\left\langle  f(\mathbf{p}_1)\epsilon_{ijk}p_{2j}\delta\omega_k \frac{\partial f(\mathbf{p}_2)}{\partial p_{2i}} \right\rangle + \left\langle  f(\mathbf{p}_2)\epsilon_{ijk}p_{1j}\delta\omega_k \frac{\partial f(\mathbf{p}_1)}{\partial p_{1i}} \right\rangle  &\simeq& \nonumber \\
&& \hspace{-9cm} -\left(\frac{\ell_1(\ell_1+1)+\ell_2(\ell_2+1)}{2}\nu_{\mathrm{r}}(x)+J(J+1)\nu_{\mathrm{c}}(x)\right)\left\langle f(\mathbf{p}_1)f(\mathbf{p}_2) \right\rangle, 
\end{eqnarray}
with $x=\hat{\mathbf{p}}_1\cdot\hat{\mathbf{p}}_2$. The parameters $\nu_{\mathrm{r}}$ and $\nu_{\mathrm{c}}$ can be seen as the relative and correlated scattering rates, which depend on the relative distance of trajectories at early times. 

This is enough to establish the stationary equation for the power spectrum coefficients. The definition of the $\langle C_\ell\rangle$ coefficients in equation~(\ref{eqn:defcl}) can be seen as following from the projection onto states with $J=0$, $M=0$, and $\ell_1=\ell_2=\ell$ of the operators $\mathbf{J}^2$ and $J_z$. Since the rate coefficients depend only on $\hat{\mathbf{p}}_1\cdot\hat{\mathbf{p}}_2$, they can be expanded onto $J=0$ and $M=0$ states. In this case, only relative rotations contribute to the stationary power spectrum, so that the contribution of $\nu_{\mathrm{c}}$ is null. The action on $\langle f(\mathbf{p}_1)f(\mathbf{p}_2)\rangle$ of the operation in the r.h.s. of equation~(\ref{eqn:bkg_f1f2}) thus receives contributions from the projection onto $J=0$ and $M=0$ only~\cite{AhlersMertsch2015}, so that the searched set of equations for the power spectrum reads as
\begin{eqnarray}
\label{eqn:eq_f1f2}
\frac{3D_{ij}}{(4\pi)^2c}\frac{\partial\langle\phi_0\rangle}{\partial x_{1j}}\frac{\partial\langle\phi_0\rangle}{\partial x_{2k}}\iint \dif\hat{\mathbf{p}}_1\dif\hat{\mathbf{p}}_2~P_\ell(\hat{\mathbf{p}}_1\cdot\hat{\mathbf{p}}_2)\hat{p}_{1i}\hat{p}_{2k} &\simeq&\nonumber \\
 && \hspace{-9cm}\iint \dif\hat{\mathbf{p}}_1\dif\hat{\mathbf{p}}_2~P_\ell(\hat{\mathbf{p}}_1\cdot\hat{\mathbf{p}}_2) \nu_{\mathrm{r}}(\hat{\mathbf{p}}_1\cdot\hat{\mathbf{p}}_2)\sum_{k\geq0} k(k+1)\frac{2k+1}{4\pi}\langle C_k \rangle P_k(\hat{\mathbf{p}}_1\cdot\hat{\mathbf{p}}_2).
\end{eqnarray}
The integration in the l.h.s. is non-zero for $\ell=1$ and then proportional to $\delta_{ik}$, so that the l.h.s. is then equivalent to a term behaving as $S\delta_{\ell 1}/4\pi$, with $S$ a source term corresponding in fact to the dipolar state of $f$ at early times, outside the sphere of last diffusion scattering. The selection $\ell=1$ ensures that only this dipolar state matters. The integration in the r.h.s. can be substituted for an integration over $x=\hat{\mathbf{p}}_1\cdot\hat{\mathbf{p}}_2$ by inserting a Dirac function $\delta(x,\hat{\mathbf{p}}_1\cdot\hat{\mathbf{p}}_2)$, the integration over $\hat{\mathbf{p}}_1$ and $\hat{\mathbf{p}}_2$ giving then a factor $8\pi^2$. In this way, the set of equations for the power spectrum simplifies to
\begin{eqnarray}
\label{eqn:eq_f1f2_bis}
S\delta_{\ell1} \simeq \sum_{k\geq0} k(k+1)\frac{2k+1}{2}\langle C_k \rangle \int_{-1}^1 \dif x~\nu_{\mathrm{r}}(x) P_\ell(x) P_k(x),
\end{eqnarray}
which can be inverted in the same way as the inverse of the matrix $\underline{M}$ was obtained in~\S~\ref{subsec:cl}:
\begin{eqnarray}
\label{eqn:final_cl}
\langle C_\ell \rangle \simeq \frac{3}{2}\frac{S}{\ell(\ell+1)}\int_{-1}^1 \dif x~\frac{xP_\ell(x)}{\nu_{\mathrm{r}}(x)}.
\end{eqnarray}
This is the searched expression, obtained in~\cite{AhlersMertsch2015}.

Assuming $\nu_{\mathrm{r}}(x)\propto(1-x)^{1/2}$ as expected from the relative velocities of two particles emitted with close momenta, and for large modes $\ell$, the $\langle C_\ell \rangle$ coefficients are found to behave as $\ell^{-3}$, which is remarkably close to the observations. Applying the diffusion approximations to the second order moments of the phase-space density is thus found to provide a workable framework to describe data. \\

Many other suggestions have been put forward to explain the structures observed on intermediate and small scales. Among these mechanisms, an interesting possibility for producing small-scale anisotropies in a narrow energy interval around TeV energies relies on the electric field induced by the motion of the heliosphere relative to the plasma rest frame where the electromagnetic field can be considered as purely magnetic (due to the high conductivity of the medium)~\cite{Drury2013}. The anisotropies would then be due to the slight changes of energies due to the retarding or accelerating electric fields acting on a distance scale such that the magnetic deflections remain negligible in a particular direction. Specifically, in the case of the heliosphere whose length scale is of the order of a hundred astronomical units, the effective potential shift due to the induction fields is of order $-\int\dif s VB$, with $B$ of the order of a few nT and a typical velocity $V$ of order of $10^{-4}$--$10^{-5}~$m~s$^{-1}$, that is, of the order a few hundreds of MV. Hence, for TeV protons almost undeflected by the magnetic field in the heliosphere, a complex pattern should result from this effect with small-scale anisotropy contrasts at the level of $10^{-4}$. The power spectrum at large modes $\ell$ measured at TeV energies is thus inline with this expectation, and CRs may thus provide relevant signatures to probe and study the electric field induced by the motion of the heliosphere relative to the plasma rest frame. However, this effect may be hard to disentangle from the previous one, unless some bump at larger modes than currently accessible can be captured on top of a general concave spectrum.

\subsection{Searches at high energies}
\label{subsec:search_cl}

\begin{wrapfigure}{L}{8. cm}
{\includegraphics[width=0.5\textwidth]{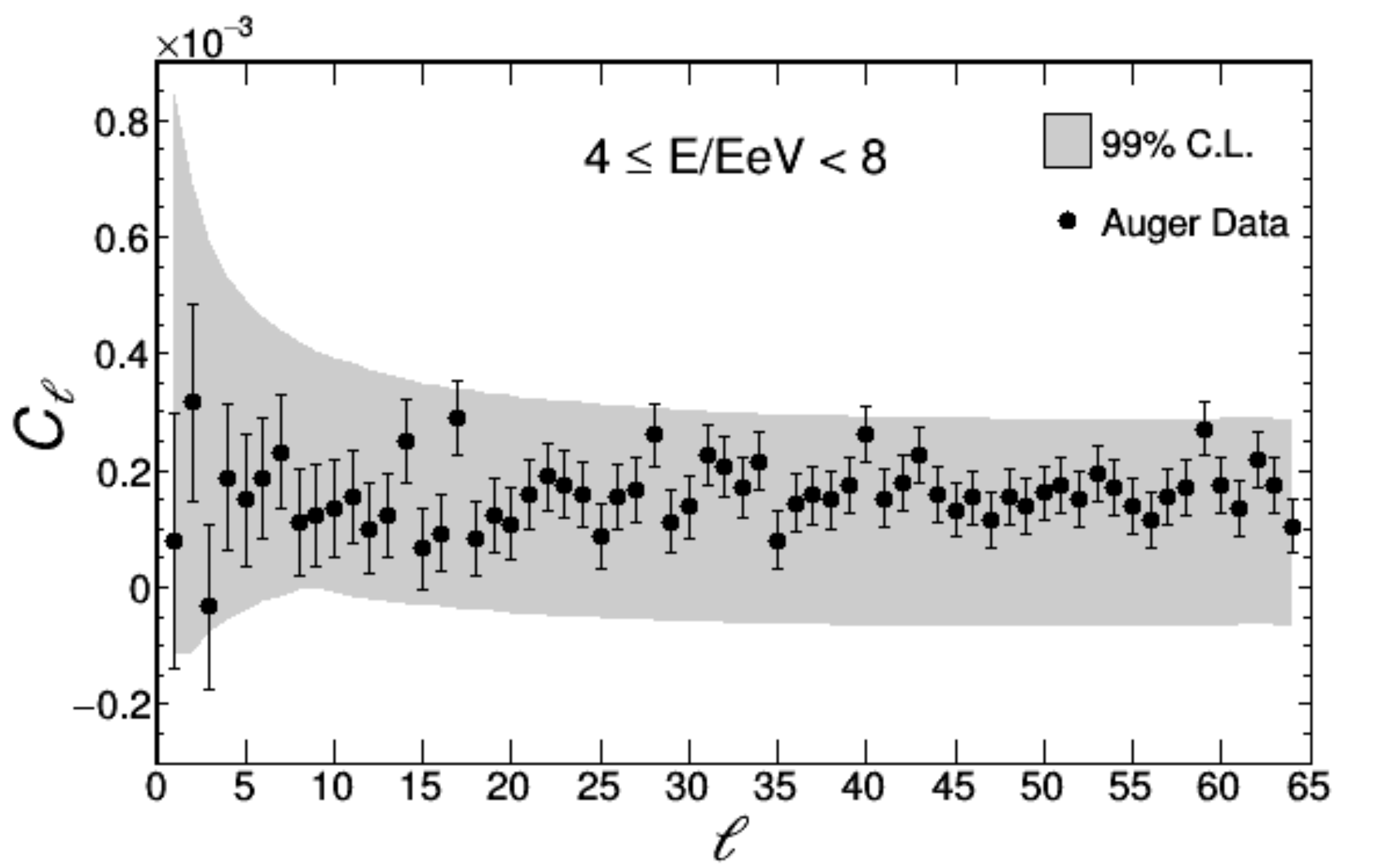}}
\caption{Angular power spectrum measured between 4 and 8~EeV at the Pierre Auger Observatory~\cite{AugerJCAP2017}.}
\label{fig:power_auger_4_8}
\end{wrapfigure}
At higher energies, searches have also been conducted to capture signals at intermediate and small scales through the measurements of angular power spectra at the Pierre Auger Observatory. One example is shown in figure~\ref{fig:power_auger_4_8}, in the energy range between 4 and 8~EeV~\cite{AugerJCAP2017}. No significant signal is currently observed. There are obviously much less events cumulated so far in this range compared to the TeV--PeV one, which explains the different range of probed power on the $y-$axis in this figure compared the figure~\ref{fig:powerspectrum_icecube}. Note that, in contrast to the results discussed previously, the directional exposure used in these studies is truly a function of both right ascension and declination (following equation~(\ref{eqn:direxpo_sat_bis})), so that the reconstructed anisotropy patterns are not absorbed along the declination. 

To improve the sensitivity of these measurements at high energy, full-sky coverage has been achieved above 10~EeV by combining data from the Auger Observatory located in the southern hemisphere and the Telescope Array located in the northern one~\cite{AugerTA2014}. Thanks to the full-sky coverage, the measurement of the power spectrum does not rely on any assumption on the underlying intensity of CRs. The combination of data from both experiments is however not as straightforward as a simple sum of events and exposures. For instance, because of the steepness of the UHECR spectrum, even a minor systematic difference in the energy scales of the two detectors can result in a significant spurious North-South anisotropy. This implies an unavoidable uncertainty in the relative exposures of the experiments. For that reason, individual exposures have to be re-weighted by some empirical factor $b$ to obtain the full-sky directional exposure:
\begin{equation}
\label{eqn:def_b}
\bar{\mu}(\nn)=\mu_1(\nn)+\bar{b}\mu_2(\nn).
\end{equation}
Written in this way, $b$ is a dimensionless parameter of order unity arbitrarily chosen to re-weight the directional exposure of one experiment relative to the other one. The parameter $b$ can thus be viewed as an effective correction which absorbs any kind of systematic uncertainties in the relative exposures, whatever the sources of these uncertainties. In practice, only an estimation $\bar{b}$ of the factor $b$ can be obtained, so that only an estimation of the directional exposure $\bar{\mu}(\nn)\equiv \mu(\nn,\bar{b})$ can be achieved. Following the methodology designed in~\cite{DiMatteo2016}, an estimation of $\bar{\mu}$ with $\bar{b}=1$ is possible. This is a sensible choice, given that the systematic uncertainties of both experiments on their absolute energy scale are much larger than the ones on their exposure for full efficiency of triggering. This method can be applied to the meta analysis of any experiments performing in the same energy range, provided that these experiments have an overlap in their field of view. 

The guiding principle relies on exploiting the wide declination band $(-16^\circ\leq\delta\leq+45^\circ)$ where the two datasets overlap~\cite{DiMatteo2016}. Regardless of the true arrival direction distribution, within a region of the sky $\Delta\Omega$ fully contained in the field of view of both observatories, the weighted sum over observed events $\sum_{i=1}^{N_{\Delta\Omega}} 1/\mu(\nn_i)$ is an unbiased estimator of the total flux $\Phi_{\Delta\Omega}$ integrated above the considered energy threshold and restricted to $\Delta\Omega$, and should be the same for both experiments except for statistical fluctuations which follow from Poisson statistics:
\begin{equation}
\label{eqn:rms_flux_domega}
\sigma_{\Phi i}\simeq\left[\frac{N_{\Delta\Omega}}{\Omega_{\Delta\Omega i}}\int_{\Delta\Omega}\frac{\dif\nn}{\mu_i(\nn)}\right]^{1/2},
\end{equation}
with $\Omega_{\Delta\Omega i}$ the total exposure of each experiment $i$ in the sky region $\Delta\Omega$. The uncertainty on $\overline{b}=1$, $\sigma_b$, is then obtained from the p.d.f. of the ratio of the measured fluxes:
\begin{equation}
\label{eqn:pdf_b}
p_B(\overline{b})=\frac{1}{2\pi\sigma_{\Delta\Omega 1}\sigma_{\Delta\Omega 2}}\int\dif x~|x|\exp{\left(-\frac{(x-\bar{\Phi}_{\Delta\Omega 1})^2\sigma^2_{\Delta\Omega 2}+(x\overline{b}-\bar{\Phi}_{\Delta\Omega 1})^2\sigma^2_{\Delta\Omega 1}}{2\sigma^2_{\Delta\Omega 1}\sigma^2_{\Delta\Omega 2}}\right)}.
\end{equation}
The uncertainty $\sigma_b$ propagates into uncertainty in the directional exposure function and thus in the inferred anisotropy parameters using the same recipe as in~\S~\ref{subsec:alm}. The first two moments of the observed angular distribution are here getting more complex due to the uncertainty on $\mu$:
\begin{equation}
\label{eqn:moments_with_b_1}
\left\langle \frac{1}{\bar{\mu}_{\mathrm{r}}(\nn)} \frac{\dif N}{\dif\nn} \right\rangle_{\mathrm{P}} \simeq \left\langle \frac{1}{\bar{\mu}_{\mathrm{r}}(\nn)} \right\rangle_{\mathrm{P}}\mu(\nn) I(\nn),
\end{equation}
\begin{equation}
\label{eqn:moments_with_b_2}
\left\langle \frac{1}{\bar{\mu}_{\mathrm{r}}(\nn)}\frac{1}{\bar{\mu}_{\mathrm{r}}(\nn')} \frac{\dif N}{\dif\nn}\frac{\dif N}{\dif\nn'} \right\rangle_{\mathrm{P}} = \left\langle \frac{1}{\bar{\mu}_{\mathrm{r}}(\nn)}\frac{1}{\bar{\mu}_{\mathrm{r}}(\nn')} \right\rangle_{\mathrm{P}} \left[\mu(\nn)\mu(\nn') I(\nn) I(\nn')+\mu(\nn) I(\nn)\delta(\nn,\nn')\right].
\end{equation}
For an unbiased estimator of $b$ with a resolution not larger than $\simeq$ 10\% (which is indeed the case in practice), the relative differences between $\langle 1/\bar{\mu}_\mathrm{r}\rangle_{\mathrm{P}}$ and $1/\mu_\mathrm{r}$ are actually not larger than $10^{-3}$. By propagation, the variance $\sigma^2_{\ell m}$ on each $a_{\ell m}$ multipole is then, to first order, the sum of a first term reflecting the usual Poisson fluctuations induced by the finite number of events and of a second term reflecting the uncertainty in the relative exposures of the two experiments~\cite{AugerTA2014}:
\begin{equation}
\label{eqn:rms_alm_augerta}
\sigma^2_{\ell m}\simeq\frac{4\pi f_1N}{\Omega_0^2}\int\dif\nn\left\langle\frac{\mu_\mathrm{r}(\nn)}{\bar{\mu}_\mathrm{r}^2(\nn)}\right\rangle_\mathrm{P} Y_{\ell m}^2(\nn)+\frac{N^2}{\Omega_0^2}\int\dif\nn\dif\nn'\left[\left\langle\frac{\mu_\mathrm{r}(\nn)\mu_\mathrm{r}(\nn')}{\bar{\mu}_\mathrm{r}(\nn)\bar{\mu}_\mathrm{r}(\nn')}\right\rangle_\mathrm{P}-1\right]Y_{\ell m}(\nn)Y_{\ell m}(\nn').
\end{equation}
The second term mainly impacts the resolution in the dipole coefficient $a_{10}$, while it has a small influence on the quadrupole coefficient $a_{20}$ and a marginal one on higher order moments $\{a_{\ell 0}\}_{\ell\geq3}$. The ``price'' of the methodology to perform the meta-analysis of different experiments is thus essentially a degraded resolution of the dipole moments compared to the resolution that would be obtained with a single experiment with full-sky coverage. 

\begin{figure}[!h]
\centering\includegraphics[width=0.49\textwidth]{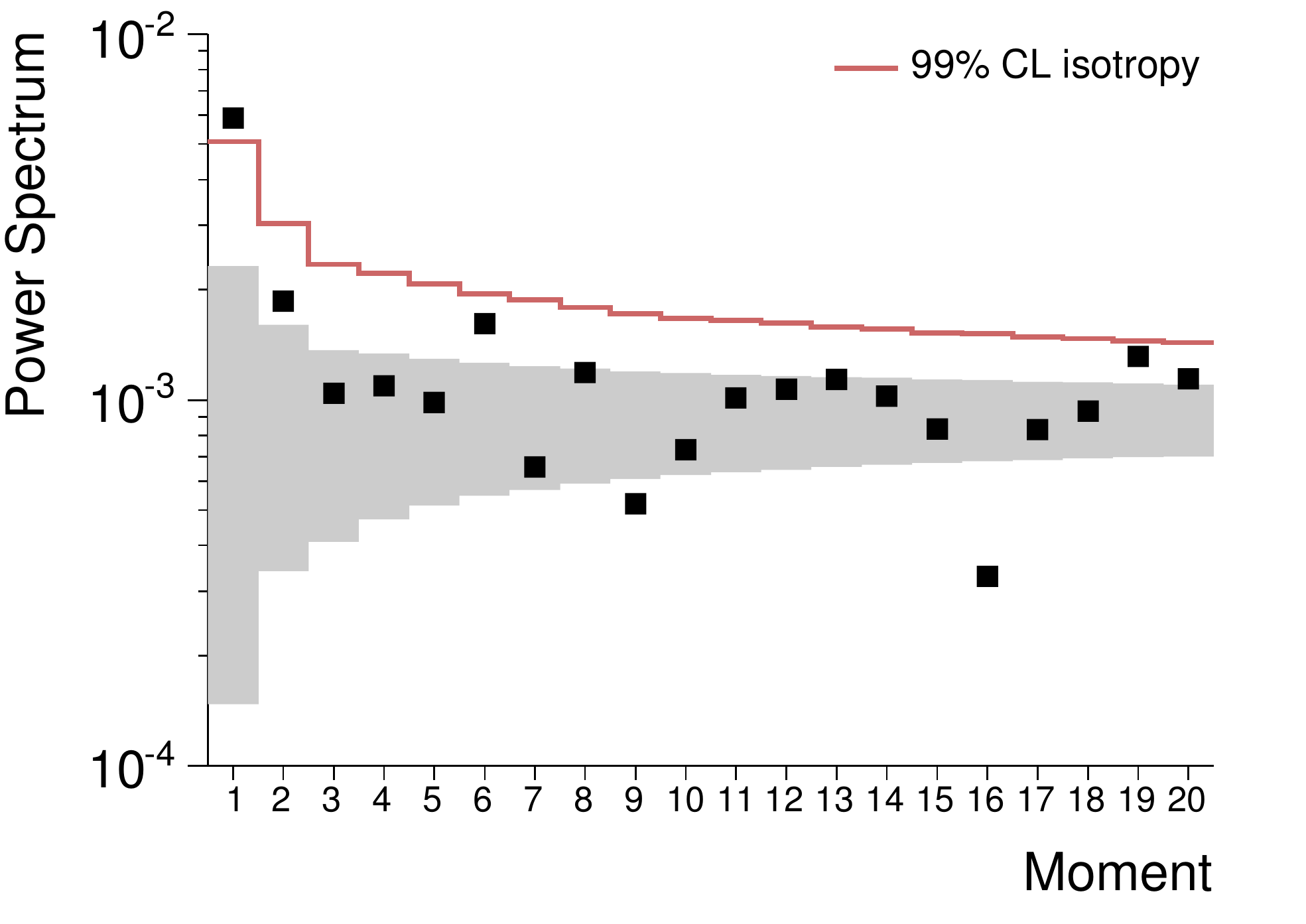}
\centering\includegraphics[width=0.49\textwidth]{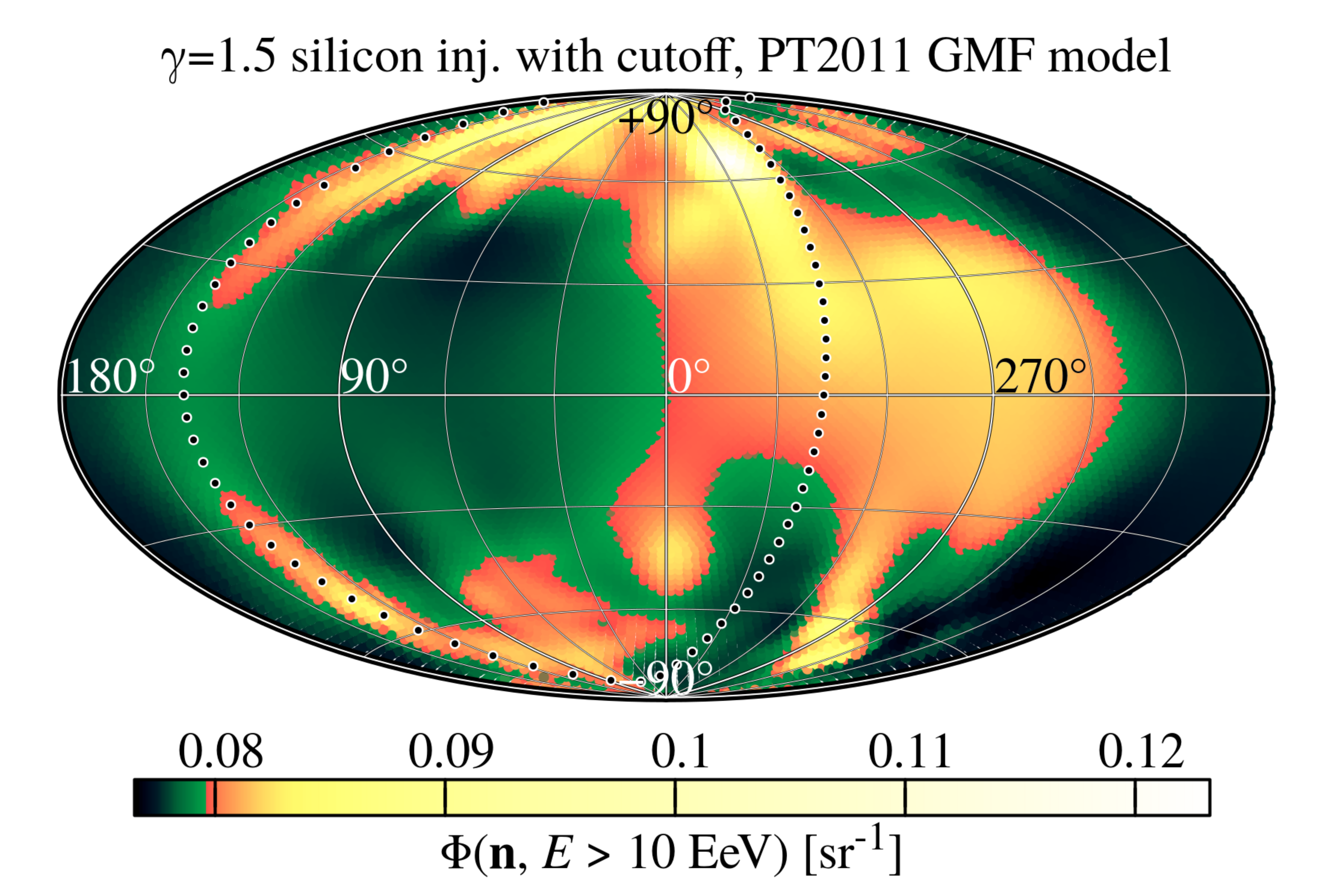}
\caption{Left: angular power spectrum as derived from Auger and Telescope Array data above $\simeq 10~$EeV~\cite{Deligny2015}. The gray band stands for the RMS of power around the mean values expected from an isotropic distribution, while the solid line stands for the 99\% confidence level upper bounds that would result from fluctuations of an isotropic distribution. Right: Some generic predictions for the sky map above 10~EeV for silicon nuclei injected as $E^{-1.5}$ and originating from the XSCz catalog~\cite{DiMatteoTinyakov2017}. The Galactic magnetic field model here follows from~\cite{Tinyakov2011}.}
\label{fig:powerspectrum_augerta}
\end{figure}

The measured power spectrum is shown in the left panel of figure~\ref{fig:powerspectrum_augerta}~\cite{Deligny2015}. The gray band stands for the RMS of power around the mean values expected from an isotropic distribution, while the solid line stands for the 99\% confidence level upper bounds that would result from fluctuations of an isotropic distribution. The dipole moment is observed to stand out from the background noise. Beyond the dipole, no other multipole deviates from expected fluctuations at 99\% CL in an isotropic flux. As already emphasised in~\S~\ref{subsec:transeev}, large-scale anisotropies of UHECRs with energies in excess of $10~$EeV are closely connected to the sources and the propagation mode of extragalactic UHECRs. The most contributing sources could, depending on the poorly-known size of the angular deflections in extragalactic magnetic fields, give rise to multipoles beyond the dipole that could help to characterise better their distribution in the sky. An example of sky map above 10~EeV expected for silicon nuclei originating from the 2MASS Galaxy Redshift Catalog~\cite{Skrutskie2006} is shown in the right panel of figure~\ref{fig:powerspectrum_augerta}~\cite{DiMatteoTinyakov2017}. In addition to a dipolar pattern similar to the one discussed in~\S~\ref{subsec:transeev}, the relatively small magnetic deflections in the extragalactic medium do not wash out the structure of the super-galactic plane in this particular example. For energies high enough, the corresponding symmetric quadrupole pattern is even expected to be reachable with the current generation of detectors. Different assumptions on the source density and/or the distance of the closest source can lead, however, to smaller quadrupole and higher-order amplitudes~\cite{WittowskiKampert2017}. 

Overall, the existing power spectrum measurements already help in constraining the needed deflections allowing a ``hot spot'' from a dominant nearby source to be smeared out to give rise to a dipolar moment only~\cite{Harari2015,Sigl2017}. The detection of significant multipole moments beyond the dipole could provide further valuable constraints on the extragalactic source and propagation of UHECRs.

\section{Extragalactic patterns and searches for sources at ultra-high energies}
\label{sec:xgal}

In this last section, special attention is given to UHECRs. The energy spectrum and chemical composition measurements at the highest energies are valuable information in inferring properties of the acceleration processes at work in the Universe. However, the identification of the sources can only be achieved by capturing in the arrival directions a pattern suggestive in an evident way of a class of astrophysical objects. The results obtained above 8--10~EeV and presented in~\S~\ref{subsec:transeev} and~\S~\ref{subsec:search_cl} already provide important constraints, but arrival directions at still higher energies potentially contain more information due to the reduction of the horizon for UHECRs combined to the smaller magnetic deflections with increasing rigidities. This remains nonetheless a difficult task because of the small UHECR intensity. 

Lots of dedicated methodologies have been designed to challenge the null hypothesis of isotropy by picking up signals at intermediate- and/or small-scales in the regime of small statistics. Only a limited selection of results are reviewed below, exclusively based on searches conducted at the Pierre Auger Observatory and at the Telescope Array, whose cumulated exposure in each hemisphere and improved instrumentation compared to past experiments have allowed a jump in sensitivity in searching for anisotropies at the highest energies.

\subsection{Searches for hot spots}
\label{subsec:hot spots}

The reduction of the horizon of the highest energy particles, whatever their nature, implies an erasure of the contribution of remote sources. This provides a natural mechanism to suppress an unresolved isotropic ``background'', so that UHECRs should originate from the nearby Universe. With only foreground sources present, clusters of events could stand out from the isotropic background, depending on the source density. Up-to-date searches for clusterings at small and intermediate angular scales are presented here.

To search for over-densities of events over the exposed sky, the widely-used technique consists in building a smoothed sky map by attributing the observed number of events $n_i$ within a circular window with some specific radius to each sampled point on the exposed sky. For a total of $N$ events, the probability $p_i(n_i,\mu_i)$ of the observed number of events in each sample point $(\alpha_i,\delta_i)$ in equatorial coordinates is then computed from the cumulative binomial distribution by estimating the expected number of events $\mu_i$ for an isotropic distribution within each circular window:
\begin{eqnarray}
\label{eqn:mu_i}
\mu_i=\frac{N}{\int_{4\pi}\dif\nn~\mu(\nn)}\int_{\Delta\Omega_i}\dif\nn~\mu(\nn),
\end{eqnarray}
with $\Delta\Omega_i$ the subtended solid angle in $(\alpha_i,\delta_i)$. A significance sky map is then derived using generally the Li-Ma estimator~\cite{LiMa1983}, which allows a mimic of a Gaussian process in an approximated way and thus allows for estimating the significance from the observed and expected number of events only\footnote{Note, however, that such a conversion can also be done in a direct way from Gaussian correspondence tables.}. Conventionally, positive (negative) significances correspond to over-densities (under-densities). 

However, by not specifying \textit{a priori} the targeted regions of the sky where the excesses are searched for as well as the angular window radius and the energy threshold, the probability/significance sky map obtained in this way suffers from the numerous performed trials. In a simple situation in which each trial would be independent from every other, obtaining a probability as low as any specified threshold could always be reached by increasing the number of trials. Hence, the number of trials needs to be accounted to establish the $p$-value of an excess and its corresponding significance. In some sense, the $p$-value is the original probability ``penalized'' for the various scans performed on the parameters intervening in the analysis. 

In most cases of interest however, each trial is however not independent of every other, so that the application of the  Fisher's combined probability test would overestimate the penalty factor to a large extent. Thus, to calculate the $p$-value of an apparent excess of events, the brute force is used. Monte-Carlo samples are generated to mimic the procedure applied to the analyzed data set, hence reproducing the various correlations between each trial. The mock samples are generated following an isotropic distribution folded into the directional exposure of the considered experiment. The same number of events as in the actual data is generated. The same energy distribution as in the actual data is used as well. On each of these mock samples, the set of scanned parameters of the actual data is optimized to capture the most significant excess anywhere on the sampled grid of the exposed sky. The searched $p$-value characterising the excess is then the number of samples yielding to more significant excess anywhere in the scanned parameter space normalized to the total number of generated samples~\cite{Tinyakov2004}.

\begin{wrapfigure}{R}{8. cm}
{\includegraphics[width=0.5\textwidth]{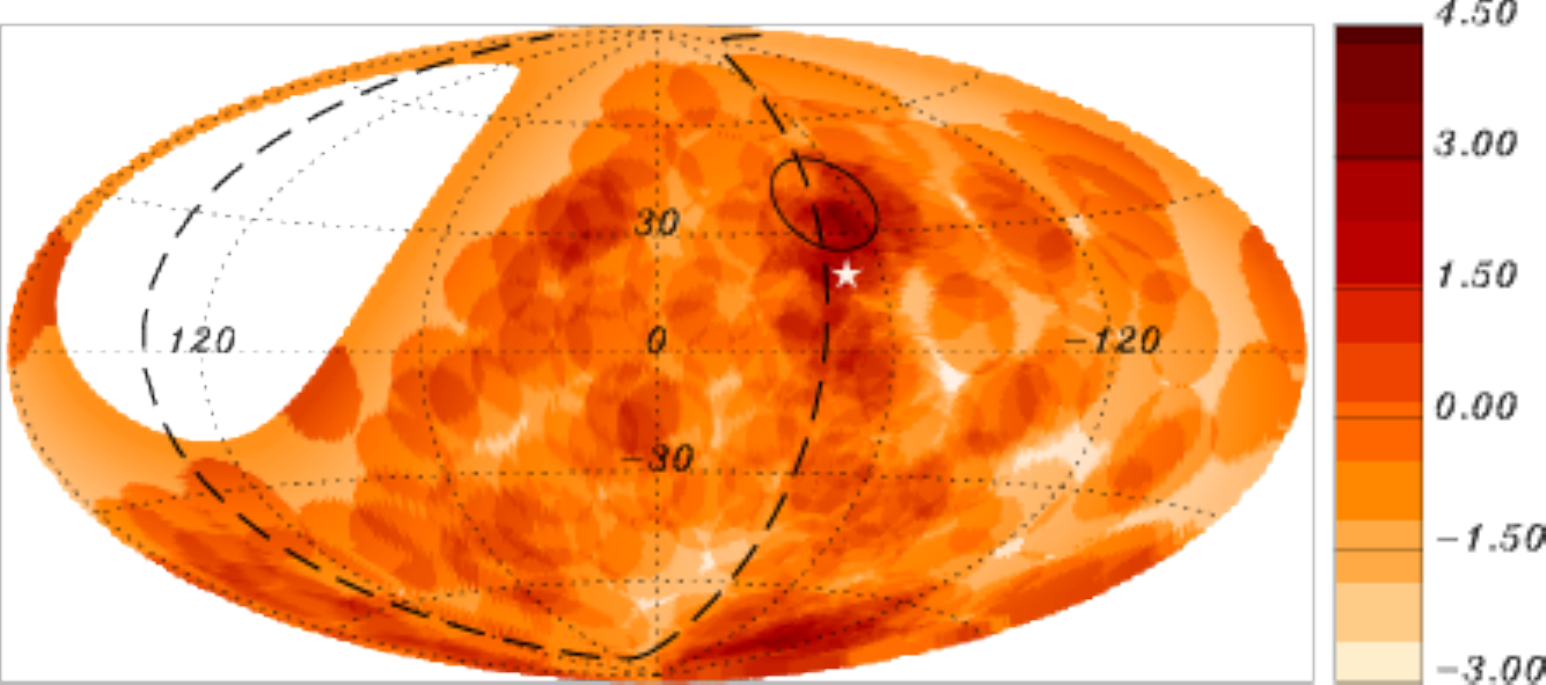}}
\caption{Map in Galactic coordinates of the (pre-trial) Li-Ma significances of over-densities in 12$^\circ$-radius windows for the events with energy in excess of 54~EeV as observed at the Pierre Auger Observatory~\cite{AugerApJ2015}.}
\label{fig:auger_hotspotsearch}
\end{wrapfigure}
The most up-to-date search for over-densities performed at the Pierre Auger Observatory has been reported in~\cite{AugerApJ2015}, based on data recorded with a total exposure of $\simeq 66,400$~km$^2$~sr~yr. The exposed sky is sampled using circular windows with radii varying from 1$^{\circ}$ up to 30$^{\circ}$ in 1$^{\circ}$ steps, while the energy thresholds is varied from 40~EeV up to 80~EeV in steps of 1~EeV. The resulting (pre-trial) significance sky map is shown in Fig.~\ref{fig:auger_hotspotsearch} for energies in excess of 54~EeV in 12$^\circ$-radius windows, parameters leading to the maximal significance. The largest departure from isotropy, indicated with a black circle, leads to a pre-trial 4.3$\sigma$ effect and is centered at $(\alpha,\delta)=(198^\circ,-25^\circ)$, where 14 events are observed against 3.23 expected from isotropy. It is close to the Supergalactic plane (shown as the dashed line) and centered at about 18$^\circ$ from the direction of Centaurus A (shown as the white star), one of the closest radio-galaxies from the Milky Way with an active nucleus. Once penalized for the trials, the probability of this excess is
found, however, to be as large as 69\% so that the observed  over-density does not provide any statistically significant evidence of anisotropy.

\begin{wrapfigure}{L}{8. cm}
{\includegraphics[width=0.5\textwidth]{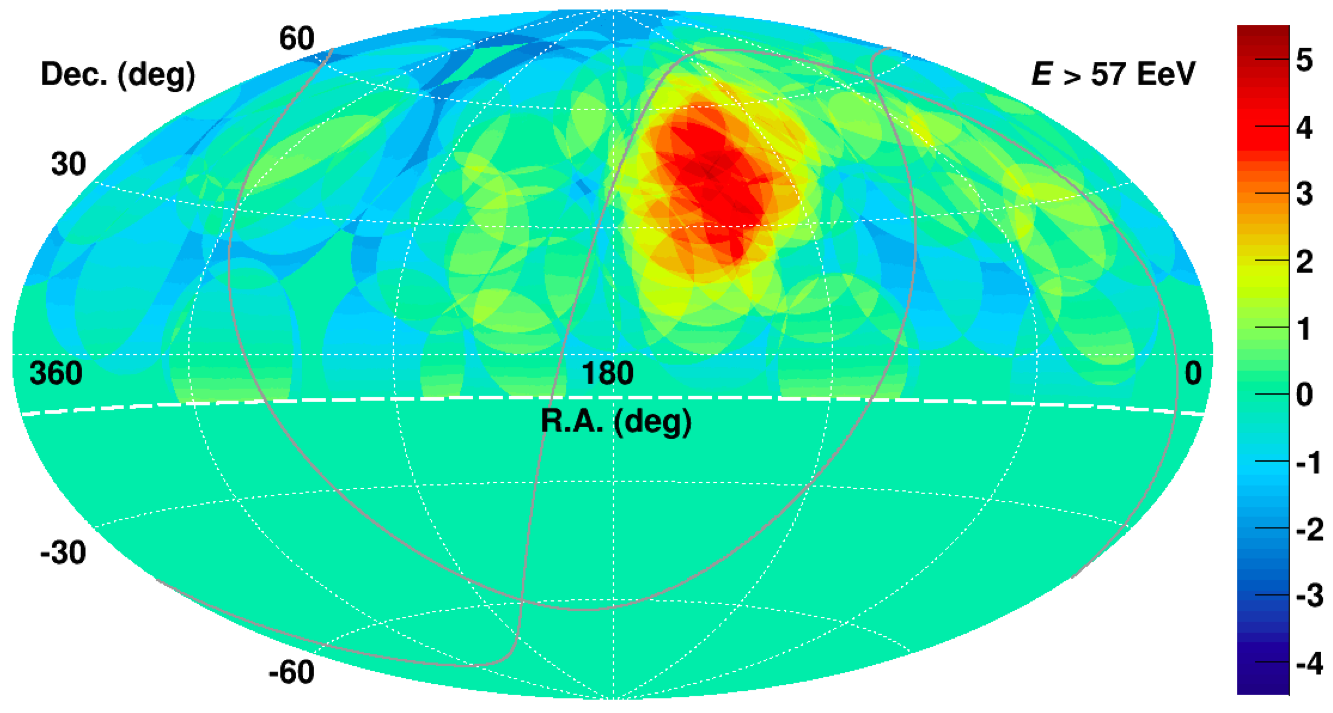}}
\caption{Map in equatorial coordinates of the (pre-trial) Li-Ma significances of over-densities in 25$^\circ$-radius windows for the events with energy in excess of 57~EeV as observed at the Telescope Array~\cite{Troitsky2017}.}
\label{fig:ta_hotspotsearch}
\end{wrapfigure}
Similar searches have been performed on data collected at the Telescope Array. One notable difference relies on the unique energy threshold used for this analysis, namely 57~EeV, selected from a previous analysis of arrival directions detected at the Pierre Auger Observatory which had initially led to establish an anisotropy at the 99\% confidence level~\cite{AugerScience2007} but which was not confirmed with subsequent data~\cite{AugerApJ2015}. In addition, only oversampling radii of 15, 20, 25, 30, and 35 degrees were used. The most up-to-date report makes use of data recorded from May 2008 to May 2017~\cite{Troitsky2017}. Out of the 143 selected events with zenith angles less than 55$^\circ$, an over-density of 34 events clustered within a circular window of 25$^{\circ}$ radius is observed around the equatorial coordinates $(\alpha,\delta)=(144.3^{\circ},40.3^{\circ})$, whereas 13.5 are expected in an isotropic distribution. The corresponding pre-trial significance for this ``hotspot'' is 5~$\sigma$, while the post-trial significance is 3~$\sigma$. A confirmation of this excess requires more statistics. In Fig.~\ref{fig:ta_hotspotsearch} is shown the corresponding (pre-trial) significance map of the excess. \\

A clustering of UHECRs at a certain angular scale might first reveal itself in the autocorrelation function of the events, which is a measure of the cumulative excess of event pairs separated by the given angular scale over the whole field of view, and not necessarily localized around a single reference point as in the approach described above. This test is potentially more sensitive than the blind search just presented in a situation when several small excesses of a similar angular size are present, because these excesses contribute then coherently in the autocorrelation function. Similarly to the blind searches, however, none of the searches could capture any significant excess~\cite{AugerApJ2015,Troitsky2017}. \\

\subsection{Correlations with nearby extragalactic matter}
\label{subsec:correlations}

Even without compelling indications for discrete sources, a correlation between the arrival directions of UHECRs and the positions of a class of astrophysical objects could more quickly reveal an anisotropy that would trace the sources. This is because even in the case of a quasi-isotropic distribution, the arrival directions could get stacked around some pre-defined directions. Numerous searches for such correlations with various catalogs of extragalactic objects have been conducted in the past. Here, following the recent report of the Pierre Auger Collaboration~\cite{AugerApJL2017}, the focus is given to a particular selection of non-thermal powerful emitters of gamma rays that were shown to be responsible for the extragalactic gamma-ray background measured by the \textit{Fermi}-LAT satellite from a hundred MeV up to hundreds of GeV~\cite{DiMauroDonato2015}. In addition, these particular objects constitute so far the sole extragalactic populations detected at TeV energies by ground-based instruments\footnote{This is excluding ``Galactic-like'' sources observed in the closeby Large Magellanic Cloud.}. The results of this particular study provide, as of today, the strongest indication for anisotropy in the UHECR arrival directions on an intermediate angular scale. 

Above 50~GeV, 360 galaxies hosting an actively-accreting super-massive black hole (commonly known as active galactic nuclei, referred to as $\gamma$AGNs hereafter) have been listed in the 2FHL catalog~\cite{Ackermann2016}. Selecting only the objects within a 250~Mpc radius results in a list of 17 bright nearby $\gamma$AGNs candidates. They mainly consist of radio-loud objects beamed towards the observer, except for the closest ones such as Cen~A and M~87. The question of the beaming of UHECR emission is uncertain. For a jet with a relativistic bulk motion (and for UHECRs produced in the relativistic jet like the gamma rays), UHECRs emitted isotropically in the bulk frame would appear to be coming out in the jet direction in our cosmic reference frame. In this case, the UHECR fluxes would be strongly enhanced in the jet direction along with the gamma rays, so that the population under study can be considered as a complete sample of $\gamma$AGNs for the potential brightest sources in the ``GZK sphere.''

The high gamma-ray luminosity of starburst galaxies is thought to be due to intense starburst episodes possibly triggered by galaxy mergers. Although not firmly established, the detected gamma rays have been argued to be produced by the decay of neutral pions originating from accelerated cosmic rays in interaction with the ambient environment~\cite{Abramowski2012}. Only a handful of starburst galaxies have been detected in the gamma-ray band. Among the 63 objects within 250~Mpc that have been searched for gamma-ray emission, only the 23 brightest objects with a radio flux larger than 0.3~Jy are considered. This is to evade as much as possible incompleteness effects. 

In addition to testing the arrival directions of UHECRs from the pre-defined positions of the objects in each catalog, the clustering in UHECRs is also tested by attributing to each source candidate a weight in proportion to its gamma-ray luminosity or its surrogate. This is reasonable for UHECRs and gamma rays originating from the same population of sources producing CRs at a similar rate from low energies to the highest ones, CRs which then undergo energy losses in calorimetric environments. Starburst galaxies are particularly good candidates to act as such ``cosmic calorimeters''~\cite{Waxman2015}, and to harbor with an increased rate cataclysmic events associated with the deaths of short-lived, massive stars, such as gamma-ray bursts, hypernovae, and magnetars. From the perspective of the energetics, and despite the small number of objects detected in the gamma-ray band, both nearby $\gamma$AGNs and starburst galaxies have a cumulated gamma-ray luminosity that safely match the required energy production rate observed in UHECRs~\cite{DermerRazzaque2010}, namely $\simeq 10^{45}~$erg~Mpc$^{-3}$~yr$^{-1}$ above 1 EeV~\cite{Farrar2015}. Hence, for the $\gamma$AGNs, the integral flux measured between 50~GeV and 2~TeV is used as a proxy for the UHECR flux. On the other hand, due to the too small number of starburst galaxies detected so far in the gamma-ray band, the continuum emission at 1.4~GHz is used for the proxy UHECR flux. This is because the gamma-ray luminosity has been observed to scale almost linearly with this continuum~\cite{Ackermann2012}. 

Each model is tested against the null hypothesis of isotropy through an unbinned maximum-likelihood analysis. Following~\cite{AugerAPP2010}, the likelihood $\mathcal{L}$ is defined as the product over the UHECR arrival directions of the model density in every UHECR directions. Smoothed density maps $F(\nn)$ are constructed through superposition of $N_{\mathrm{cat}}$ Fisher-Von Mises distributions $\mathcal{V}(\nn,\nn_i;\Theta)$ centered at each object position $\nn_i$ with an angular width $\Theta$ and weighted by the electromagnetic flux $\Phi_i$ discussed above and by the directional exposure of the Pierre Auger Observatory:
\begin{equation}
\label{eqn:likelihood}
F(\nn;f_{\mathrm{sig}},\Theta)\propto \mu(\nn)\left((1-f_{\mathrm{sig}})+f_{\mathrm{sig}}\sum_{i=1}^{N_{\mathrm{cat}}}\Phi_iw(z_i)\mathcal{V}(\nn,\nn_i;\Theta)\right).
\end{equation}
There are two parameters left free: $\Theta$, which is a search radius common to all sources that accounts in an effective way for the magnetic deflections, and $f_{\mathrm{sig}}$, which is the searched signal fraction pinpointing, if significantly non-zero, that a contribution from the considered astrophysical sources is preferred to a purely diffuse distribution. The fraction $1-f_{\mathrm{sig}}$ which is left stands for either faint unresolved objects absent from the considered catalog or from highly deflected nuclei; $f_{\mathrm{sig}}=0$ corresponds to the null hypothesis, that is, to the density map of isotropy. The weights $w(z_i)$ attributed to the $i$th source located at redshift $z_i$ stand for an attenuation factor due to the GZK suppression, evaluated as the fraction of the events produced above a given energy threshold which are able to reach the Earth from a source at a redshift $z$ with an energy still above that same threshold. These weights depend on the assumed composition and injection spectrum at the sources; models that have been found to reproduce the composition and spectral constraints are used~\cite{AugerJCAP2017b}. Finally, an overall constant guarantees that $\int\dif\nn~F=1$. The test statistic (TS) for deviation from isotropy is then the likelihood ratio test between any UHECR sky model and the null hypothesis, $2\ln{(\mathcal{L}/\mathcal{L}_0)}$. The TS is maximized as a function of the two free parameters and of the energy threshold ranging from 20 to 80~EeV. For a given energy threshold, the TS for isotropy follows a $\chi^2$ distribution with two degrees of freedom (Wilks theorem). The scan in energy thresholds induces a penalty factor which is estimated by means of Monte-Carlo simulations.
 
\begin{figure}[!h]
\centering\includegraphics[width=0.99\textwidth]{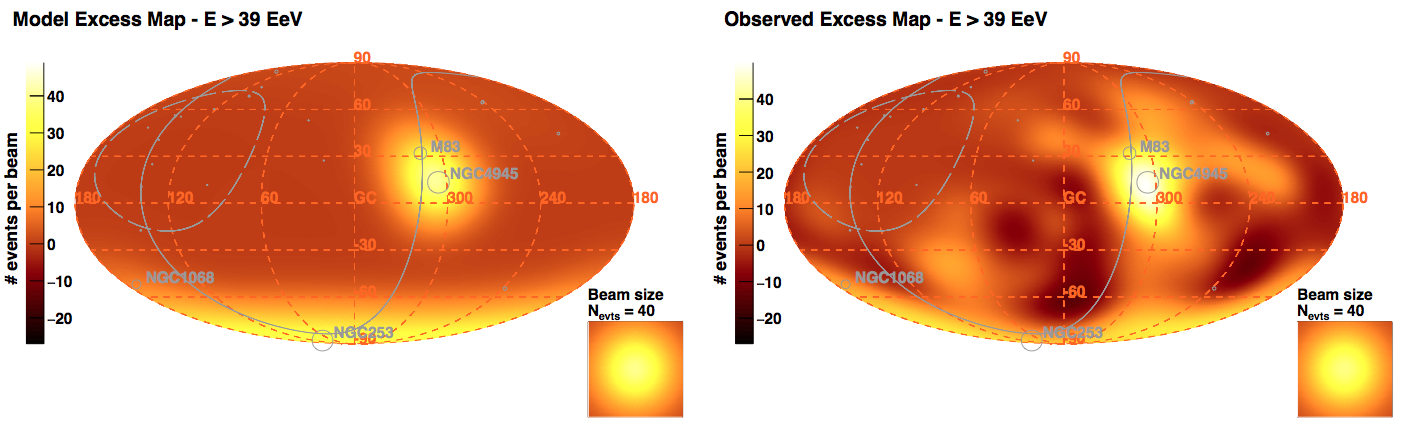}
\caption{Model excess map (left) and observed excess map (right) in Galactic coordinates for the best-fit parameters obtained with starburst galaxies above 39~EeV. The maps are background subtracted and smeared at the best-fit angular scale.}
\label{fig:skymap_sbg_auger}
\end{figure}
For $\gamma$AGNs, a 2.7$\sigma$ excess is found above 60~EeV at an angular scale of $6.8^{+4.0\circ}_{-2.2}$ and for a $6.7^{+4.5}_{-3.5}$\% fraction of anisotropic events. A stronger deviation from isotropy, significant with 4.0~$\sigma$ confidence, is captured in the case of starburst galaxies above 39~EeV at an intermediate angular scale of $12.9^{+4.1\circ}_{-3.0}$, corresponding to a fraction of $9.7^{+3.9}_{-3.8}$\% of UHECR events. The different attenuation factors corresponding to the different composition/injection spectrum scenarios have a pronounced effect for $\gamma$AGNs, moving the 2.7$\sigma$ excess down. In contrast, since the brightest starburst galaxies contributing to the signal are located 3 to 20~Mpc away, the different attenuation factors have a mild impact on the maximum deviation from isotropy. This model matches particularly well the UHECR clusters, as shown in figure~\ref{fig:skymap_sbg_auger}: the excess of events hinted in figure~\ref{fig:auger_hotspotsearch} usually attributed to Cen~A receives here joint contributions from M~83 and NGC~4945\footnote{And from the Circinus galaxy, absent from the catalog used here but classified as a starburst galaxy, when added to the list.}; and the excesses close to the Galactic south pole receive contributions from NGC~1068 and NGC~253.

The starburst galaxy model thus provides the most significant indication that UHECRs are not isotropically distributed on an intermediate angular scale. Additional contests between different scenarios favor the starburst model over the $\gamma$AGN model, a model including all AGNs detected by \textit{Swift}, or a model including all galaxies listed in the 2MASS redshift survey catalog~\cite{AugerApJL2017}. A few caveats are however important to be kept in mind at this stage. First, the reported significance pertains only to the specific test with starburst galaxies. This significance should be, in principle, penalised for the other tested scenarios with data from the Pierre Auger Observatory. Unavoidably, this would lead to some dilution of the 4$\sigma$ effect. A comprehensive reproduction of all the tests that were applied to the data in recognizing an intriguing pattern is impossible. There is consequently no rigorous way to correct the \textit{a posteriori} 4$\sigma$ effect by evaluating such a statistical penalty factor~\cite{Clay2003,Sommers2008}. Secondly, the strategy to account for magnetic deflections through a unique search radius centered on the source positions is advantageous for minimising the number of free parameter to one, but is not fully satisfactory given that the large-scale component of the Galactic magnetic field is expected to induce some offset between the brighter source positions and the UHECR hotspots. The search radius picked up in this analysis can be bracketed in simulations of light and intermediate nuclei originating from starburst galaxies and propagated into Galactic magnetic field models~\cite{AugerApJL2017}, but an unambiguous identification of the sources requires going beyond the strategy adopted here. Improving the density maps by including deflection models is however a difficult task~\cite{FarrarSutherland2017}. Finally, the signal fraction requires to be explained. For low-energy CRs (10--100~GeV), the energy release of CRs per logarithmic interval of energy is of order $10^{47}~$erg per solar mass of star formation in both the Milky Way and starburst galaxies, environments where the star formation rate differs by order of magnitude~\cite{Katz2013}. Interestingly, for redshifts $z<2$, starburst galaxies are responsible for $\simeq$ 15\% of the total star formation rate~\cite{Rodighiero2011,Sargent2012}. Proportionality between UHECR production rate and star formation rate may thus constitute an interesting framework to test the consistency of scenarios; further investigations are underway~\cite{Biteau2018}. 

\subsection{Perspectives}
\label{subsec:perspectives}

From an observational point of view, there are several perspectives to further investigate the scenarios underlying the production of UHECRs. 

Full-sky coverage is obviously advantageous to probe all possible sources. A recent status of the future orbital experiments that will cover the whole sky can be found in~\cite{Panasyuk2015}. First attempts to conduct such surveys have however already been undertaken by combining data from ground-based observatories~\cite{DiMatteo2016}. In the starburst scenario explored in~\cite{AugerApJL2017}, the galaxy M~82, located in the northernmost quarter of the (equatorial) sky well covered by the Telescope Array experiment, is expected to be one of the dominant sources. Interestingly, the excess of events shown in figure~\ref{fig:ta_hotspotsearch} has some overlap with the position of this starburst galaxy. 

In addition to the additional exposure that will be cumulated in the near future by the Auger and extended Telescope Array observatories, an instrumentation upgrade of the Pierre Auger Observatory, currently deployed, will provide mass-sensitive observables for each shower enabling charge-discriminated studies with a duty cycle of nearly 100\%~\cite{AugerTDR}. This could improve the sensitivity of the correlation searches, by excluding an eventual large fraction of highly charged nuclei from the analysis -- and thus by eliminating a quasi-isotropic ``background.''

Besides, multi-messenger approaches could also provide important insights. Several searches for correlations between the directions of UHECRs observed at the Pierre Auger Observatory and at the Telescope Array and those of very high-energy neutrino candidates detected by IceCube have been conducted~\cite{MM2016,AlSamarai2017}. No significant correlation is found, but the largest excess is at angular scales of $\simeq20^\circ$, arising mostly from pairs of events in the region of the sky where the Telescope Array has detected an excess of events and in regions close to the super-Galactic plane in correspondence with the largest excess observed in Auger data. Further insight will potentially arise from increased statistics and eventually with the inclusion of CR composition information that may become available so as to better model possible effects of magnetic field deflections. This may help to understand if there is a contribution in the astrophysical neutrino signal observed by IceCube correlated to the sources of the observed UHECRs. For transient events responsible for the production of UHECRs embedded in calorimetric environments such as starburst galaxies, however, an unambiguous identification of the sources through the association of a high-energy neutrino with an electromagnetic counterpart requires a tenfold increase in neutrino detector mass together with a wide monitoring program of transient events~\cite{Waxman2015}.

\section{Conclusion}
\label{sec:conclusion}

During the past decade, important observational results have been reported on the angular distributions of TeV--PeV CRs. While only dipolar excesses were expected, the myriad of reported anisotropies has led to important progresses on the understanding of the propagation regime of low-energy Galactic CRs. Overall, the information encompassed in the angular distributions appear today as a tool allowing a possible probe of the local magnetic field environments.

In contrast, the quest for finding UHECR sources is more difficult than expected a decade ago. Recent correlations with nearby extragalactic objets look promising. Future work will profit from the increased statistics and ability to perform anisotropy searches with distinction based on the mass of the primaries as anticipated with the upgraded instrumentation at the Pierre Auger Observatory. However, another jump in statistics appears necessary, keeping similar observable resolutions.








\end{document}